\documentclass[twocolumn, amsmath, amssymb, aps, notitlepage, nofootinbib,superscriptaddress]{revtex4-1}
\usepackage[normalem]{ulem}
\usepackage{bm,color,float,graphicx}
\usepackage{lipsum}
\usepackage{simplewick}
\usepackage{aas_macros}

\newcommand{\tn}[1]{\textcolor{magenta}{#1}}

\def\t{\tilde}
\def\m{\mathrm}

\def\h{\hat}
\newcommand{\expect}[1]{\langle #1\rangle}

\date{\today}
\begin{document}
\title{Impact of Anisotropic Birefringence on Measuring Cosmic Microwave Background Lensing}

\author{Hongbo Cai}
\email{hoc34@pitt.edu}
\affiliation{Department of Physics and Astronomy, University of Pittsburgh, Pittsburgh, PA, USA 15260}

\author{Yilun Guan}
\email{yilun.guan@dunlap.utoronto.ca}
\affiliation{Department of Physics and Astronomy, University of Pittsburgh, Pittsburgh, PA, USA 15260}
\affiliation{Dunlap Institute for Astronomy and Astrophysics, University of Toronto, Toronto, ON M5S 3H4, Canada}

\author{Toshiya Namikawa}
\affiliation{Kavli IPMU (WPI), UTIAS, University of Tokyo, Kashiwa, 277-8583, Japan}

\author{Arthur Kosowsky}
\affiliation{Department of Physics and Astronomy, University of Pittsburgh, Pittsburgh, PA, USA 15260}

\begin{abstract}
The power spectrum of cosmic microwave background lensing is a powerful tool for constraining fundamental physics such as the sum of neutrino masses and the dark energy equation of state.
Current lensing measurements primarily come from distortions to the microwave background temperature field, but the polarization lensing signal will
dominate upcoming experiments with greater sensitivity.
Cosmic birefringence refers to the rotation of the linear polarization direction of microwave photons propagating from the last scattering surface to us, which can be induced by parity-violating physics such as axion-like dark matter or primordial magnetic fields. We find that, for an upcoming CMB-S4-like experiment, scale-invariant anisotropic birefringence with an amplitude corresponding to the current $95\%$ upper bound can bias the measured lensing power spectrum by up to a factor of a few at small scales, $L\agt 1000$.
We show that the bias scales linearly with the amplitude of the scale-invariant birefringence spectrum.
The signal-to-noise of the contribution from anisotropic birefringence is larger than unity even if the birefringence amplitude decreases to around $5\%$ of the current upper bound.
Our results indicate that measurement and characterization of possible anisotropic birefringence is important for lensing analysis in future low-noise polarization experiments.
\end{abstract}

\maketitle

\section{Introduction}
\label{sec:introduction}
Cosmic microwave background (CMB) lensing refers to the deflection of CMB photons by the
matter distribution along their path from the last scattering surface
to the observer. This effect creates a displacement field on the microwave sky,
causing subtle distortions in both the CMB temperature
and polarization maps.
This displacement leads to correlations of CMB temperature and polarization fluctuations on a wide range of angular scales, which
can be used to reconstruct the lensing displacement field \cite{Hu:2002,Okamoto:2003}.
Precise measurements of CMB lensing probes the
matter distribution of the universe
and places constraints on cosmological parameters such as the amplitude of matter density fluctuations, $\sigma_8$,
and the energy density of matter, $\omega_m$, to which CMB lensing is
particularly sensitive.

CMB lensing has been measured with
increasing significance by several experiments, including Planck
\cite{PlanckCollaboration:2020,PlanckCollaboration:2014, carron:2022:lensing}, ACT \cite{vanEngelen:2015}, BICEP \cite{BICEP2Collaboration:2016}, Polarbear \cite{PolarbearCollaboration:2014,POLARBEARCollaboration:2017}, and SPT
\cite{vanEngelen:2012,Story:2015}, and will be measured with higher precision in the upcoming experiments like Simons Observatory \cite{SO:2019:SciGoal} and CMB-S4 \cite{S4:2016:SciBook}.
To date, lensing detection from temperature maps dominates the signal-to-noise for all of these experiments. As map noise levels decrease,
CMB polarization maps are eventually expected to provide higher signal-to-noise in upcoming surveys such as CMB-S4 \cite{2016arXiv161002743A}.

Compared to temperature lensing, polarization lensing is much less affected by various contaminants such as the thermal and kinematic Sunyaev–Zeldovich (tSZ, kSZ) effects, which may significantly bias temperature-based lensing reconstruction \cite{ferraro18_bias_to_cmb_lensin_recon, Cai:2021hnb} at  arcminute angular scales.
However, polarization-based lensing reconstruction can also be biased by effects such as instrumental systematics \cite{Mirmelstein:2020pfk,Nagata:2021}.
Another potential bias comes from cosmic birefringence, which refers to a rotation of CMB linear polarization plane caused by some parity-violating physics such as primordial magnetic fields or axion-like particles.  Cosmic birefringence can potentially be isotropic, with a constant rotation independent
of direction on the sky, or a direction-dependent anisotropic rotation.
Several recent analyses of Planck polarization data find a tantalizing hint of isotropic birefringence \cite{Minami:2020odp,Diego-Palazuelos:2022dsq,Eskilt:2022cff}. Cosmic birefringence therefore has gained growing interest; for a recent review see, e.g., \cite{Komatsu:2022nvu}).
Anticipating such a discovery, multiple recent papers have explored the potential of future CMB experiments to constrain the mass of axion-like particles \cite{Sherwin:2021,Nakatsuka:2022epj,Lee:2022udm,Gasparotto:2022uqo} and to probe models which produce both isotropic and anisotropic cosmic birefringence \cite{Capparelli:2020,Fujita:2020,Kitajima:2022jzz,Jain:2022jrp}.

To leading order, an anisotropic rotation field induced by cosmic birefringence
does not bias the reconstructed lensing field, thanks to the orthogonality between the lensing and rotation response in the CMB polarization field \cite{Kamionkowski:2009}. However, an anisotropic rotation field may potentially bias the reconstructed lensing power spectrum, 
a possibility which has not previously been considered in detail.
In this paper we examine such a bias in the presence of a scale-invariant rotation field.

The paper is structured as follows. In Sec.~\ref{sec:lensing and rotation} we
summarize the basics of CMB lensing and cosmic birefringence. In Sec.~\ref{sec:CMB lensing reconstruction} we review CMB lensing reconstruction on full-sky using the EB estimator and explain the motivation for investigating the bias to the reconstructed CMB lensing power spectrum.
Sec.~\ref{sec:simulation} presents simulation procedures and numerical results. We discuss the results in Sec.~\ref{sec:discussion} and conclude in Sec.~\ref{sec:conclusion}.

\section{Lensed and Rotated CMB Power Spectra}
\label{sec:lensing and rotation}
Both CMB lensing and anisotropic cosmic birefringence affect the CMB polarization field in a similar way. In this section we first give an overview of their effects on the CMB polarization field in \ref{subsec:overview}, followed by brief reviews of CMB lensing and cosmic birefringence in \ref{subsec:lensing} and \ref{subsec:rotation}, respectively. We then summarize some relevant results about power spectra in \ref{subsec:power spectra}.

\subsection{Overview}\label{subsec:overview}
Linear polarization of the CMB can be described by the Stokes parameters $Q(\bm{\hat{n}})$ and $U(\bm{\hat{n}})$, measured with respect to a set of local orthogonal polarizers, with $\bm{\hat{n}}$ an angular coordinate on the sky. The sky direction dependence of $Q$ and $U$ can be decomposed into spin-weighted spherical harmonics \cite{Goldberg:1967:spin} to obtain rotation-invariant quantities $E$ and $B$ as \cite{Kamionkowski:1996zd,Zaldarriaga:1997}
\begin{equation}
  \label{eq:EB lm}
  E_{\ell m} \pm iB_{\ell m} = -\int d^2\bm{\hat{n}}~{}_{\pm 2}Y^{*}_{\ell m} (Q \pm i U)(\bm{\hat{n}}),
\end{equation}
with $E_{\ell m}$ and $B_{\ell m}$ the multipole moments of E-mode and B-mode polarization, respectively. Their power spectra are defined as
\begin{equation}
  \begin{aligned}
    &\langle E_{\ell m}E^*_{\ell' m'}  \rangle = \delta_{\ell \ell'} \delta_{m m^{\prime}} C^{\m{EE}}_{\ell}\\
    &\langle B_{\ell m}B^*_{\ell' m'}  \rangle = \delta_{\ell \ell'} \delta_{m m^{\prime}} C^{\m{BB}}_{\ell}\\
        &\langle E_{\ell m}B^*_{\ell' m'}  \rangle = \delta_{\ell \ell'} \delta_{m m^{\prime}} C^{\m{EB}}_{\ell},\\
    \end{aligned}
\end{equation}
where the $\langle ... \rangle$ is taken over different statistical realizations of the CMB. The primary CMB polarization fields have no off-diagonal covariance between different $\ell$ and $m$ values.

The effects of both CMB lensing and cosmic birefringence can be interpreted as small perturbations added to the primary CMB field, effected by the lensing potential, $\phi(\bm{\hat{n}})$, for CMB lensing, and a rotation field, $\alpha(\bm{\hat{n}})$, for cosmic birefringence. 
In the presence of both CMB lensing and cosmic birefringence, the CMB polarization fields can be described as
\begin{eqnarray}
  \label{eq:rot lens E}
    \t{E}'_{\ell m}&=&E_{\ell m}+\delta \t{E}_{\ell m}+ \delta E'_{\ell m} + \mathcal{O}(\phi^{n_{1}} \alpha^{n_{2}}), \\
    \label{eq:rot lens B}
   \t{B}'_{\ell m}&=&\delta \t{B}_{\ell m}+ \delta B'_{\ell m} + \mathcal{O}(\phi^{n_{1}} \alpha^{n_{2}}),
\end{eqnarray}
where $\delta \t{E}_{\ell m}$, $\delta \t{B}_{\ell m}$ denote the first-order perturbation from CMB lensing, $\delta E'_{\ell m}$, $\delta B'_{\ell m}$ denote that from the rotation field\footnote{Note that we will follow this notational convention, denoting rotation induced quantities with prime and lensing-induced quantities with a tilde, throughout this paper.}, and $\mathcal{O}(\phi^{n_{1}} \alpha^{n_{2}})$, with $n_1+n_2>1$, represents the higher-order terms which mix the lensing and rotation effects. In the subsequent subsections we will discuss each of the terms in detail.

\subsection{CMB lensing}
\label{subsec:lensing}
CMB lensing distortion occurs when photons are deflected by gravitational
potentials and measures the integrated mass distribution along the
trajectories of photons (for a review, see, e.g. \cite{Lewis:2006fu}). Lensing distortion results in an effective displacement field $\bm{d}(\bm{\hat{n}})$ acting on the primary CMB fields,
\begin{equation}
\begin{split}\label{eq:lensing basic}
  \t{T}(\bm{\hat{n}}) &= T(\bm{\hat{n}}+\bm{d}(\bm{\hat{n}})),\\
  (\t{Q}\pm i\t{U})(\bm{\hat{n}}) &= (Q\pm iU)(\bm{\hat{n}}+\bm{d}(\bm{\hat{n}})).
\end{split}
\end{equation}
Under Born's approximation and the assumption that all CMB photons come from the last scattering surface, and neglecting non-linear effects, the displacement field $\bm{d}(\bm{\hat{n}})$ can be expressed as a pure gradient,
\begin{equation}
  \label{eq:phi and d}
  \bm{d}(\bm{\hat{n}}) = \nabla \phi(\bm{\hat{n}}),
\end{equation}
with $\phi(\bm{\hat{n}})$ the lensing potential field given by
\begin{equation}
  \label{eq:phi real}
  \phi(\bm{\hat{n}}) = -2 \int_{0}^{\chi_{\ast}} d\chi \frac{\chi_{\ast} -\chi}{\chi_{\ast}\chi} \Psi(\chi\bm{\hat{n}}, \eta_0-\chi)
\end{equation}
in a flat universe, with $\chi_{\ast}$ the conformal distance to the last scattering surface, $\eta_0$ the conformal time today,
and $\Psi$ the (Weyl) gravitational potential.

One can similarly decompose the lensing potential field in spherical harmonics as
\begin{equation}
  \label{eq:phi multipole}
\phi(\hat{\mathbf{n}})=\sum_{L M} \phi_{L M} Y_{LM}(\hat{\mathbf{n}}),
\end{equation}
with the multipole moments $\phi_{LM}$ related to its power spectrum as
\begin{equation}
  \label{eq:phi ps}
\left\langle\phi_{LM} \phi_{L'M'}^{*}\right\rangle=\delta_{L L'} \delta_{M M'} C_{L}^{\phi \phi}.
\end{equation}
Using the Limber approximation, the power spectrum $C_L^{\phi\phi}$ can be expressed as
\begin{equation}\label{eq:clpp}
C_{L}^{\phi \phi} \approx \frac{8\pi^2}{L^3} \int_{0}^{\chi_{\ast}} \chi d\chi \left(\frac{\chi_{\ast} -\chi}{\chi_{\ast}\chi}\right)^2 P_{\Psi}\left(\frac{L}{\chi};\eta_0-\chi\right).
\end{equation}
Here $P_{\Psi}(k;\eta)$ is the power spectrum of the gravitational potential, which is connected to the matter power spectrum $P_m(k;\eta)$ by
\begin{equation}
  P_{\Psi}(k;\eta) = \frac{9\Omega^2_{m}(\eta)H^4(\eta)}{8\pi^2} \frac{P_{m}(k;\eta)}{k},
\end{equation}
with $\Omega_{m}(\eta)$ the fractional matter energy density and $H(\eta)$ the Hubble parameter at conformal time $\eta$.
Hence, CMB lensing encodes information about the matter distribution, especially at late times (as the integrand in Eq.~\eqref{eq:clpp} peaks at $z \approx 2$). 
Measuring CMB lensing therefore constrains cosmological parameters such as the matter fluctuation amplitude $\sigma_8$, the matter density $\Omega_{\m{m}}$, the spatial curvature of the universe $\Omega_k$, the sum of neutrino masses $\sum m_{\nu}$, and the dark energy equation of state $w$, as previously demonstrated in, e.g., \cite{Sherwin:2017, vanEngelen:2012}.

A promising way to measure CMB lensing is through its effect on CMB polarization field \cite{Hu:2002, Okamoto:2003}. Following from Eq.~\eqref{eq:lensing basic}, CMB lensing leads to perturbations in CMB E-mode and B-mode polarization fields, given by
\begin{equation}
  \label{eq:1st E lens}
  \begin{aligned}
\delta \t{E}_{\ell m}=\sum_{LM} \sum_{\ell' m'}(-1)^{m} \phi_{LM} &\left(\begin{array}{ccc}
\ell & L & \ell' \\
-m & M & m'
\end{array}\right)\\
&\times {}_{2}F^{\mathrm{\phi}}_{\ell L \ell'}\epsilon_{\ell L \ell'}E_{\ell' m'}
\end{aligned}
\end{equation}
and
\begin{equation}
   \label{eq:1st B lens}
   \begin{aligned}
\delta \t{B}_{\ell m}=\sum_{LM} \sum_{\ell' m'}(-1)^{m} \phi_{LM} &\left(\begin{array}{ccc}
\ell & L & \ell' \\
-m & M & m'
\end{array}\right) \\
&\times {}_{2}F^{\phi}_{\ell L \ell'} \beta_{\ell L \ell'}E_{\ell' m'}
\end{aligned}
 \end{equation}
to leading order in $\phi$, where $\epsilon_{\ell L \ell'}$ and $\beta_{\ell L \ell'}$ are parity terms defined as
\begin{equation}
  \label{eq:parity}
  \begin{aligned}
&\epsilon_{\ell L \ell'}\equiv\frac{1+(-1)^{\ell+L+\ell'}}{2}, \\
\quad &\beta_{\ell L \ell'}\equiv\frac{1-(-1)^{\ell+L+\ell'}}{2 i},
\end{aligned}
\end{equation}
and the function ${}_{2} F^{\phi}_{\ell L \ell'}$ is defined as 
\begin{equation}
  \label{}
  \begin{aligned}
~{}_{2} F^{\phi}_{\ell L \ell'}\equiv&\left[-\ell\left(\ell+1\right)+L(L+1)+\ell'\left(\ell'+1\right)\right]\\
\times &\sqrt{\frac{\left(2 \ell+1\right)(2L+1)\left(2 \ell'+1\right)}{16 \pi}}\left(\begin{array}{ccc}
\ell & L & \ell' \\
2 & 0 & -2
\end{array}\right).
\end{aligned}
\end{equation}
The perturbation to the CMB polarization fields introduces off-diagonal covariance in and between E-mode and B-mode polarization fields and therefore allows measuring the lensing potential by optimally weighting quadratic combinations of the CMB polarization fields (known as the quadratic estimator approach) \cite{Okamoto:2003}. In the next part we will see that rotations of the CMB polarization field have similar effects.

\subsection{Cosmic Birefringence}
\label{subsec:rotation}
In addition to the lensing distortion, CMB linear polarization may
also undergo rotation in propagating from the last scattering surface to us. This phenomenon is called cosmic birefringence. Cosmic birefringence can be caused by parity-violating physics in the early universe, such as axion-like particles coupling to photons through Chern-Simons interaction \cite{Carroll:1998, Li:2008tma, Marsh:2015xka}, more general Lorentz-violating physics beyond the Standard Model \cite{Leon:2016kvt}, and primordial magnetic fields through frequency-dependent Faraday rotation 
\cite{Kosowsky:1996yc, Harari:1996ac,Kosowsky:2004zh,2012PhRvD..86l3009Y,De:2013dra,Pogosian:2013dya}.
The rotated linear polarization field can be expressed as
\begin{equation}
\label{eq:rotation}
(Q'\pm iU')(\bm{\hat{n}}) = e^{\pm 2i\alpha(\bm{\hat{n}})}(Q \pm i U)(\bm{\hat{n}}),
\end{equation}
with $\alpha(\bm{\hat{n}})$ the rotation field. We use primes to represent the rotated quantities.

As a specific example, cosmic birefringence is induced by a Chern-Simons-type interaction between photons and axion-like particles in the early universe with a Lagrangian given by
\begin{equation}
  \label{eq:cs term}
  \mathcal{L}_{c s}=\frac{g_{a\gamma}a}{4} F^{\mu \nu} \tilde{F}_{\mu \nu},
\end{equation}
where $g_{a\gamma}$ is a dimensionless coupling constant, $a$ is the axion field, and $F^{\mu \nu}$ is the electromagnetic tensor with $\tilde{F}_{\mu \nu}$ being its dual. This term modifies the Euler-Lagrange equations for the electromagnetic field and induces a
rotation in the polarization direction of a photon if $a$ varies along its propagation path \cite{1997PhRvD..55.6760C, 1998PhRvD..58k6002C,Leon:2017}, with the rotation angle given by
\begin{equation}
  \label{eq:alpha and phi}
  \alpha(\hat{\bm{n}})=\frac{g_{a\gamma}}{2}\Delta a,
\end{equation}
where $\Delta a$ is the change of $a$ along the photon path from the last scattering surface to us.

A generic rotation field can be separated into an isotropic constant piece $\bar{\alpha}$ and an 
anisotropic part $\delta \alpha(\hat{\bm{n}})$ as
\begin{equation}
  \label{}
\alpha(\hat{\bm{n}})=\bar{\alpha}+\delta \alpha(\hat{\bm{n}})
\end{equation}
with
\begin{equation}
  \label{eq:rotation parts}
  \expect{\delta \alpha(\hat{\bm{n}})}=0.
\end{equation}
For example, some quintessence models predict both isotropic and anisotropic cosmic birefringence \cite{Caldwell:2011}, while some massless scalar fields do not necessarily induce isotropic cosmic birefringence \cite{Gluscevic:2009}.
Isotropic cosmic birefringence violates global parity symmetry and induces odd-parity CMB TB and EB power spectra, but 
does not produce any correlations between CMB polarization fields at different angular scales.

In the subsequent analysis we consider only anisotropic cosmic birefringence with $\bar{\alpha}=0$, as it behaves similarly to CMB lensing which is also anisotropic. 
Similar to Eq.~\eqref{eq:phi multipole}, we can decompose the rotation field into multipole moments using spherical harmonics as
\begin{equation}
  \label{}
\alpha(\hat{\mathbf{n}})=\sum_{L M} \alpha_{L M} Y_{LM}(\hat{\mathbf{n}}),
\end{equation}
and the rotation field power spectrum is given by
\begin{equation}
  \label{}
\left\langle\alpha_{LM} \alpha_{L'M'}^{*}\right\rangle=\delta_{L L'} \delta_{M M'} C_{L}^{\alpha \alpha}.
\end{equation}
Anisotropies in the rotation field can result from inhomogeneities in the axion field, for example. If the axion field is seeded during inflation, it leads to a Gaussian random rotation field with a nearly scale-invariant spectrum at large scales ($L\lesssim 100$), with an amplitude connected to the inflationary Hubble parameter $H_{\m{I}}$:
\begin{equation}
  \label{eq:claa}
\frac{L(L+1) C_{L}^{\alpha \alpha}}{2 \pi}=\left(\frac{H_{\m{I}} g_{a \gamma}}{4 \pi}\right)^{2}\equiv A_{\mathrm{CB}},
\end{equation}
where we have defined a dimensionless amplitude $A_{\rm CB}$ of the scale-invariant rotation power spectrum. Current best constraints on $A_{\rm CB}$ are given by ACTPol \cite{Namikawa:2020} and SPTPol \cite{SPT:2020cxx} corresponding to a $2\sigma$ upper bound \footnote{Note that $A_{\rm{CB}}$ defined in this paper is $10^{-4}$ times of that in \cite{Namikawa:2020} and \cite{SPT:2020cxx}.} $A_{\m{CB}} \leq 10^{-5}$. With improved sensitivity in the next-generation ground-based CMB experiments, we expect the constraint of $A_{\rm CB}$ to reach the level of $10^{-7}$ \cite{CMB-HD:2022bsz, Pogosian:2019jbt, Mandal:2022tqu}.

In this paper we focus only on scale-invariant rotation field for conceptual simplicity, but the methodology is generally applicable to rotation field with any power spectrum. We expect the conclusion from other rotational power spectra to be qualitatively similar, but we leave a detailed quantitative
analysis of such cases to future work.

Similar to CMB lensing in Eq.~\eqref{eq:1st E lens}, a rotation field leads to perturbations in the CMB E-mode and B-mode polarization fields given by \cite{Kamionkowski:2009, Gluscevic_2009}
\begin{equation}
  \label{eq:1st E rotation}
  \begin{aligned}
  \delta E'_{\ell m} = -2\sum_{LM}\sum_{\ell'm'} (-1)^{m}\alpha_{LM} &\left(\begin{array}{ccc}
\ell & L & \ell' \\
-m & M & m'
\end{array}\right)\\
&\times{}_{2} F^{\alpha}_{\ell L \ell'} \beta_{\ell L \ell'} E_{\ell' m'},
\end{aligned}
\end{equation}
\begin{equation}
  \label{eq:1st B rotation}
  \begin{aligned}
  \delta B'_{\ell m} = 2\sum_{LM}\sum_{\ell'm'} (-1)^{m}\alpha_{LM} &\left(\begin{array}{ccc}
\ell & L & \ell' \\
-m & M & m'
\end{array}\right)\\
&\times{}_{2} F^{\alpha}_{\ell L \ell'} \epsilon_{\ell L \ell'} E_{\ell' m'},
\end{aligned}
\end{equation}
to leading order in $\alpha$ from Eq.~\eqref{eq:rotation}\footnote{A non-perturbative treatment also exists; see \cite{Li:2013vga} for an introduction.}, where we have assumed that $C_l^{\rm BB}$ is negligible before rotation. $\beta_{\ell L \ell'}$ and $\epsilon_{\ell L \ell'}$ are defined in Eq.~\eqref{eq:parity},
and ${}_{2} F^{\alpha}_{\ell L \ell'}$ is defined as
\begin{equation}
{}_{2} F^{\alpha}_{\ell L \ell'} \equiv \sqrt{\frac{\left(2 \ell+1\right)(2L+1)\left(2 \ell'+1\right)}{4\pi}}\left(\begin{array}{ccc}
\ell & L & \ell' \\
2 & 0 & -2
\end{array}\right).
\end{equation}
Note that the parity indicators in Eq.~\eqref{eq:1st E rotation} and Eq.~\eqref{eq:1st B rotation} are opposite to the ones in Eq.~\eqref{eq:1st E lens} and Eq.~\eqref{eq:1st B lens}, i.e., $\delta \t{E}$ and $\delta B'_{\ell m}$ are only non-zero when $\ell + L + \ell'$ is even and $\delta \t{B}$ and $\delta E'_{\ell m}$ are only non-zero when $\ell + L + \ell'$ is odd. In this sense, we say that CMB lensing and anisotropic cosmic birefringence are orthogonal at leading order \cite{Kamionkowski:2009}.

\begin{figure}[t]
\includegraphics[width=0.5\textwidth]{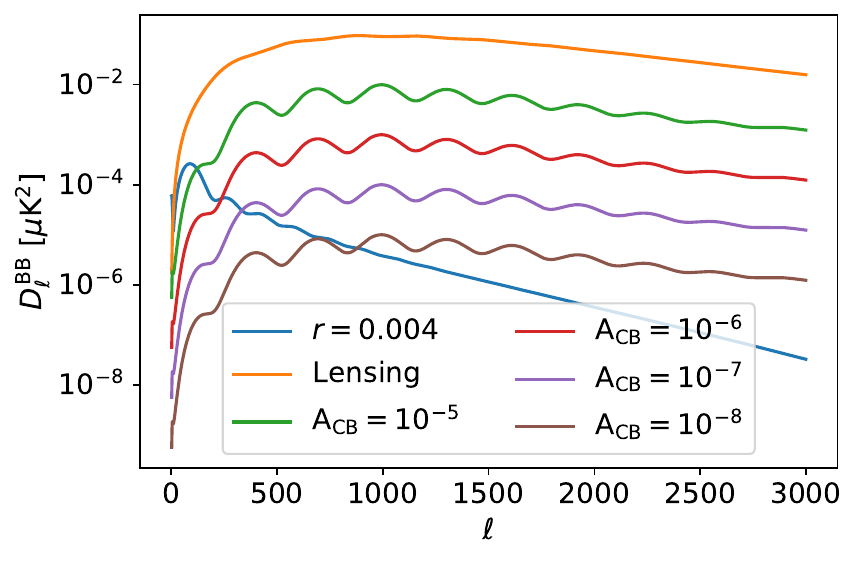}
\caption{
B-mode power spectrum contributions from a scale-invariant tensor mode ($r=0.004$), gravitational lensing, and 1st-order scale-invariant anisotropic birefringence with different amplitudes ($A_{\m{CB}}=10^{-5}$, $A_{\m{CB}}=10^{-6}$, $A_{\m{CB}}=10^{-7}$ and $A_{\m{CB}}=10^{-8}$) are shown, where we define $D^{BB}_\ell\equiv\ell(\ell+1)C_\ell^{BB}/2\pi$. All the power spectra shown above are generated by \texttt{class\_rot}  \cite{Cai:2021zbb}.}
\label{fig:cl_bb_all}
\centering
\end{figure}

\subsection{Summary remarks}
\label{subsec:power spectra}
Both CMB lensing and anisotropic cosmic birefringence lead to perturbations in the CMB E-mode and B-mode polarization fields. The resulting B-mode power spectrum reveals their relative contributions. In Fig.~\ref{fig:cl_bb_all}, we compare the B-mode power spectrum from a scale-invariant primordial tensor mode with $r=0.004$, gravitational lensing, and scale-invariant anisotropic rotation fields with different amplitudes ($A_{\m{CB}}=10^{-5}$, $A_{\m{CB}}=10^{-6}$, $A_{\m{CB}}=10^{-7}$, and $A_{\m{CB}}=10^{-8}$). Like CMB lensing, scale-invariant anisotropic rotation contributes 
predominantly to small-scale anisotropies in B-mode polarization, with a power spectrum generally orders of magnitude below the CMB lensing signal
for observationally allowed amplitudes.

In addition to contributing to the CMB polarization power spectra, both CMB lensing and rotation lead to off-diagonal covariance between E-mode and B-mode polarization fields \cite{Gluscevic_2009},
\begin{equation}
\label{eq:1st-order EB cross}
    \begin{aligned}
        {\langle E_{\ell m} \delta\t{B}_{\ell' m'} \rangle}_{\m{CMB}} &= \sum_{LM} \begin{pmatrix}
  \ell & \ell' & L\\
  m & m' & M
\end{pmatrix} f^{\phi}_{\ell L \ell'} \phi^{*}_{LM}, \\
\langle E_{\ell m}\delta B'_{\ell' m'} \rangle_{\mathrm{CMB}} &= \sum_{LM} \begin{pmatrix}
  \ell & \ell' & L\\
  m & m' & M
\end{pmatrix} f^{\alpha}_{\ell L \ell'} \alpha^{*}_{LM},
    \end{aligned}
\end{equation}
for $\ell \neq \ell'$ and $m \neq -m'$. $f^{\phi}_{\ell L \ell'}$ and $f^{\alpha}_{\ell L \ell'}$ are weight functions for $\phi$ and $\alpha$, given by
\begin{equation}
  \label{eq:phi and alpha weights}
  \begin{aligned}
      f^{\phi}_{\ell L \ell'} &= -\beta_{\ell L \ell'} ~{}_{2} F^{\phi}_{\ell L \ell'} C^{\m{EE}}_{\ell},\\
    f^{\alpha}_{\ell L \ell'} &= -2 \epsilon_{\ell L \ell'}~{}_{2} F^{\alpha}_{\ell L \ell'}C^{\m{EE}}_{\ell}.
  \end{aligned}
\end{equation}
Here $\langle ... \rangle_{\m{CMB}}$ is an ensemble average over different realizations of the primary CMB, with a fixed realization of both $\phi$ and $\alpha$.
It then follows that
\begin{equation}
  \label{eq:rotated-lensed average}
  \begin{aligned}
\langle \t{E}'_{\ell m}\t{B}'_{\ell' m'} \rangle_{\mathrm{CMB}} = &\sum_{LM} \begin{pmatrix}
  \ell & \ell' & L\\
  m & m' & M
\end{pmatrix} f^{\phi}_{\ell L \ell'} \phi^{*}_{LM} \\
+&\sum_{LM} \begin{pmatrix}
  \ell & \ell' & L\\
  m & m' & M
\end{pmatrix} f^{\alpha}_{\ell L \ell'} \alpha^{*}_{LM},
  \end{aligned}
\end{equation}
for $\ell \neq \ell'$ and $m \neq -m'$. Eq.~\eqref{eq:rotated-lensed average} shows that the off-diagonal covariance between E-mode and B-mode polarization is contributed by a sum of lensing and rotation contributions, to leading order\footnote{Note that Eq.~\eqref{eq:rotated-lensed average} does not contradict the fact that lensing and anisotropic rotation do not induce parity-odd power spectra, because the average is taken with fixed realizations of $\phi$ and $\alpha$; if, however, we also average over $\phi$ and $\alpha$ in Eq.~\eqref{eq:rotated-lensed average}, the resulting EB power spectrum will be zero.}.

\section{CMB Lensing reconstruction bias due to a rotation field}
\label{sec:CMB lensing reconstruction}
In this section, we introduce the CMB lensing reconstruction pipeline considered in this work. In particular, we focus on the EB estimator, as it is expected to be a dominant source of statistical power for lensing reconstruction in upcoming experiments \cite{Okamoto:2003}. We then discuss the motivation and our methodology for evaluating the bias to the estimated lensing power spectrum from a rotation field.

\subsection{Lensing reconstruction}
The lensing potential can be reconstructed from the off-diagonal covariance generated by CMB lensing in Eq.~\eqref{eq:rotated-lensed average}
using a quadratic estimator approach. An estimator for lensed E and B maps is given by \cite{Okamoto:2003}
\begin{equation}
  \label{eq:phi estimator}
  \begin{aligned}
\h{\phi}_{LM}=A_{L} \sum_{\ell m} \sum_{\ell' m'}(-1)^{M} &\left(\begin{array}{ccc}
\ell & \ell' & L \\
m & m' & -M
\end{array}\right) \\
&\times (f^{\phi}_{\ell L \ell'})^* \frac{\t{E}_{\ell m}}{\hat{C}^{\m{EE}}_{\ell}} \frac{\t{B}_{\ell' m'}}{\hat{C}^{\m{BB}}_{\ell'}},
\end{aligned}
\end{equation}
where $A_{L}$ is a normalization factor ensuring $\hat{\phi}_{LM}$ is unbiased, given by
\begin{equation}
  \label{}
A_{L}=\left(\frac{1}{2 L+1}\sum_{\ell \ell'} \frac{|f^{\phi}_{\ell L \ell'}|^2}{\hat{C}^{\m{EE}}_{\ell}\hat{C}^{\m{BB}}_{\ell'}}\right)^{-1},
\end{equation}
and $\h{C}^{\m{EE}}_{\ell}$ and $\h{C}^{\m{BB}}_{\ell}$ are the total observed EE and BB power spectra with noise power spectrum included. The quadratic estimator $\hat{\phi}_{LM}$ collects all the off-diagonal correlations of a CMB map realization and average them by a weight function $f^{\phi}_{\ell L \ell'}$.

{}From the reconstructed lensing potential, its power spectrum can be estimated as \cite{Sherwin:2017}
\begin{equation}
  \label{eq:phi ps estimation}
\h{C}_{L}^{\phi \phi} =  C_{L}^{\h{\phi} \h{\phi}} - {}^{(\m{RD})}N^{(0)}_{L} - N^{(1)}_{L} - N^{(\m{MC})}_{L} - N^{(\m{FG})}_{L},
\end{equation}
with $C_{L}^{\h{\phi} \h{\phi}}$ given by
\begin{equation}
  \label{eq:bandpower}
  C_{L}^{\h{\phi}\h{\phi}}=\frac{1}{2L+1} \sum_{M=-L}^{L}\left|\h{\phi}_{LM}\right|^{2}.
\end{equation}
The remaining terms in Eq.~\eqref{eq:bandpower} are noise biases that need to be subtracted: ${}^{(\m{RD})}N^{(0)}_{L}$ is the realization-dependent Gaussian noise (RDN0) \cite{Namikawa_2013} (Appendix~\ref{sec:appendix A} discusses the motivation to use this instead of the standard Gaussian noise $N^{(0)}_L$). $N^{(1)}_{L}$ represents the bias from connected terms in the CMB four-point function which contain the first-order lensing potential $C^{\phi\phi}_{L}$.
$N^{(\m{MC})}_{L}$ is the ``Monte-Carlo''(MC) noise encapsulating biases not accounted for otherwise, such as higher-order reconstruction noise estimated from MC simulations, and $N^{(\m{FG})}_{L}$ is the modeled foreground bias from extragalactic and galactic foregrounds like galactic dust, galaxy clusters and the cosmic infrared background.

\subsection{Bias to $C_L^{\phi\phi}$ from rotation}
Now consider a scenario in which we perform lensing reconstruction on a set of polarization maps which have been \textit{unknowingly} rotated: how will this affect our estimated $C_L^{\phi\phi}$? To understand this, we can define an effective estimator
\begin{equation}
  \label{eq:phi estimator}
  \begin{aligned}
\h{\phi'}_{LM}=A_{L} \sum_{\ell m} \sum_{\ell' m'}(-1)^{M} &\left(\begin{array}{ccc}
\ell & \ell' & L \\
m & m' & -M
\end{array}\right) \\
&\times (f^{\phi}_{\ell L \ell'})^* \frac{\t{E}'_{\ell m}}{\hat{C}^{\m{EE}}_{\ell}} \frac{\t{B}'_{\ell' m'}}{\hat{C}^{\m{BB}}_{\ell'}},
\end{aligned}
\end{equation}
which uses the rotated-lensed quantities $\t{E}'_{\ell m}$, $\t{B}'_{\ell m}$ in place of the lensed-only ones.

This estimator does not bias the reconstructed lensing potential, up to leading order in $\phi$ and $\alpha$ (also see Appendix~B of \cite{Gluscevic_2009} for a similar discussion). This can be demonstrated by taking the ensemble average of $\hat{\phi}'_{LM}$ over different CMB realizations,
\begin{equation}
    \label{eq:rotated unbias full}
    \begin{aligned}
        \langle \h{\phi}'_{LM} \rangle_{\m{CMB}} = &A_{L} \sum_{\ell m} \sum_{\ell' m'} \sum_{L'M'}(-1)^{M}  \\
        &\times \left(\begin{array}{ccc}
\ell & \ell' & L \\
m & m' & -M
\end{array}\right) \begin{pmatrix}
  \ell & \ell' & L'\\
  m & m' & M'
\end{pmatrix}  \\
&\times \left((f^{\phi}_{\ell L \ell'})^*f^{\phi}_{\ell L' \ell'}\phi^{*}_{L'M'} + (f^{\phi}_{\ell L \ell'})^*f^{\alpha}_{\ell L' \ell'}\alpha^{*}_{L'M'}\right).
    \end{aligned}
\end{equation}
Applying the orthogonality relation of the Wigner-3j symbols,
\begin{equation}
    \label{eq:orthogonality wigner-3j}
            \sum_{mm'} \left(\begin{array}{ccc}
\ell & \ell' & L \\
m & m' & -M
\end{array}\right) \begin{pmatrix}
  \ell & \ell' & L'\\
  m & m' & M'
\end{pmatrix} = \frac{1}{2L+1}\delta_{LL'}\delta_{M,-M'},
\end{equation}
and the orthogonality of the parity indicators
\begin{equation}
  \label{eq:parity orthogonality}
  \epsilon_{\ell L \ell'} \beta_{\ell L \ell'} = 0,
\end{equation}
we can see that the leading-order rotation contribution disappears due to parity, and we get
\begin{equation}
    \label{eq:rotated unbias}
    \begin{aligned}
    \langle \h{\phi}'_{LM} \rangle_{\m{CMB}} &= A_{L} \sum_{\ell \ell'} \frac{1}{2 L+1}\frac{|f^{\phi}_{\ell L \ell'}|^2}{\hat{C}^{\m{EE}}_{\ell}\hat{C}^{\m{BB}}_{\ell'}} \phi_{LM} \\
    & = \phi_{LM},
    \end{aligned}
\end{equation}
which shows that our estimator for the lensing potential $\phi$ remains unbiased.

However, we should emphasize that an unbiased $\phi$ does not necessarily imply an unbiased $C_L^{\phi\phi}$. Similar to the fact that $\langle x\rangle=0$ does not imply $\langle x^2 \rangle = 0$, the estimator, $C_L^{\h{\phi}\h{\phi}}$ contains quadratic terms in $\hat{\phi}$ and thus may be biased. In addition, higher-order terms in Eq.~\eqref{eq:1st E rotation} and Eq.~\eqref{eq:1st B rotation} may also break the orthogonality between CMB lensing and anisotropic rotation and induce bias to $\h{C}_{L}^{\phi \phi}$.

Based on Eq.~\eqref{eq:phi ps estimation}, we define the bias from rotation
to the reconstructed CMB lensing power spectrum as
\begin{equation}
    \label{eq:ps bias raw}
    \Delta(\hat{C}_{L}^{\phi \phi})_{\mathrm{rot}} \equiv \langle \h{C'}_{L}^{\phi \phi} \rangle- \langle \h{C}_{L}^{\phi \phi} \rangle
\end{equation}
where $\langle ... \rangle$ represents the average over the
reconstructed lensing power spectra from different realizations of the sky, and $\h{C'}_{L}^{\phi \phi}$ refers to applying the power spectrum estimator given
in Eq.~\eqref{eq:phi ps estimation} to rotated-lensed CMB maps.

Among the noise biases in Eq.~\eqref{eq:phi ps estimation}, we assume that anisotropic rotation does not affect $N^{(1)}_{L}$, $N^{(\m{MC})}_{L}$ and $N^{(\m{FG})}_{L}$ on full sky.
RDN0, on the other hand, is calculated based on the observed CMB power spectrum and therefore will be affected
by rotation as it changes the CMB polarization power spectra
(as shown in Fig.~\ref{fig:cl_bb_all}). RDN0 is calculated based on
both ``data'' and MC simulation, encapsulating all the disconnected terms in the CMB four-point correlation \cite{Namikawa_2013, Madhavacheril:2020ido}. Consequently, when compared with a standard MC $\m{N}^{(0)}_{L}$,
RDN0 automatically mitigates the biases from small changes to
the CMB covariance\footnote{The difference between
${}^{(\m{RD})}\m{N}^{(0)}_{L}$ and $\m{N}^{(0)}_{L}$
has been introduced in \cite{Mirmelstein:2020pfk}.}, such as that arising from rotation in the context of this paper. See Appendix.~\ref{sec:appendix A} for further details.

We can then write Eq.~\eqref{eq:ps bias raw} as
\begin{equation}
    \label{eq:ps bias estimator}
    \begin{aligned}
      \Delta(\hat{C}_{L}^{\phi \phi})_{\mathrm{rot}}
      = &\langle C^{\h{\phi'}\h{\phi'}}_{L} \rangle - \langle C^{\h{\phi}\h{\phi}}_{L} \rangle\\ - &(\langle {}^{(\m{RD})}\m{N}'^{(0)}_{L}  \rangle
 - \langle {}^{(\m{RD})}\m{N}^{(0)}_{L} \rangle),
 \end{aligned}
\end{equation}
where $\langle \hat{C}'^{\m{\phi}\m{\phi}}_{L} \rangle$ and $\langle \hat{C}^{\m{\phi}\m{\phi}}_{L} \rangle$ represent the average over the two sets of reconstructed lensing potential power spectra using rotated lensed and unrotated lensed CMB maps respectively, and $\langle {}^{(\m{RD})}\m{N}'^{(0)}_{L} \rangle$ and $\langle {}^{(\m{RD})}\m{N}^{(0)}_{L} \rangle$ represent the corresponding average RDN0.  We will show the computation of RDN0 in the following Section.

\begin{figure*}[t]
\includegraphics[width=0.8\textwidth]{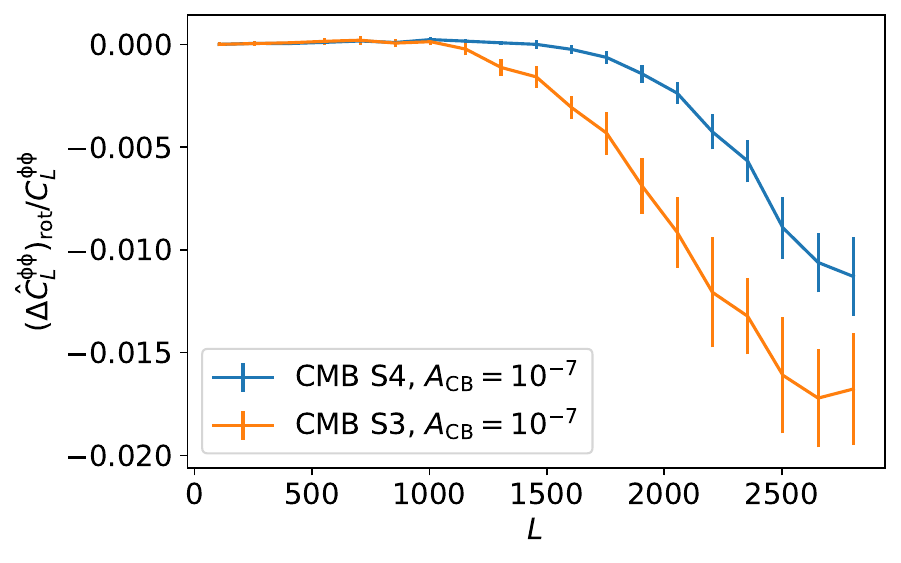}
\caption{The fractional bias of the reconstructed lensing potential power spectrum from a scale-invariant anisotropic rotation field with amplitude  $A_{\m{CB}}=10^{-7}$ for a CMB-S3-like experiment and a CMB-S4-like experiment. We apply full-sky CMB lensing reconstruction using the EB estimator, for the CMB multipole range from $\ell_{\mathrm{min}} = 30$ to $\ell_{\mathrm{max}}=3000$. The curves are binned by $\Delta \ell =150$, and the error bars on $\Delta(\hat{C}_{L}^{\phi \phi})_{\mathrm{rot}}$ are estimated from the scatter among the differences of the reconstructed CMB lensing power spectra using rotated-lensed CMB polarization realizations and using lensed CMB polarization realizations. }
\label{fig:bias.pdf}
\centering
\end{figure*}

\section{Simulation and Results}
\label{sec:simulation}
As discussed in Sec.~\ref{sec:CMB lensing reconstruction}, to estimate the bias $\Delta(\hat{C}_{L}^{\phi \phi})_{\mathrm{rot}}$ in Eq.~\eqref{eq:ps bias estimator}, we need two sets of simulated CMB polarization maps: one set with both CMB lensing and rotation, and the other set with CMB lensing only. The first set of simulations is generated by rotating the lensed simulations in pixel space with a random realization of rotation field. All simulations in this work are generated in CAR pixelization on the full sky using \texttt{pixell}\footnote{\url{https://github.com/simonsobs/pixell}} \cite{2021ascl.soft02003N}. We then use \texttt{cmblensplus}\footnote{\url{https://github.com/toshiyan/cmblensplus}} to perform the CMB lensing reconstruction on the full sky which avoids unnecessary complications due to partial sky coverage\footnote{There can be extra mean-field bias for cut-sky CMB lensing reconstruction \cite{Namikawa_2013}.}. Below we summarize our simulation steps:
\begin{enumerate}
\item
  We generate 10 realizations of lensed CMB polarization maps \{$\t{X}(\h{\bm{n}})$\} ($X\in \{E, B\}$) as our mock data, based on a fiducial power spectrum that does not include the cosmic birefringence effect, $C_l^{XX}\vert_{\rm fid}$, given by the best-fit cosmology from Planck 2018 \cite{Planck2018:VI:CP}.
\item
  We generate 10 Gaussian realizations of an anisotropic rotation field, \{$\alpha(\h{\bm{n}})$\}, assuming a scale-invariant power spectrum with $A_{\m{CB}}=10^{-7}$ as defined in Eq.~\eqref{eq:claa}, which corresponds to a $1\sigma$ upper limit for CMB-S3-like experiments \cite{2019PhRvD.100b3507P}. This amplitude corresponds to an rms rotation of around 0.05 degrees.
\item
  We perform a pixel-wise polarization rotation on each $\t{X}(\bm{\hat{n}})$ with the rotation field $\alpha(\hat{\bm{n}})$ to make a set of 10 rotated-lensed CMB polarization maps, denoted as \{$\t{X}'(\bm{\hat{n}})$\}. Both $\{\t{X}(\bm{\hat{n}})\}$ and $\{\t{X}'(\bm{\hat{n}})\}$ are used as \textit{mock data}.

\item
  We perform full-sky CMB lensing reconstruction using \texttt{cmblensplus} \cite{2021ascl.soft04021N} on both sets of \textit{mock data}, \{$\t{X}(\bm{\hat{n}})$\} and \{$\t{X}'(\bm{\hat{n}})$\}, and obtain an average lensing power spectrum for each set of simulations, denoted as $\langle C^{\hat{\mathrm{\phi}}\hat{\mathrm{\phi}}}_{L}\rangle$ and $ \langle C^{\hat{\mathrm{\phi'}}\hat{\mathrm{\phi'}}}_{L}\rangle$. We restrict lensing reconstruction to multipoles between $\ell_{\m{min}} = 30$ to $\ell_{\m{max}} = 3000$. We use the 
  Knox formula for including detector noise in the power spectrum \cite{PhysRevD.52.4307},
  \begin{equation}
      \hat{C}_\ell^{XX} = C_\ell^{XX}\vert_{\rm fid} + N_\ell,
  \end{equation}
  with the homogeneous detector noise power spectrum for polarization given by
  \begin{equation}
      N_\ell = \Delta^2_{\m{P}} e^{\ell(\ell+1) \theta^2_{\m{FWHM}}/(8\ln2)},
  \end{equation}
  where $\Delta_{\m{P}}$ is the polarization noise level of the experiment and $\theta_{\m{FWHM}}$ is the full-width at half maximum (FWHM) of the beam in radians. Note that although experimental noise is modeled at the power spectrum level in $\h{C}_l^{XX}$, Gaussian noise in the map does not affect the estimation of lensing bias but introduces additional scatter in the result, so we choose not to include it in our simulated maps.
\item
            Following \cite{Namikawa_2013, Madhavacheril:2020ido}, we calculate RDN0 for each map in $\{\t{X}\}$ and $\{\t{X}'\}$ by making two additional set of lensed CMB simulations, $\{\t{X}^{S_1}\}$ and $\{\t{X}^{S_2}_i\}$, based on the fiducial model without cosmic birefringence as in step 1,  each containing 20 simulations. These two sets of independent realizations, $\{\t{X}^{S_1}\}$ and $\{\t{X}^{S_2}_i\}$, are needed in calculating RDN0 as a way to minimize errors from statistical uncertainties in covariance matrix estimation or mismodeling (see Appendix~\ref{sec:appendix A} for a more detailed discussion). For a given map, $\t{X}(\bm{\hat{n}})$, we calculate its RDN0 as
    \begin{equation}
        \label{eq:RND0}
        \begin{aligned}
            {}^{(\m{RD})}\m{N}^{(0)}_{L} = &\langle C_{L}^{\h{\phi}\h{\phi}}[\t{E}\t{B}^{S_1}, \t{E}\t{B}^{S_1}] + C_{L}^{\h{\phi}\h{\phi}}[\t{E}^{S_1}\t{B}, \t{E}\t{B}^{S_1}] \\
            &+ C_{L}^{\h{\phi}\h{\phi}}[\t{E}^{S_1}\t{B}, \t{E}^{S_1}\t{B}] + C_{L}^{\h{\phi}\h{\phi}}[\t{E}\t{B}^{S_1}, \t{E}^{S_1}\t{B}] \\
            &- C_{L}^{\h{\phi}\h{\phi}}[\t{E}^{S_1}\t{B}^{S_2}, \t{E}^{S_1}\t{B}^{S_2}] \\
            &- C_{L}^{\h{\phi}\h{\phi}}[\t{E}^{S_1}\t{B}^{S_2}, \t{E}^{S_2}\t{B}^{S_1}] \rangle_{S_1,S_2},
        \end{aligned}
    \end{equation}
and similarly we calculate RDN0 for a given rotated-lensed map $\t{X}'$ as
\begin{equation}
  \label{eq:RND0 rot}
  \begin{aligned}
    {}^{(\m{RD})}\m{N}'^{(0)}_{L} = &\langle C_{L}^{\h{\phi}\h{\phi}}[\t{E}'\t{B}^{S_1}, \t{E}'\t{B}^{S_1}] + C_{L}^{\h{\phi}\h{\phi}}[\t{E}^{S_1}\t{B}', \t{E}'\t{B}^{S_1}] \\
    &+ C_{L}^{\h{\phi}\h{\phi}}[\t{E}^{S_1}\t{B}', \t{E}^{S_1}\t{B}'] + C_{L}^{\h{\phi}\h{\phi}}[\t{E}'\t{B}^{S_1}, \t{E}^{S_1}\t{B}'] \\
    &- C_{L}^{\h{\phi}\h{\phi}}[\t{E}^{S_1}\t{B}^{S_2}, \t{E}^{S_1}\t{B}^{S_2}] \\
    &- C_{L}^{\h{\phi}\h{\phi}}[\t{E}^{S_1}\t{B}^{S_2}, \t{E}^{S_2}\t{B}^{S_1}] \rangle_{S_1,S_2},
  \end{aligned}
\end{equation}
where the combinations in $\left[...\right]$ represent the input maps for lensing reconstruction, and $\langle ... \rangle_{S_1,S_2}$ refers to an average over the two sets of MC simulations ($S_1$, $S_2$) \footnote{In practice, due to memory constraints, we split the set, $S_{1}$, further into two subsets, $S_{11}$ and $S_{12}$, evenly; Similarly $S_{2}$ is split into $S_{21}$ and $S_{22}$. The average over $S_1$ and $S_2$ is calculated by averaging over the combinations of $\{S_{11}, S_{21}\}$, $\{S_{11}, S_{22}\}$, $\{S_{12}, S_{21}\}$, and $\{S_{12}, S_{22}\}$.}. Using Eq.~\eqref{eq:RND0} and \eqref{eq:RND0 rot}, we calculate RDN0 for each simulation in $\{\t{X}(\bm{\hat{n}})\}$ and $\{\t{X}'(\bm{\hat{n}})\}$ and calculate the average within each set of simulations, denoted as $\langle {}^{(\m{RD})}\m{N}^{(0)}_{L}  \rangle$ and $\langle {}^{(\m{RD})}\m{N}'^{(0)}_{L} \rangle$ respectively. Note that, as mentioned in the last step, simulated CMB maps in Eq.~\eqref{eq:RND0} and \eqref{eq:RND0 rot} do not contain map noise, so the calculated RDN0 only contains contribution from cosmic variance.
\item
  We then estimate the lensing bias, $\Delta(\hat{C}_{L}^{\phi \phi})_{\mathrm{rot}}$ using Eq.~\eqref{eq:ps bias estimator}, and calculate fractional bias defined as $\Delta(\hat{C}_{L}^{\phi \phi})_{\mathrm{rot}}/C^{\phi\phi}_{L}$, with $C^{\phi\phi}_{L}$ the power spectrum of the CMB lensing potential from the fiducial model.
\item
  We repeat the above procedures for two sets of experimental configurations: CMB-S3-like and CMB-S4-like. The experimental configurations including the noise level and beam size are list in Table \ref{tab:expts}.
\end{enumerate}

In Fig.~\ref{fig:bias.pdf}, we show the fractional bias, $\Delta(\hat{C}_{L}^{\phi \phi})_{\mathrm{rot}}/C^{\phi\phi}_{L}$, obtained from the procedure described above for both CMB-S3-like and CMB-S4-like experiments.
We find that a rotation field with $A_{\m{CB}}=10^{-7}$, which is well below the current experimental constraint, is capable of introducing a percent-level bias to the reconstructed lensing power spectrum. The bias is most evident at small scales ($L\gtrsim 1000$), reaching up to $1.5\%$ for CMB-S3-like experiments and slightly lower for CMB-S4-like experiments.

The signal-to-noise ratio (SNR) of the contribution from the anisotropic birefringence is
\begin{equation}\label{eq:snr}
    {\rm SNR} = \left( \sum_L \frac{f_{\rm sky}L \left(\Delta(\hat{C}_{L}^{\phi \phi})_{\mathrm{rot}}\right)^2}{ \left(C_L^{\phi\phi}+N^{(0)}_L\right)^2}\right)^{1/2},
\end{equation}
where $N^{(0)}$ is calculated theoretically using \texttt{cmblensplus} that contains contributions from both cosmic variance and map noise, and $f_{\rm sky}=0.4$, accounting for the partial sky coverage of a ground-based experiment. Using Eq.~\eqref{eq:snr}, we estimate that the contribution of cosmic birefringence with $A_{\rm CB}=10^{-7}$ has a SNR of $0.2$ for a CMB-S4-like experiment.

\begin{table}[tp]
\centering
\begin{tabular}{|p{2.4cm}|p{1.4cm}|p{1.4cm}|}
 \hline
 Expt & ${\Delta}_{T}~[\mathrm{\mu K '}]$ & ${\theta}_{\mathrm{FWHM}} [']$\\[0.5ex]
  \hline
  CMB-S3-like & 7 & 1.4 \\
  CMB-S4-like & 1 & 1.4 \\
 \hline
\end{tabular}
\caption{Experiments configurations considered in this study.} 
\label{tab:expts}
\end{table}

\section{The Source of Bias}
\label{sec:discussion}

To understand the observed bias, we compare the four-point correlation function contained in $C_L^{\hat{\phi}'\hat{\phi}'}$,
\begin{equation}
  \label{eq:trispec1}
  \langle \t{E}'_{\ell_1 m_1} \t{B}'_{\ell_2 m_2} \t{E}'_{\ell_3 m_3} \t{B}'_{\ell_4 m_4}\rangle,
\end{equation}
to the four-point correlation function in $C_L^{\h{\phi}\h{\phi}}$,
\begin{equation}
  \label{eq:trispec2}
  \langle \t{E}_{\ell_1 m_1} \t{B}_{\ell_2 m_2} \t{E}_{\ell_3 m_3} \t{B}_{\ell_4 m_4}\rangle.
\end{equation}
To leading order, a dominant contribution to their difference is given by
\begin{equation}
\begin{split}\label{eq:bias main term}
&\:\:\:\:\:\:\langle E_{\ell_1m_1}\delta B'_{\ell_2m_2}E_{\ell_3m_3}\delta B'_{\ell_4m_4}\rangle \\
& \propto \langle E_{\ell_1m_1}(E_{\ell'_2m'_2}\alpha_{LM})E_{\ell_3m_3} (E_{\ell'_4m'_4}\alpha_{L'M'})\rangle,
\end{split}
\end{equation}
where we have applied Eq.~\eqref{eq:1st B rotation} and used parentheses to indicate useful groupings of terms. 
As we have assumed $\alpha$ to be a Gaussian random field, Eq.~\eqref{eq:bias main term} can be simplified using Wick's theorem into products of two-point correlation functions that fall into two classes: disconnected terms and connected terms.

A disconnected term can be expressed as
\begin{equation}
  \label{eq:disconnected term}
  {\contraction {\langle} {E}{_{\ell_1 m_1} (E_{\ell'_2 m'_2}\alpha_{LM}) }{E}
    \contraction [2ex] {\langle E_{\ell_1 m_1}(} {E}{_{\ell'_2 m'_2}\alpha_{LM})E_{\ell_3 m_3}(} {E}
    \bcontraction {\langle E_{\ell_1 m_1} (E_{\ell'_2 m'_2}}{\alpha}{_{LM}) E_{\ell_3 m_3} (E_{\ell'_4 m'_4}}{\alpha}
    \langle E_{\ell_1 m_1} (E_{\ell'_2 m'_2}\alpha_{LM}) E_{\ell_3 m_3} (E_{\ell'_4 m'_4}\alpha_{L' M'})\rangle},
\end{equation}
where we have used Wick contraction notation to denote products of two-point correlation functions. This term is classified as a disconnected term of the CMB four-point correlation function, because fields within the same ``group'', e.g., $E_{\ell'_2 m’_2}$ and $\alpha_{LM}$ (as indicated by the parentheses), follow the same contraction behaviour, i.e., contracting with a common group, $E_{\ell'_4m'_4}$ and $\alpha_{L'M'}$. If, on the other hand, contraction behavior of fields in the same group is different, we classify the term as connected. For example, the term
\begin{equation}
  \label{eq:connected term}
  {\contraction {\langle}{E}{_{\ell_1 m_1} (E_{\ell'_2 m'_2}\alpha_{LM}) E_{\ell_3 m_3}(}{E}
    \contraction [2ex] {\langle E_{\ell_1 m_1}(} {E}{_{\ell'_2 m'_2}\alpha_{LM})} {E}
    \bcontraction {\langle E_{\ell_1 m_1} (E_{\ell'_2 m'_2}}{\alpha}{_{LM}) E_{\ell_3 m_3} (E_{\ell'_4 m'_4}}{\alpha}
  \langle E_{\ell_1 m_1} (E_{\ell'_2 m'_2}\alpha_{LM}) E_{\ell_3 m_3} (E_{\ell'_4 m'_4}\alpha_{L'M'})\rangle},
\end{equation}
is connected, and so is 
\begin{equation}
  \label{eq:connected term 1}
  {\contraction {\langle}{E}{_{\ell_1 m_1(}}  {E}
    \contraction {\langle E_{\ell_1 m_1}({E}{_{\ell'_2 m'_2}\alpha_{LM})}} {E}{_{\ell_3 m_3}(}{E}
    \bcontraction {\langle E_{\ell_1 m_1} (E_{\ell'_2 m'_2}}{\alpha}{_{LM}) E_{\ell_3 m_3} (E_{\ell'_4 m'_4}}{\alpha}
  \langle E_{\ell_1 m_1} (E_{\ell'_2 m'_2}\alpha_{LM}) E_{\ell_3 m_3} (E_{\ell'_4 m'_4}\alpha_{L'M'})\rangle}.
\end{equation}
In both cases $E_{\ell'_2m'_2}$ and $\alpha_{LM}$ do not contract with a common group. The distinction of disconnected and connected terms is important because the RDN0 technique that we apply mitigates the leading-order  biases in the disconnected terms of the four-point correlation, such as that in Eq.~\eqref{eq:disconnected term}, but cannot mitigate biases in the connected terms. Hence we expect that a dominant contribution to the observed bias in Sec.~\ref{sec:simulation} comes from the accumulated effect of all connected terms, such as Eq.~\eqref{eq:connected term} and \eqref{eq:connected term 1}.

\begin{figure}[t]
\includegraphics[width=0.5\textwidth]{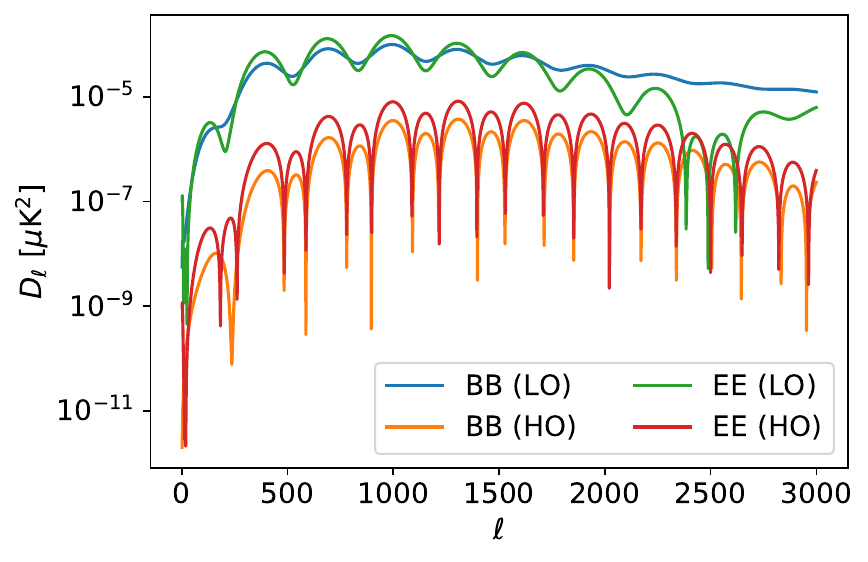}
\caption{
B-mode power spectrum contributions from the leading-order (LO) contribution and higher-order (HO) contribution of scale-invariant anisotropic cosmic birefringence with an amplitude of $A_{\m{CB}}=10^{-7}$. The former is about two order of magnitude larger than the latter. 
The power spectra shown here are generated by \texttt{class\_rot} \cite{Cai:2021zbb}.}
\label{fig:cl_bb_rot}
\centering
\end{figure}

In the above discussion we have neglected the effect of higher-order terms which mix lensing and rotation. Higher-order terms break the orthogonality between lensing and rotation, leading to bias in $\hat{\phi}$; they also contribute additional connected terms to $\h{C}_l^{\phi\phi}$. In Fig.~\ref{fig:cl_bb_rot} we show the contributions from leading-order and higher-order terms to the power spectrum, calculated using \texttt{class\_rot}\footnote{The higher-order contribution is estimated by subtracting the leading-order lensing and rotation contributions from a non-perturbative calculation with both lensing and rotation.}. We find that for $A_{\rm CB}=10^{-7}$, higher-order terms generally contribute a few percent of the leading-order power spectrum bias. 

\begin{figure}[t]
\includegraphics[width=\linewidth]{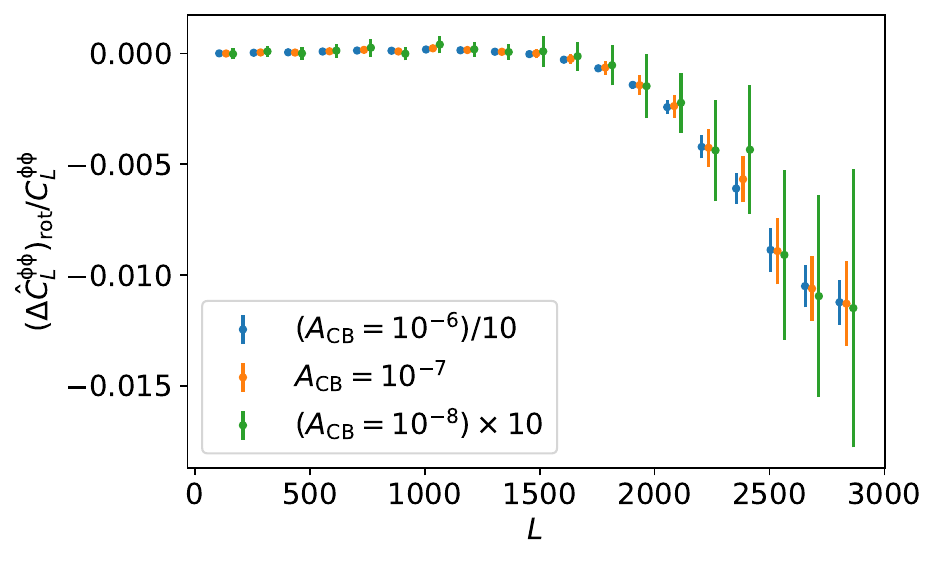}
\caption{Comparing biases from different $A_{\rm CB}$ in a CMB-S4-like experiment. For visualization we have scaled both data and error bar by a factor of
10 and 0.1 for $A_{\rm CB}=10^{-8}$ and $10^{-6}$, respectively. Similarly the horizontal displacements between the three data series are for visualization purpose only.}
\label{fig:bias scaling}
\centering
\end{figure}

If the contributions from connected terms, such as Eq.~\eqref{eq:connected term} and \eqref{eq:connected term 1}, are dominant and the higher-order terms can be neglected, we expect the size of the observed bias to scale linearly with $C_L^{\alpha\alpha}$ and thus linearly with $A_{\rm CB}$. To verify this, we repeat the steps in Sec.~\ref{sec:simulation} for $A_{\rm CB}=10^{-6}$ and $10^{-8}$ and compare the resulting biases in Fig.~\ref{fig:bias scaling}. It is evident that the observed bias does indeed scale linearly with $A_{\rm CB}$. This also suggests that, if a rotation field with $A_{\rm CB}= 10^{-6}$, which is well below the current observational constraint, were present but not accounted for in CMB lensing reconstruction, our estimated $C_L^{\phi\phi}$ may be biased low by $\gtrsim 10\%$ at the small scales, and as shown in Fig.~\ref{fig:bias.pdf}, the bias gets marginally worse at higher noise levels.

\begin{figure}[t]
    \centering
    \includegraphics[width=\linewidth]{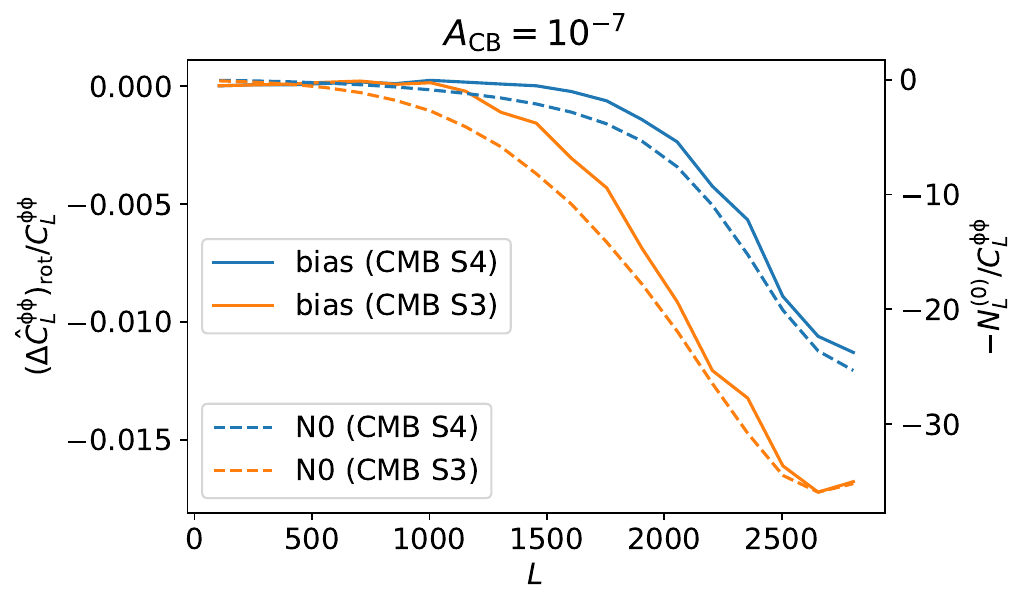}
    \caption{Comparing the observed bias (solid lines) with $-N^{(0)}_L/C_L^{\phi\phi}$ (dashed lines), where $N_L^{(0)}$ is obtained by averaging the ${}^{\rm (RD)}N^{(0)}$ from the lensed simulations (without rotation). Note that as mentioned in Sec.~\ref{sec:simulation}, ${}^{\rm (RD)}N^{(0)}$ only contains contribution from cosmic variance of CMB maps. The minus sign in front of $N_L^{(0)}$ is added to aid visual comparison.}
    \label{fig:bias vs n0}
\end{figure}

We also find that the bias, $\Delta (\hat{C}_L^{\phi\phi})_{\rm rot}/C_L^{\phi\phi}$, has a shape that roughly follows that of the term $-N_L^{(0)}/C_L^{\phi\phi}$, as shown in Fig.~\ref{fig:bias vs n0}. Terms in Eq.~\eqref{eq:connected term} and \eqref{eq:connected term 1} are quadratic in $\alpha$ which are similar to the $N_L^{(1)}$ bias in CMB lensing, and the $N_L^{(1)}$ bias (quadratic in $\phi$) is in general not linear to $N_L^{(0)}$ in the context of CMB lensing (see, e.g., \cite{Kesden:2003,PhysRevD.83.043005}). The apparent linearity between the observed bias and $N_L^{(0)}$ seen in Fig.~\ref{fig:bias vs n0} is therefore surprising and may be a consequence of the scale-invariance of the rotation field considered here, or coincidental. We leave a detailed investigation of the shape of the bias to future work.

We have only considered the EB estimator for lensing reconstruction, but we expect a similar effect to be present with other polarization-based quadratic estimators, such as the EE estimators. As we expect polarization-based estimators to contribute substantial statistical power in CMB lensing reconstruction in a CMB-S3-like experiment and to dominate the statistical power in a CMB-S4-like experiment \cite{S4:2016:SciBook}, rotation-induced bias is thus an important factor to account for. On the other hand, lensing reconstruction with the TT estimator should remain unaffected by rotation because it only affects CMB polarization fields; this suggests that an easy diagnostic of the rotation-induced bias is to compare the reconstructed lensing power spectrum obtained from TT and EB estimators and look for a reduction of power in the latter.
Multiple methods for mitigating biases in CMB lensing measurements have been proposed. For example, the bias-hardening approach \cite{Namikawa_2013,Namikawa:2013:BHE-pol,Osborne:2013nna,Sailer:2020lal} has been used to mitigate the mean-field bias in the reconstructed lensing map. In our case, however, this approach does not apply since the leading-order rotation fields do not produce mean-field bias in the reconstructed lensing map (see Eq.~\eqref{eq:rotated unbias}).
A simple approach to mitigating the rotation-induced bias arising from the connected four-point correlation is to first reconstruct the rotation fields using the quadratic estimator similar to the lensing reconstruction \cite{Gluscevic_2009,Yadav:2010} and then to de-rotate polarization maps using the reconstructed rotation fields, as originally proposed by \cite{Kamionkowski:2009}.
We leave a detailed analysis of the effectiveness of this approach to future work.

\section{Conclusion}
\label{sec:conclusion}

The CMB lensing power spectrum encodes a wealth of information of matter fluctuations over a broad range of redshifts. Bias to the CMB lensing power spectrum may thus lead to bias in the cosmological parameters extracted from it. For example, the sum of neutrino masses can be constrained using CMB lensing power spectrum as massive neutrino can suppress CMB lensing power spectrum by several percent due to its suppression effect on the matter power spectrum below the neutrino free-streaming scale \cite{DeBernardis:2009di,TopicalConvenersKNAbazajianJECarlstromATLee:2013bxd}; it has also been shown in \cite{Planck:2015mym} that bias in the high-$L$ ($L\gtrsim 1000$) may lead to bias in $\sigma_{8} \Omega_{\m{m}}^{0.25}$ at $1\sigma$ level.
Therefore, bias to the CMB lensing power spectrum needs to be carefully accounted for when constraining cosmological parameters.

In this paper we investigated the bias from anisotropic polarization rotation to the reconstructed CMB lensing power spectrum using the EB estimator.
This bias has previously been investigated in the context of instrumental systematics \cite{Mirmelstein:2020pfk,Nagata:2021}. 
In this work we focus instead on a particular class of physically motivated models that produce anisotropic cosmic birefringence, such as axion-like particles coupling to photons through Chern-Simons interaction or primordial magnetic field. In addition, we aim to identify the dominant terms responsible for the rotation-induced bias to CMB lensing. 

Our results show that scale-invariant anisotropic cosmic birefringence with $A_{\rm CB}=10^{-7}$ can induce a percent-level bias to reconstructed CMB lensing power spectrum at small scales ($L>1000$) for both CMB-S3-like and CMB-S4-like experiments, and the bias scales linearly with $A_{\rm CB}$.
This effect can be detectable (i.e. SNR$\geq 1$) if $A_{\rm CB}=5\times 10^{-7}$ with CMB-S4-like noise level.

We also found that the observed bias likely arises from the the connected terms in the CMB four-point correlation function induced by the leading-order perturbation from rotation; this explains the linear scaling of the bias with $A_{\rm CB}$.
Polarization-based lensing reconstruction is expected to dominate or contribute substantially to the statistical power in CMB lensing reconstruction in the next generation CMB experiments. Rotation-induced bias to the CMB lensing power spectrum, therefore, may lead to non-negligible bias in the constraints of cosmological parameters such as the sum of neutrino masses and $\sigma_8$ and poses a significant challenge to achieving the science goals in future CMB lensing analysis. Thus, a possible rotation field is an important factor to account for and mitigate during lensing reconstruction. We argued that the bias-hardened estimator approach to mitigate bias does not work for rotation-induced bias, because of the orthogonality between rotation and CMB lensing. A promising mitigation method is to first reconstruct the anisotropic rotation field and then de-rotate the CMB polarization field with the reconstructed rotation field before proceeding to lensing reconstruction, but its effectiveness in the presence of galactic foreground and partial sky coverage remains to be seen.

As we have demonstrated in this paper, the reconstructed lensing power spectrum is highly sensitive to the level of anisotropic cosmic birefringence. Although this lensing bias is by itself not significant enough (SNR $\approx 0.2~\sigma$ for $A_{\m{CB}}=10^{-7}$ with CMB-S4-like noise level) to be used as a probe for cosmic birefringence signal, jointly estimating cosmic birefringence and CMB lensing once maps get to low enough noise levels may still provide a useful consistency check should a cosmic birefringence signal be detected in the future.


\section*{Acknowledgments}
We thank Blake Sherwin, Alexander van Engelen and Colin Hill for useful discussions. This work uses resources of the National Energy Research Scientific Computing Center (NERSC), and open source software including \texttt{healpy} \cite{2019JOSS....4.1298Z},
\texttt{pixell} \cite{2021ascl.soft02003N} and \texttt{cmblensplus} \cite{2021ascl.soft04021N}.
TN acknowledges support from JSPS KAKENHI Grant No. JP20H05859 and No. JP22K03682.

\appendix
\section{Realization-dependent Gaussian noise (RDN0)}
\label{sec:appendix A}
Here we review the realization-dependent Gaussian noise (RDN0) applied in Eq.~\eqref{eq:RND0} and \eqref{eq:RND0 rot}. RND0 is introduced in \cite{Namikawa_2013, Namikawa:2013xka} as a robust way to evaluate the reconstruction noise to CMB lensing power spectrum sourced by the disconnected terms of CMB four-point function (Gaussian noise). It has been applied in the CMB lensing power spectrum measurements of different experiments including \cite{PlanckCollaboration:2014, vanEngelen:2015, BICEP2Collaboration:2016}). In this section, we revisit the derivation of RDN0. We will adopt the flat-sky approximation in \cite{Namikawa_2013}, but the result applies to both flat-sky and full-sky analyses.

We start by defining the Gaussian probability distribution function (PDF) of the unlensed CMB multipoles as
\begin{equation}
  \label{eq:Gaussian likelihood}
  P_{\mathrm{g}}=\frac{1}{\sqrt{(2 \pi)^N \det C}} \exp \left(-\frac{1}{2} \sum_{a b} \sum_{\bm{\ell} \bm{\ell}'} a_{\bm{\ell}}\left(\mathrm{C}^{-1}\right)^{a_{\bm{\ell}}, b_{\bm{\ell}'}} b_{\bm{\ell}'}\right),
\end{equation}
where $C$ is the covariance matrix between CMB multipoles, $a_{\bm{\ell}}, b_{\bm{\ell}'}$ represent the CMB multipoles, and $a, b \in \{T, E, B\}$.

When CMB fields are distorted by weak lensing, CMB multipoles become weakly non-Gaussian. The PDF of the lensed CMB multipoles can then be approximated using the Edgeworth expansion as \cite{10.1093/mnras/283.3.983, PhysRevD.82.023520} 
\begin{equation}
  \label{eq:approximate likelihood}
    \begin{aligned}
      P &\approx \left[1+\sum_{a b c d} \sum_{\{\bm{\ell}_i\}}\left\langle a_{\bm{\ell}_1} b_{\bm{\ell}_2} c_{\bm{\ell}_3} d_{\bm{\ell}_4}\right\rangle_c \frac{\partial}{\partial a_{\bm{\ell}_1}} \frac{\partial}{\partial b_{\bm{\ell}_2}} \frac{\partial}{\partial c_{\bm{\ell}_3}} \frac{\partial}{\partial d_{\bm{\ell}_4}}\right] P_{\mathrm{g}},
    \end{aligned}
  \end{equation}
where $\left\langle a_{\bm{\ell}_1} b_{\bm{\ell}_2} c_{\bm{\ell}_3} d_{\bm{\ell}_4}\right\rangle_c$ represents the connected terms (also known as cumulants or trispectrum) contributed by CMB lensing, given by \cite{Namikawa:2013xka}
\begin{equation}
  \label{eq:abcd cumulants}
\begin{aligned}
\left\langle a_{\bm{\ell}_1} b_{\bm{\ell}_2} c_{\bm{\ell}_3} d_{\bm{\ell}_4}\right\rangle_{\mathrm{c}} \approx &~f_{\bm{\ell}_{12}, \bm{\ell}_1}^{a b} f_{-\bm{\ell}_{12}, \bm{\ell}_3}^{c d} C_{\left|\bm{\ell}_{12}\right|}^{\phi \phi} \delta_{\bm{\ell}_{12},-\bm{\ell}_{34}} \\
&+f_{\bm{\ell}_{13}, \bm{\ell}_1}^{a c} f_{-\bm{\ell}_{13}, \bm{\ell}_2}^{b d} C_{\left|\bm{\ell}_{13}\right|}^{\phi \phi}\delta_{\bm{\ell}_{13},-\bm{\ell}_{24}} \\
& +f_{\bm{\ell}_{14}, \bm{\ell}_1}^{a d} f_{-\bm{\ell}_{14}, \bm{\ell}_2}^{b c} C_{\left|\bm{\ell}_{14}\right|}^{\phi \phi} \delta_{\bm{\ell}_{14},-\bm{\ell}_{23}},
\end{aligned}
\end{equation}
to the leading order of $C_L^{\phi\phi}$, where $\bm{\ell}_{ij} \equiv \bm{\ell}_{i} + \bm{\ell}_{j}$, and $f_{\bm{\ell},\bm{\ell}'}^{a b}$ represents the weight functions for lensing potential as in \cite{Namikawa:2013xka}.

We can then obtain an optimal estimator, $\widehat{C}_{L}^{\phi \phi}$, by maximizing the log-likelihood, ${\mathcal L}\equiv \ln P$, with
\begin{equation}
  \label{eq:}
\frac{\partial \mathcal{L}}{\partial C^{\phi \phi}_{L}} \left(\widehat{C}_{L}^{\phi \phi}\right)=0,
\end{equation}
and get \cite{Namikawa:2013xka}
\begin{equation}
  \label{eq:clpp estimator operator}
\widehat{C}_{L}^{\phi \phi} \propto \sum_{a b c d} \sum_{\vert \bm{L}\vert=L}P_g^{-1}\left(\widehat{f}_{\bm{L}}^{a b} \widehat{f}_{-\bm{L}}^{c d}+\widehat{f}_{\bm{L}}^{a c} \widehat{f}_{-\bm{L}}^{b d}+\widehat{f}_{\bm{L}}^{a d} \widehat{f}_{-\bm{L}}^{b c}\right) P_{\mathrm{g}},
\end{equation}
where we have defined
\begin{equation}
  \label{eq:}
\widehat{f}_{\bm{L}}^{a b} \equiv \sum_{\bm{\ell}'} f_{\bm{L},\bm{\ell}'}^{a b} \frac{\partial}{\partial a_{\bm{L}}} \frac{\partial}{\partial b_{\bm{L} - \bm{\ell}'}},
\end{equation}
and have omitted the normalization. One can further show that 
\begin{equation}
    \label{eq:operator}
    \widehat{f}_{\bm{L}}^{a b} P_{\mathrm{g}} = \left(\bar{x}_{\bm{L}}^{a b}-\left\langle\bar{x}_{\bm{L}}^{a b}\right\rangle\right)P_{\mathrm{g}},
\end{equation}
where 
\begin{equation}
\label{eq:unnormalized estimator}
\bar{x}_{\bm{L}}^{a b}\equiv\sum_{\bm{\ell}} f_{\bm{L}, \bm{\ell}}^{a b} \bar{a}_{\bm{\ell}} \bar{b}_{\bm{L}-\bm{\ell}},
\end{equation}
in which $\bar{a}_{\bm{\ell}} = \sum_{a',\bm{\ell'}} C^{-1}_{a_{\bm{\ell}} a'_{\bm{\ell}'}}a'_{\bm{\ell}'}$ represents the inverse-variance filtered CMB multipoles. 
Note that $\bar{x}_{\bm{L}}^{a b}$ resembles the unnormalized quadratic estimator for CMB lensing, and $\left\langle\bar{x}_{\bm{L}}^{a b}\right\rangle$ the mean-field bias\footnote{In the real experiments, the mean-field bias is estimated by averaging the reconstructed maps using simulated CMB realizations based on the fiducial model. The mean-field bias does not affect the result in this paper.}.

One can further show that
\begin{equation}
    \label{eq:clpp estimator full}
    \begin{aligned}
        P_g^{-1} \widehat{f}_{\bm{L}}^{a b} \widehat{f}_{-\bm{L}}^{c d} P_{\mathrm{g}}\propto \left[\left(\bar{x}_{\bm{L}}^{a b}-\left\langle\bar{x}_{\bm{L}}^{a b}\right\rangle \right)\left(\bar{x}_{\tn{-}\bm{L}}^{c d}-\left\langle\bar{x}_{\tn{-}\bm{L}}^{c d}\right\rangle\right)- n_{\bm{L}}^{a b, c d}\right],
    \end{aligned}
\end{equation}
with
\begin{equation}
  \label{eq:n abcd}
  \begin{aligned}
    n_{\bm{L}}^{a b, c d} \equiv& \sum_{\bm{\ell}_1, \bm{\ell}_2} f_{\bm{L}, \bm{\ell}_1}^{a b} f_{\bm{-L}, \bm{\ell}_2}^{c d} \times \\
     (&\bar{b}_{\bm{L}-\bm{\ell}_1} \bar{d}_{\bm{-L}-\bm{\ell}_1} C^{-1}_{a_{\bm{\ell}_1},c_{\bm{\ell}_2}}  + \bar{b}_{\bm{L}-\bm{\ell}_1} \bar{c}_{\bm{-L}-\bm{\ell}_1} C^{-1}_{a_{\bm{\ell}_1},d_{\bm{\ell}_2}} \\
      +  &\bar{a}_{\bm{L}-\bm{\ell}_1} \bar{d}_{\bm{-L}-\bm{\ell}_1} C^{-1}_{b_{\bm{\ell}_1},c_{\bm{\ell}_2}}
      + \bar{a}_{\bm{L}-\bm{\ell}_1} \bar{c}_{\bm{-L}-\bm{\ell}_1} C^{-1}_{b_{\bm{\ell}_1},d_{\bm{\ell}_2}})\\
      - &\frac{1}{2}\sum_{\bm{\ell}_1, \bm{\ell}_2} f_{\bm{L}, \bm{\ell}_1}^{a b} f_{\bm{-L}, \bm{\ell}_2}^{c d} \times\\
      (&C^{-1}_{a_{\bm{\ell}_1},c_{\bm{\ell}_2}} C^{-1}_{b_{\bm{L}-\bm{\ell}_1},d_{\bm{-L}-\bm{\ell}_2}} + C^{-1}_{a_{\bm{\ell}_1},d_{\bm{\ell}_2}} C^{-1}_{b_{\bm{L}-\bm{\ell}_1},c_{\bm{-L}-\bm{\ell}_2}}\\
      + &C^{-1}_{b_{\bm{\ell}_1},c_{\bm{\ell}_2}} C^{-1}_{a_{\bm{L-\ell}_1},d_{\bm{-L-\ell}_2}} + C^{-1}_{b_{\bm{\ell}_1},d_{\bm{\ell}_2}} C^{-1}_{a_{\bm{L-\ell}_1},c_{\bm{-L-\ell}_2}}),
  \end{aligned}
\end{equation}
where we have used 
\begin{equation}
  \label{eq:average of C^-1}
  C^{-1}_{a_{\bm{\ell}} a'_{\bm{\ell}'}} = \langle \bar{a}_{\bm{\ell}}\bar{a}'_{\bm{\ell}'} \rangle.
\end{equation}
With Eq.~\eqref{eq:clpp estimator full}, the optimal estimator for CMB lensing potential in Eq.~\eqref{eq:clpp estimator operator} becomes
\begin{equation}
\begin{aligned}
\widehat{C}_{L}^{\phi \phi} \propto & \sum_{a b c d} \sum_{\vert \bm{L}\vert=L}\bigg(
\left[\left(\bar{x}_{\bm{L}}^{ab}-\left\langle\bar{x}_{\bm{L}}^{a b}\right\rangle \right)\left(\bar{x}_{\tn{-}\bm{L}}^{c d}-\left\langle\bar{x}_{\tn{-}\bm{L}}^{c d}\right\rangle\right)- n_{\bm{L}}^{a b, c d}\right] \\
&+ [b \leftrightarrow c] + [b \leftrightarrow d]\bigg).
\end{aligned}
\end{equation}

Note that $\bar{a}, \bar{b}, \bar{c}, \bar{d}$ in Eq.~\eqref{eq:n abcd} correspond to filtered CMB multipoles from \textit{data}, whereas the covariance matrices, e.g., $C_{a_{\bm{\ell}_1},c_{\bm{\ell}_2}}$, are to be estimated from \textit{simulations}. In particular, we apply two sets of Monte-Carlo CMB realizations based on the same fiducial cosmological model to estimate the covariance matrix, denoted as $S_1$ and $S_2$.  Specifically, $S_1$ is used to estimate the covariance matrices in the first parentheses of Eq.~\eqref{eq:n abcd}; each product of covariance matrices in the second parenthesis of Eq.~\eqref{eq:n abcd} is estimated using S1 and S2 respectively (i.e., $C^{-1}_{S1}C^{-1}_{S2}$). This way of estimating the covariance matrix is less sensitive to errors in the covariance matrix from statistical uncertainties as well as mis-modeling. To see that, we can express the estimated covariance matrix as $\hat{C}=\bar{C} + \Delta C$, with $\bar{C}$ the true covariance matrix and $\Delta C$ the error. 
It is then easy to check that, when averaging over different CMB realizations, the error in covariance matrix, $\Delta C$, does not contribute to $\langle n_{\bm{L}}^{a b, c d} \rangle$ in the leading order. Since $S_1$ and $S_2$ are assumed to be independent, statistical uncertainties in $\hat{C}$ will also be eliminated up to $O((\Delta C)^2)$. 

Incorporating the estimation of covariance matrices using $S_1$ and $S_2$, Eq.~\eqref{eq:n abcd} can be expressed more succinctly as
\begin{equation}
  \label{eq:n abcd s1 s2}
  \begin{aligned}
n_{\bm{L}}^{a b, c d} &= \left\langle\left(\bar{x}_{\bm{L}}^{a^{S_1} b}+\bar{x}_{\bm{L}}^{a b^{S_1}}\right)\left(\bar{x}_{-\bm{L}}^{c^{S_1} d}+\bar{x}_{-\bm{L}}^{c d^{S_1}}\right)\right\rangle_{S_1} -  \\
& \frac{1}{2}\left\langle\left(\bar{x}_{\bm{L}}^{a^{S_1} b^{S_2}}+\bar{x}_{\bm{L}}^{a^{S_2} b^{S_1}}\right)\left(\bar{x}_{-\bm{L}}^{c^{S_1} d^{S_2}}+\bar{x}_{-\bm{L}}^{c^{S_2} d^{S_1}}\right)\right\rangle_{S_1,S_2},
\end{aligned}
\end{equation}
where the superscripts $S_i$ denote the CMB simulation provided by the set of $S_i$, and $\langle ... \rangle_{S_i}$ represents the average over the realizations of $S_i$. 
Adding back the omitted normalization in Eq.~\eqref{eq:n abcd s1 s2},
and the omitted normalization, we obtain the realization-dependent Gaussian noise (RDN0) with a given set of $\{ab, cd\}$ as
\begin{equation}
  \label{eq:RDN0 1}
  \begin{aligned}
    {}^{(\m{RD})}\m{N}^{(0)ab,cd}_{L} \propto & \sum_{\vert \bm{L}\vert=L} n_{\bm{L}}^{a b, c d} \\
     =&~\langle C_{L}^{\h{\phi}\h{\phi}}[ab^{S_1}, cd^{S_1}] + C_{L}^{\h{\phi}\h{\phi}}[a^{S_1}b, ab^{S_1}] \\
            &+ C_{L}^{\h{\phi}\h{\phi}}[a^{S_1}b, c^{S_1}d] + C_{L}^{\h{\phi}\h{\phi}}[ab^{S_1}, c^{S_1}d] \\
            &- C_{L}^{\h{\phi}\h{\phi}}[a^{S_1}b^{S_2}, c^{S_1}d^{S_2}] \\
            &- C_{L}^{\h{\phi}\h{\phi}}[a^{S_1}b^{S_2}, c^{S_2}d^{S_1}] \rangle_{S_1,S_2}.
  \end{aligned}
\end{equation}
In the context of this paper, we set $a=c=E$, $b=d=B$. $S_1$ and $S_2$ are two sets of independent lensed CMB realizations based on the fiducial cosmological model which does not include the cosmic birefringence effect. As a result, for the lensed-only \textit{data} maps, the RDN0 is given by Eq.~\eqref{eq:RND0}; for the rotated-lensed \textit{data} maps, the RDN0 is given by Eq.~\eqref{eq:RND0 rot}.

Note that by averaging ${}^{\rm{(RD)}}\m{N}^{(0)}_L$ over many CMB realisations, we reproduce the naive Gaussian reconstruction noise $\m{N}^{(0)}_L$. However, one can see that ${}^{\rm{(RD)}}\m{N}^{(0)}_L$ is more optimal because it preserves the critical realisation-dependent information which is lost in $\m{N}^{(0)}_L$ after averaging.
In addition, as previously discussed, RDN0 is also less sensitive to the errors in the covariance matrix such as the statistical uncertainties from the limited set of simulations used to calculate the covariance matrix and errors in the underlying fiducial model used to generate the simulations. It has been demonstrated in \cite{Madhavacheril:2020ido} that such an error in covariance matrix may lead to a catastrophic error (see Fig.~2 of \cite{Madhavacheril:2020ido}) in the estimated lensing power spectrum if the naive $N_L^{(0)}$ is used, and it is significantly reduced with RDN0.
\vspace{1.5cm}
\bibliography{birefringence,lensing,cite}

\begin{thebibliography}{79}%
\makeatletter
\providecommand \@ifxundefined [1]{%
 \@ifx{#1\undefined}
}%
\providecommand \@ifnum [1]{%
 \ifnum #1\expandafter \@firstoftwo
 \else \expandafter \@secondoftwo
 \fi
}%
\providecommand \@ifx [1]{%
 \ifx #1\expandafter \@firstoftwo
 \else \expandafter \@secondoftwo
 \fi
}%
\providecommand \natexlab [1]{#1}%
\providecommand \enquote  [1]{``#1''}%
\providecommand \bibnamefont  [1]{#1}%
\providecommand \bibfnamefont [1]{#1}%
\providecommand \citenamefont [1]{#1}%
\providecommand \href@noop [0]{\@secondoftwo}%
\providecommand \href [0]{\begingroup \@sanitize@url \@href}%
\providecommand \@href[1]{\@@startlink{#1}\@@href}%
\providecommand \@@href[1]{\endgroup#1\@@endlink}%
\providecommand \@sanitize@url [0]{\catcode `\\12\catcode `\$12\catcode
  `\&12\catcode `\#12\catcode `\^12\catcode `\_12\catcode `\%12\relax}%
\providecommand \@@startlink[1]{}%
\providecommand \@@endlink[0]{}%
\providecommand \url  [0]{\begingroup\@sanitize@url \@url }%
\providecommand \@url [1]{\endgroup\@href {#1}{\urlprefix }}%
\providecommand \urlprefix  [0]{URL }%
\providecommand \Eprint [0]{\href }%
\providecommand \doibase [0]{http://dx.doi.org/}%
\providecommand \selectlanguage [0]{\@gobble}%
\providecommand \bibinfo  [0]{\@secondoftwo}%
\providecommand \bibfield  [0]{\@secondoftwo}%
\providecommand \translation [1]{[#1]}%
\providecommand \BibitemOpen [0]{}%
\providecommand \bibitemStop [0]{}%
\providecommand \bibitemNoStop [0]{.\EOS\space}%
\providecommand \EOS [0]{\spacefactor3000\relax}%
\providecommand \BibitemShut  [1]{\csname bibitem#1\endcsname}%
\let\auto@bib@innerbib\@empty
\bibitem [{\citenamefont {{Hu}}\ and\ \citenamefont
  {{Okamoto}}(2002)}]{Hu:2002}%
  \BibitemOpen
  \bibfield  {author} {\bibinfo {author} {\bibfnamefont {W.}~\bibnamefont
  {{Hu}}}\ and\ \bibinfo {author} {\bibfnamefont {T.}~\bibnamefont
  {{Okamoto}}},\ }\href {\doibase 10.1086/341110} {\bibfield  {journal}
  {\bibinfo  {journal} {\apj}\ }\textbf {\bibinfo {volume} {574}},\ \bibinfo
  {pages} {566} (\bibinfo {year} {2002})},\ \Eprint
  {http://arxiv.org/abs/astro-ph/0111606} {arXiv:astro-ph/0111606 [astro-ph]}
  \BibitemShut {NoStop}%
\bibitem [{\citenamefont {{Okamoto}}\ and\ \citenamefont
  {{Hu}}(2003)}]{Okamoto:2003}%
  \BibitemOpen
  \bibfield  {author} {\bibinfo {author} {\bibfnamefont {T.}~\bibnamefont
  {{Okamoto}}}\ and\ \bibinfo {author} {\bibfnamefont {W.}~\bibnamefont
  {{Hu}}},\ }\href {\doibase 10.1103/PhysRevD.67.083002} {\bibfield  {journal}
  {\bibinfo  {journal} {\prd}\ }\textbf {\bibinfo {volume} {67}},\ \bibinfo
  {eid} {083002} (\bibinfo {year} {2003})},\ \Eprint
  {http://arxiv.org/abs/astro-ph/0301031} {arXiv:astro-ph/0301031 [astro-ph]}
  \BibitemShut {NoStop}%
\bibitem [{\citenamefont {{Planck Collaboration}}\ \emph
  {et~al.}(2020)\citenamefont {{Planck Collaboration}}, \citenamefont
  {{Aghanim}}, \citenamefont {{Akrami}}, \citenamefont {{Ashdown}},
  \citenamefont {{Aumont}}, \citenamefont {{Baccigalupi}}, \citenamefont
  {{Ballardini}}, \citenamefont {{Banday}}, \citenamefont {{Barreiro}},
  \citenamefont {{Bartolo}}, \citenamefont {{Basak}}, \citenamefont
  {{Benabed}}, \citenamefont {{Bernard}}, \citenamefont {{Bersanelli}},
  \citenamefont {{Bielewicz}}, \citenamefont {{Bock}}, \citenamefont {{Bond}},
  \citenamefont {{Borrill}}, \citenamefont {{Bouchet}}, \citenamefont
  {{Boulanger}}, \citenamefont {{Bucher}}, \citenamefont {{Burigana}},
  \citenamefont {{Calabrese}}, \citenamefont {{Cardoso}}, \citenamefont
  {{Carron}}, \citenamefont {{Challinor}}, \citenamefont {{Chiang}},
  \citenamefont {{Colombo}}, \citenamefont {{Combet}}, \citenamefont {{Crill}},
  \citenamefont {{Cuttaia}}, \citenamefont {{de Bernardis}}, \citenamefont {{de
  Zotti}}, \citenamefont {{Delabrouille}}, \citenamefont {{Di Valentino}},
  \citenamefont {{Diego}}, \citenamefont {{Dor{\'e}}}, \citenamefont
  {{Douspis}}, \citenamefont {{Ducout}}, \citenamefont {{Dupac}}, \citenamefont
  {{Efstathiou}}, \citenamefont {{Elsner}}, \citenamefont {{En{\ss}lin}},
  \citenamefont {{Eriksen}}, \citenamefont {{Fantaye}}, \citenamefont
  {{Fernandez-Cobos}}, \citenamefont {{Finelli}}, \citenamefont {{Forastieri}},
  \citenamefont {{Frailis}}, \citenamefont {{Fraisse}}, \citenamefont
  {{Franceschi}}, \citenamefont {{Frolov}}, \citenamefont {{Galeotta}},
  \citenamefont {{Galli}}, \citenamefont {{Ganga}}, \citenamefont
  {{G{\'e}nova-Santos}}, \citenamefont {{Gerbino}}, \citenamefont {{Ghosh}},
  \citenamefont {{Gonz{\'a}lez-Nuevo}}, \citenamefont {{G{\'o}rski}},
  \citenamefont {{Gratton}}, \citenamefont {{Gruppuso}}, \citenamefont
  {{Gudmundsson}}, \citenamefont {{Hamann}}, \citenamefont {{Handley}},
  \citenamefont {{Hansen}}, \citenamefont {{Herranz}}, \citenamefont {{Hivon}},
  \citenamefont {{Huang}}, \citenamefont {{Jaffe}}, \citenamefont {{Jones}},
  \citenamefont {{Karakci}}, \citenamefont {{Keih{\"a}nen}}, \citenamefont
  {{Keskitalo}}, \citenamefont {{Kiiveri}}, \citenamefont {{Kim}},
  \citenamefont {{Knox}}, \citenamefont {{Krachmalnicoff}}, \citenamefont
  {{Kunz}}, \citenamefont {{Kurki-Suonio}}, \citenamefont {{Lagache}},
  \citenamefont {{Lamarre}}, \citenamefont {{Lasenby}}, \citenamefont
  {{Lattanzi}}, \citenamefont {{Lawrence}}, \citenamefont {{Le Jeune}},
  \citenamefont {{Levrier}}, \citenamefont {{Lewis}}, \citenamefont
  {{Liguori}}, \citenamefont {{Lilje}}, \citenamefont {{Lindholm}},
  \citenamefont {{L{\'o}pez-Caniego}}, \citenamefont {{Lubin}}, \citenamefont
  {{Ma}}, \citenamefont {{Mac{\'\i}as-P{\'e}rez}}, \citenamefont {{Maggio}},
  \citenamefont {{Maino}}, \citenamefont {{Mandolesi}}, \citenamefont
  {{Mangilli}}, \citenamefont {{Marcos-Caballero}}, \citenamefont {{Maris}},
  \citenamefont {{Martin}}, \citenamefont {{Mart{\'\i}nez-Gonz{\'a}lez}},
  \citenamefont {{Matarrese}}, \citenamefont {{Mauri}}, \citenamefont
  {{McEwen}}, \citenamefont {{Melchiorri}}, \citenamefont {{Mennella}},
  \citenamefont {{Migliaccio}}, \citenamefont {{Miville-Desch{\^e}nes}},
  \citenamefont {{Molinari}}, \citenamefont {{Moneti}}, \citenamefont
  {{Montier}}, \citenamefont {{Morgante}}, \citenamefont {{Moss}},
  \citenamefont {{Natoli}}, \citenamefont {{Pagano}}, \citenamefont
  {{Paoletti}}, \citenamefont {{Partridge}}, \citenamefont {{Patanchon}},
  \citenamefont {{Perrotta}}, \citenamefont {{Pettorino}}, \citenamefont
  {{Piacentini}}, \citenamefont {{Polastri}}, \citenamefont {{Polenta}},
  \citenamefont {{Puget}}, \citenamefont {{Rachen}}, \citenamefont
  {{Reinecke}}, \citenamefont {{Remazeilles}}, \citenamefont {{Renzi}},
  \citenamefont {{Rocha}}, \citenamefont {{Rosset}}, \citenamefont {{Roudier}},
  \citenamefont {{Rubi{\~n}o-Mart{\'\i}n}}, \citenamefont {{Ruiz-Granados}},
  \citenamefont {{Salvati}}, \citenamefont {{Sandri}}, \citenamefont
  {{Savelainen}}, \citenamefont {{Scott}}, \citenamefont {{Sirignano}},
  \citenamefont {{Sunyaev}}, \citenamefont {{Suur-Uski}}, \citenamefont
  {{Tauber}}, \citenamefont {{Tavagnacco}}, \citenamefont {{Tenti}},
  \citenamefont {{Toffolatti}}, \citenamefont {{Tomasi}}, \citenamefont
  {{Trombetti}}, \citenamefont {{Valiviita}}, \citenamefont {{Van Tent}},
  \citenamefont {{Vielva}}, \citenamefont {{Villa}}, \citenamefont
  {{Vittorio}}, \citenamefont {{Wandelt}}, \citenamefont {{Wehus}},
  \citenamefont {{White}}, \citenamefont {{White}}, \citenamefont {{Zacchei}},\
  and\ \citenamefont {{Zonca}}}]{PlanckCollaboration:2020}%
  \BibitemOpen
  \bibfield  {author} {\bibinfo {author} {\bibnamefont {{Planck
  Collaboration}}}, \bibinfo {author} {\bibfnamefont {N.}~\bibnamefont
  {{Aghanim}}}, \bibinfo {author} {\bibfnamefont {Y.}~\bibnamefont {{Akrami}}},
  \bibinfo {author} {\bibfnamefont {M.}~\bibnamefont {{Ashdown}}}, \bibinfo
  {author} {\bibfnamefont {J.}~\bibnamefont {{Aumont}}}, \bibinfo {author}
  {\bibfnamefont {C.}~\bibnamefont {{Baccigalupi}}}, \bibinfo {author}
  {\bibfnamefont {M.}~\bibnamefont {{Ballardini}}}, \bibinfo {author}
  {\bibfnamefont {A.~J.}\ \bibnamefont {{Banday}}}, \bibinfo {author}
  {\bibfnamefont {R.~B.}\ \bibnamefont {{Barreiro}}}, \bibinfo {author}
  {\bibfnamefont {N.}~\bibnamefont {{Bartolo}}}, \bibinfo {author}
  {\bibfnamefont {S.}~\bibnamefont {{Basak}}}, \bibinfo {author} {\bibfnamefont
  {K.}~\bibnamefont {{Benabed}}}, \bibinfo {author} {\bibfnamefont {J.~P.}\
  \bibnamefont {{Bernard}}}, \bibinfo {author} {\bibfnamefont {M.}~\bibnamefont
  {{Bersanelli}}}, \bibinfo {author} {\bibfnamefont {P.}~\bibnamefont
  {{Bielewicz}}}, \bibinfo {author} {\bibfnamefont {J.~J.}\ \bibnamefont
  {{Bock}}}, \bibinfo {author} {\bibfnamefont {J.~R.}\ \bibnamefont {{Bond}}},
  \bibinfo {author} {\bibfnamefont {J.}~\bibnamefont {{Borrill}}}, \bibinfo
  {author} {\bibfnamefont {F.~R.}\ \bibnamefont {{Bouchet}}}, \bibinfo {author}
  {\bibfnamefont {F.}~\bibnamefont {{Boulanger}}}, \bibinfo {author}
  {\bibfnamefont {M.}~\bibnamefont {{Bucher}}}, \bibinfo {author}
  {\bibfnamefont {C.}~\bibnamefont {{Burigana}}}, \bibinfo {author}
  {\bibfnamefont {E.}~\bibnamefont {{Calabrese}}}, \bibinfo {author}
  {\bibfnamefont {J.~F.}\ \bibnamefont {{Cardoso}}}, \bibinfo {author}
  {\bibfnamefont {J.}~\bibnamefont {{Carron}}}, \bibinfo {author}
  {\bibfnamefont {A.}~\bibnamefont {{Challinor}}}, \bibinfo {author}
  {\bibfnamefont {H.~C.}\ \bibnamefont {{Chiang}}}, \bibinfo {author}
  {\bibfnamefont {L.~P.~L.}\ \bibnamefont {{Colombo}}}, \bibinfo {author}
  {\bibfnamefont {C.}~\bibnamefont {{Combet}}}, \bibinfo {author}
  {\bibfnamefont {B.~P.}\ \bibnamefont {{Crill}}}, \bibinfo {author}
  {\bibfnamefont {F.}~\bibnamefont {{Cuttaia}}}, \bibinfo {author}
  {\bibfnamefont {P.}~\bibnamefont {{de Bernardis}}}, \bibinfo {author}
  {\bibfnamefont {G.}~\bibnamefont {{de Zotti}}}, \bibinfo {author}
  {\bibfnamefont {J.}~\bibnamefont {{Delabrouille}}}, \bibinfo {author}
  {\bibfnamefont {E.}~\bibnamefont {{Di Valentino}}}, \bibinfo {author}
  {\bibfnamefont {J.~M.}\ \bibnamefont {{Diego}}}, \bibinfo {author}
  {\bibfnamefont {O.}~\bibnamefont {{Dor{\'e}}}}, \bibinfo {author}
  {\bibfnamefont {M.}~\bibnamefont {{Douspis}}}, \bibinfo {author}
  {\bibfnamefont {A.}~\bibnamefont {{Ducout}}}, \bibinfo {author}
  {\bibfnamefont {X.}~\bibnamefont {{Dupac}}}, \bibinfo {author} {\bibfnamefont
  {G.}~\bibnamefont {{Efstathiou}}}, \bibinfo {author} {\bibfnamefont
  {F.}~\bibnamefont {{Elsner}}}, \bibinfo {author} {\bibfnamefont {T.~A.}\
  \bibnamefont {{En{\ss}lin}}}, \bibinfo {author} {\bibfnamefont {H.~K.}\
  \bibnamefont {{Eriksen}}}, \bibinfo {author} {\bibfnamefont {Y.}~\bibnamefont
  {{Fantaye}}}, \bibinfo {author} {\bibfnamefont {R.}~\bibnamefont
  {{Fernandez-Cobos}}}, \bibinfo {author} {\bibfnamefont {F.}~\bibnamefont
  {{Finelli}}}, \bibinfo {author} {\bibfnamefont {F.}~\bibnamefont
  {{Forastieri}}}, \bibinfo {author} {\bibfnamefont {M.}~\bibnamefont
  {{Frailis}}}, \bibinfo {author} {\bibfnamefont {A.~A.}\ \bibnamefont
  {{Fraisse}}}, \bibinfo {author} {\bibfnamefont {E.}~\bibnamefont
  {{Franceschi}}}, \bibinfo {author} {\bibfnamefont {A.}~\bibnamefont
  {{Frolov}}}, \bibinfo {author} {\bibfnamefont {S.}~\bibnamefont
  {{Galeotta}}}, \bibinfo {author} {\bibfnamefont {S.}~\bibnamefont {{Galli}}},
  \bibinfo {author} {\bibfnamefont {K.}~\bibnamefont {{Ganga}}}, \bibinfo
  {author} {\bibfnamefont {R.~T.}\ \bibnamefont {{G{\'e}nova-Santos}}},
  \bibinfo {author} {\bibfnamefont {M.}~\bibnamefont {{Gerbino}}}, \bibinfo
  {author} {\bibfnamefont {T.}~\bibnamefont {{Ghosh}}}, \bibinfo {author}
  {\bibfnamefont {J.}~\bibnamefont {{Gonz{\'a}lez-Nuevo}}}, \bibinfo {author}
  {\bibfnamefont {K.~M.}\ \bibnamefont {{G{\'o}rski}}}, \bibinfo {author}
  {\bibfnamefont {S.}~\bibnamefont {{Gratton}}}, \bibinfo {author}
  {\bibfnamefont {A.}~\bibnamefont {{Gruppuso}}}, \bibinfo {author}
  {\bibfnamefont {J.~E.}\ \bibnamefont {{Gudmundsson}}}, \bibinfo {author}
  {\bibfnamefont {J.}~\bibnamefont {{Hamann}}}, \bibinfo {author}
  {\bibfnamefont {W.}~\bibnamefont {{Handley}}}, \bibinfo {author}
  {\bibfnamefont {F.~K.}\ \bibnamefont {{Hansen}}}, \bibinfo {author}
  {\bibfnamefont {D.}~\bibnamefont {{Herranz}}}, \bibinfo {author}
  {\bibfnamefont {E.}~\bibnamefont {{Hivon}}}, \bibinfo {author} {\bibfnamefont
  {Z.}~\bibnamefont {{Huang}}}, \bibinfo {author} {\bibfnamefont {A.~H.}\
  \bibnamefont {{Jaffe}}}, \bibinfo {author} {\bibfnamefont {W.~C.}\
  \bibnamefont {{Jones}}}, \bibinfo {author} {\bibfnamefont {A.}~\bibnamefont
  {{Karakci}}}, \bibinfo {author} {\bibfnamefont {E.}~\bibnamefont
  {{Keih{\"a}nen}}}, \bibinfo {author} {\bibfnamefont {R.}~\bibnamefont
  {{Keskitalo}}}, \bibinfo {author} {\bibfnamefont {K.}~\bibnamefont
  {{Kiiveri}}}, \bibinfo {author} {\bibfnamefont {J.}~\bibnamefont {{Kim}}},
  \bibinfo {author} {\bibfnamefont {L.}~\bibnamefont {{Knox}}}, \bibinfo
  {author} {\bibfnamefont {N.}~\bibnamefont {{Krachmalnicoff}}}, \bibinfo
  {author} {\bibfnamefont {M.}~\bibnamefont {{Kunz}}}, \bibinfo {author}
  {\bibfnamefont {H.}~\bibnamefont {{Kurki-Suonio}}}, \bibinfo {author}
  {\bibfnamefont {G.}~\bibnamefont {{Lagache}}}, \bibinfo {author}
  {\bibfnamefont {J.~M.}\ \bibnamefont {{Lamarre}}}, \bibinfo {author}
  {\bibfnamefont {A.}~\bibnamefont {{Lasenby}}}, \bibinfo {author}
  {\bibfnamefont {M.}~\bibnamefont {{Lattanzi}}}, \bibinfo {author}
  {\bibfnamefont {C.~R.}\ \bibnamefont {{Lawrence}}}, \bibinfo {author}
  {\bibfnamefont {M.}~\bibnamefont {{Le Jeune}}}, \bibinfo {author}
  {\bibfnamefont {F.}~\bibnamefont {{Levrier}}}, \bibinfo {author}
  {\bibfnamefont {A.}~\bibnamefont {{Lewis}}}, \bibinfo {author} {\bibfnamefont
  {M.}~\bibnamefont {{Liguori}}}, \bibinfo {author} {\bibfnamefont {P.~B.}\
  \bibnamefont {{Lilje}}}, \bibinfo {author} {\bibfnamefont {V.}~\bibnamefont
  {{Lindholm}}}, \bibinfo {author} {\bibfnamefont {M.}~\bibnamefont
  {{L{\'o}pez-Caniego}}}, \bibinfo {author} {\bibfnamefont {P.~M.}\
  \bibnamefont {{Lubin}}}, \bibinfo {author} {\bibfnamefont {Y.~Z.}\
  \bibnamefont {{Ma}}}, \bibinfo {author} {\bibfnamefont {J.~F.}\ \bibnamefont
  {{Mac{\'\i}as-P{\'e}rez}}}, \bibinfo {author} {\bibfnamefont
  {G.}~\bibnamefont {{Maggio}}}, \bibinfo {author} {\bibfnamefont
  {D.}~\bibnamefont {{Maino}}}, \bibinfo {author} {\bibfnamefont
  {N.}~\bibnamefont {{Mandolesi}}}, \bibinfo {author} {\bibfnamefont
  {A.}~\bibnamefont {{Mangilli}}}, \bibinfo {author} {\bibfnamefont
  {A.}~\bibnamefont {{Marcos-Caballero}}}, \bibinfo {author} {\bibfnamefont
  {M.}~\bibnamefont {{Maris}}}, \bibinfo {author} {\bibfnamefont {P.~G.}\
  \bibnamefont {{Martin}}}, \bibinfo {author} {\bibfnamefont {E.}~\bibnamefont
  {{Mart{\'\i}nez-Gonz{\'a}lez}}}, \bibinfo {author} {\bibfnamefont
  {S.}~\bibnamefont {{Matarrese}}}, \bibinfo {author} {\bibfnamefont
  {N.}~\bibnamefont {{Mauri}}}, \bibinfo {author} {\bibfnamefont {J.~D.}\
  \bibnamefont {{McEwen}}}, \bibinfo {author} {\bibfnamefont {A.}~\bibnamefont
  {{Melchiorri}}}, \bibinfo {author} {\bibfnamefont {A.}~\bibnamefont
  {{Mennella}}}, \bibinfo {author} {\bibfnamefont {M.}~\bibnamefont
  {{Migliaccio}}}, \bibinfo {author} {\bibfnamefont {M.~A.}\ \bibnamefont
  {{Miville-Desch{\^e}nes}}}, \bibinfo {author} {\bibfnamefont
  {D.}~\bibnamefont {{Molinari}}}, \bibinfo {author} {\bibfnamefont
  {A.}~\bibnamefont {{Moneti}}}, \bibinfo {author} {\bibfnamefont
  {L.}~\bibnamefont {{Montier}}}, \bibinfo {author} {\bibfnamefont
  {G.}~\bibnamefont {{Morgante}}}, \bibinfo {author} {\bibfnamefont
  {A.}~\bibnamefont {{Moss}}}, \bibinfo {author} {\bibfnamefont
  {P.}~\bibnamefont {{Natoli}}}, \bibinfo {author} {\bibfnamefont
  {L.}~\bibnamefont {{Pagano}}}, \bibinfo {author} {\bibfnamefont
  {D.}~\bibnamefont {{Paoletti}}}, \bibinfo {author} {\bibfnamefont
  {B.}~\bibnamefont {{Partridge}}}, \bibinfo {author} {\bibfnamefont
  {G.}~\bibnamefont {{Patanchon}}}, \bibinfo {author} {\bibfnamefont
  {F.}~\bibnamefont {{Perrotta}}}, \bibinfo {author} {\bibfnamefont
  {V.}~\bibnamefont {{Pettorino}}}, \bibinfo {author} {\bibfnamefont
  {F.}~\bibnamefont {{Piacentini}}}, \bibinfo {author} {\bibfnamefont
  {L.}~\bibnamefont {{Polastri}}}, \bibinfo {author} {\bibfnamefont
  {G.}~\bibnamefont {{Polenta}}}, \bibinfo {author} {\bibfnamefont {J.~L.}\
  \bibnamefont {{Puget}}}, \bibinfo {author} {\bibfnamefont {J.~P.}\
  \bibnamefont {{Rachen}}}, \bibinfo {author} {\bibfnamefont {M.}~\bibnamefont
  {{Reinecke}}}, \bibinfo {author} {\bibfnamefont {M.}~\bibnamefont
  {{Remazeilles}}}, \bibinfo {author} {\bibfnamefont {A.}~\bibnamefont
  {{Renzi}}}, \bibinfo {author} {\bibfnamefont {G.}~\bibnamefont {{Rocha}}},
  \bibinfo {author} {\bibfnamefont {C.}~\bibnamefont {{Rosset}}}, \bibinfo
  {author} {\bibfnamefont {G.}~\bibnamefont {{Roudier}}}, \bibinfo {author}
  {\bibfnamefont {J.~A.}\ \bibnamefont {{Rubi{\~n}o-Mart{\'\i}n}}}, \bibinfo
  {author} {\bibfnamefont {B.}~\bibnamefont {{Ruiz-Granados}}}, \bibinfo
  {author} {\bibfnamefont {L.}~\bibnamefont {{Salvati}}}, \bibinfo {author}
  {\bibfnamefont {M.}~\bibnamefont {{Sandri}}}, \bibinfo {author}
  {\bibfnamefont {M.}~\bibnamefont {{Savelainen}}}, \bibinfo {author}
  {\bibfnamefont {D.}~\bibnamefont {{Scott}}}, \bibinfo {author} {\bibfnamefont
  {C.}~\bibnamefont {{Sirignano}}}, \bibinfo {author} {\bibfnamefont
  {R.}~\bibnamefont {{Sunyaev}}}, \bibinfo {author} {\bibfnamefont {A.~S.}\
  \bibnamefont {{Suur-Uski}}}, \bibinfo {author} {\bibfnamefont {J.~A.}\
  \bibnamefont {{Tauber}}}, \bibinfo {author} {\bibfnamefont {D.}~\bibnamefont
  {{Tavagnacco}}}, \bibinfo {author} {\bibfnamefont {M.}~\bibnamefont
  {{Tenti}}}, \bibinfo {author} {\bibfnamefont {L.}~\bibnamefont
  {{Toffolatti}}}, \bibinfo {author} {\bibfnamefont {M.}~\bibnamefont
  {{Tomasi}}}, \bibinfo {author} {\bibfnamefont {T.}~\bibnamefont
  {{Trombetti}}}, \bibinfo {author} {\bibfnamefont {J.}~\bibnamefont
  {{Valiviita}}}, \bibinfo {author} {\bibfnamefont {B.}~\bibnamefont {{Van
  Tent}}}, \bibinfo {author} {\bibfnamefont {P.}~\bibnamefont {{Vielva}}},
  \bibinfo {author} {\bibfnamefont {F.}~\bibnamefont {{Villa}}}, \bibinfo
  {author} {\bibfnamefont {N.}~\bibnamefont {{Vittorio}}}, \bibinfo {author}
  {\bibfnamefont {B.~D.}\ \bibnamefont {{Wandelt}}}, \bibinfo {author}
  {\bibfnamefont {I.~K.}\ \bibnamefont {{Wehus}}}, \bibinfo {author}
  {\bibfnamefont {M.}~\bibnamefont {{White}}}, \bibinfo {author} {\bibfnamefont
  {S.~D.~M.}\ \bibnamefont {{White}}}, \bibinfo {author} {\bibfnamefont
  {A.}~\bibnamefont {{Zacchei}}}, \ and\ \bibinfo {author} {\bibfnamefont
  {A.}~\bibnamefont {{Zonca}}},\ }\href {\doibase 10.1051/0004-6361/201833886}
  {\bibfield  {journal} {\bibinfo  {journal} {\aap}\ }\textbf {\bibinfo
  {volume} {641}},\ \bibinfo {eid} {A8} (\bibinfo {year} {2020})},\ \Eprint
  {http://arxiv.org/abs/1807.06210} {arXiv:1807.06210 [astro-ph.CO]}
  \BibitemShut {NoStop}%
\bibitem [{\citenamefont {{Planck Collaboration}}\ \emph
  {et~al.}(2014)\citenamefont {{Planck Collaboration}}, \citenamefont {{Ade}},
  \citenamefont {{Aghanim}}, \citenamefont {{Armitage-Caplan}}, \citenamefont
  {{Arnaud}}, \citenamefont {{Ashdown}}, \citenamefont {{Atrio-Barandela}},
  \citenamefont {{Aumont}}, \citenamefont {{Baccigalupi}}, \citenamefont
  {{Banday}},\ and\ \citenamefont {et~al.}}]{PlanckCollaboration:2014}%
  \BibitemOpen
  \bibfield  {author} {\bibinfo {author} {\bibnamefont {{Planck
  Collaboration}}}, \bibinfo {author} {\bibfnamefont {P.~A.~R.}\ \bibnamefont
  {{Ade}}}, \bibinfo {author} {\bibfnamefont {N.}~\bibnamefont {{Aghanim}}},
  \bibinfo {author} {\bibfnamefont {C.}~\bibnamefont {{Armitage-Caplan}}},
  \bibinfo {author} {\bibfnamefont {M.}~\bibnamefont {{Arnaud}}}, \bibinfo
  {author} {\bibfnamefont {M.}~\bibnamefont {{Ashdown}}}, \bibinfo {author}
  {\bibfnamefont {F.}~\bibnamefont {{Atrio-Barandela}}}, \bibinfo {author}
  {\bibfnamefont {J.}~\bibnamefont {{Aumont}}}, \bibinfo {author}
  {\bibfnamefont {C.}~\bibnamefont {{Baccigalupi}}}, \bibinfo {author}
  {\bibfnamefont {A.~J.}\ \bibnamefont {{Banday}}}, \ and\ \bibinfo {author}
  {\bibnamefont {et~al.}},\ }\href {\doibase 10.1051/0004-6361/201321543}
  {\bibfield  {journal} {\bibinfo  {journal} {\aap}\ }\textbf {\bibinfo
  {volume} {571}},\ \bibinfo {eid} {A17} (\bibinfo {year} {2014})},\ \Eprint
  {http://arxiv.org/abs/1303.5077} {arXiv:1303.5077 [astro-ph.CO]} \BibitemShut
  {NoStop}%
\bibitem [{\citenamefont {{Carron}}\ \emph {et~al.}(2022)\citenamefont
  {{Carron}}, \citenamefont {{Mirmelstein}},\ and\ \citenamefont
  {{Lewis}}}]{carron:2022:lensing}%
  \BibitemOpen
  \bibfield  {author} {\bibinfo {author} {\bibfnamefont {J.}~\bibnamefont
  {{Carron}}}, \bibinfo {author} {\bibfnamefont {M.}~\bibnamefont
  {{Mirmelstein}}}, \ and\ \bibinfo {author} {\bibfnamefont {A.}~\bibnamefont
  {{Lewis}}},\ }\href@noop {} {\bibfield  {journal} {\bibinfo  {journal} {arXiv
  e-prints}\ ,\ \bibinfo {eid} {arXiv:2206.07773}} (\bibinfo {year} {2022})},\
  \Eprint {http://arxiv.org/abs/2206.07773} {arXiv:2206.07773 [astro-ph.CO]}
  \BibitemShut {NoStop}%
\bibitem [{\citenamefont {{van Engelen}}\ \emph {et~al.}(2015)\citenamefont
  {{van Engelen}}, \citenamefont {{Sherwin}}, \citenamefont {{Sehgal}},
  \citenamefont {{Addison}}, \citenamefont {{Allison}}, \citenamefont
  {{Battaglia}}, \citenamefont {{de Bernardis}}, \citenamefont {{Bond}},
  \citenamefont {{Calabrese}}, \citenamefont {{Coughlin}}, \citenamefont
  {{Crichton}}, \citenamefont {{Datta}}, \citenamefont {{Devlin}},
  \citenamefont {{Dunkley}}, \citenamefont {{D{\"u}nner}}, \citenamefont
  {{Gallardo}}, \citenamefont {{Grace}}, \citenamefont {{Gralla}},
  \citenamefont {{Hajian}}, \citenamefont {{Hasselfield}}, \citenamefont
  {{Henderson}}, \citenamefont {{Hill}}, \citenamefont {{Hilton}},
  \citenamefont {{Hincks}}, \citenamefont {{Hlozek}}, \citenamefont
  {{Huffenberger}}, \citenamefont {{Hughes}}, \citenamefont {{Koopman}},
  \citenamefont {{Kosowsky}}, \citenamefont {{Louis}}, \citenamefont {{Lungu}},
  \citenamefont {{Madhavacheril}}, \citenamefont {{Maurin}}, \citenamefont
  {{McMahon}}, \citenamefont {{Moodley}}, \citenamefont {{Munson}},
  \citenamefont {{Naess}}, \citenamefont {{Nati}}, \citenamefont {{Newburgh}},
  \citenamefont {{Niemack}}, \citenamefont {{Nolta}}, \citenamefont {{Page}},
  \citenamefont {{Pappas}}, \citenamefont {{Partridge}}, \citenamefont
  {{Schmitt}}, \citenamefont {{Sievers}}, \citenamefont {{Simon}},
  \citenamefont {{Spergel}}, \citenamefont {{Staggs}}, \citenamefont
  {{Switzer}}, \citenamefont {{Ward}},\ and\ \citenamefont
  {{Wollack}}}]{vanEngelen:2015}%
  \BibitemOpen
  \bibfield  {author} {\bibinfo {author} {\bibfnamefont {A.}~\bibnamefont {{van
  Engelen}}}, \bibinfo {author} {\bibfnamefont {B.~D.}\ \bibnamefont
  {{Sherwin}}}, \bibinfo {author} {\bibfnamefont {N.}~\bibnamefont {{Sehgal}}},
  \bibinfo {author} {\bibfnamefont {G.~E.}\ \bibnamefont {{Addison}}}, \bibinfo
  {author} {\bibfnamefont {R.}~\bibnamefont {{Allison}}}, \bibinfo {author}
  {\bibfnamefont {N.}~\bibnamefont {{Battaglia}}}, \bibinfo {author}
  {\bibfnamefont {F.}~\bibnamefont {{de Bernardis}}}, \bibinfo {author}
  {\bibfnamefont {J.~R.}\ \bibnamefont {{Bond}}}, \bibinfo {author}
  {\bibfnamefont {E.}~\bibnamefont {{Calabrese}}}, \bibinfo {author}
  {\bibfnamefont {K.}~\bibnamefont {{Coughlin}}}, \bibinfo {author}
  {\bibfnamefont {D.}~\bibnamefont {{Crichton}}}, \bibinfo {author}
  {\bibfnamefont {R.}~\bibnamefont {{Datta}}}, \bibinfo {author} {\bibfnamefont
  {M.~J.}\ \bibnamefont {{Devlin}}}, \bibinfo {author} {\bibfnamefont
  {J.}~\bibnamefont {{Dunkley}}}, \bibinfo {author} {\bibfnamefont
  {R.}~\bibnamefont {{D{\"u}nner}}}, \bibinfo {author} {\bibfnamefont
  {P.}~\bibnamefont {{Gallardo}}}, \bibinfo {author} {\bibfnamefont
  {E.}~\bibnamefont {{Grace}}}, \bibinfo {author} {\bibfnamefont
  {M.}~\bibnamefont {{Gralla}}}, \bibinfo {author} {\bibfnamefont
  {A.}~\bibnamefont {{Hajian}}}, \bibinfo {author} {\bibfnamefont
  {M.}~\bibnamefont {{Hasselfield}}}, \bibinfo {author} {\bibfnamefont
  {S.}~\bibnamefont {{Henderson}}}, \bibinfo {author} {\bibfnamefont {J.~C.}\
  \bibnamefont {{Hill}}}, \bibinfo {author} {\bibfnamefont {M.}~\bibnamefont
  {{Hilton}}}, \bibinfo {author} {\bibfnamefont {A.~D.}\ \bibnamefont
  {{Hincks}}}, \bibinfo {author} {\bibfnamefont {R.}~\bibnamefont {{Hlozek}}},
  \bibinfo {author} {\bibfnamefont {K.~M.}\ \bibnamefont {{Huffenberger}}},
  \bibinfo {author} {\bibfnamefont {J.~P.}\ \bibnamefont {{Hughes}}}, \bibinfo
  {author} {\bibfnamefont {B.}~\bibnamefont {{Koopman}}}, \bibinfo {author}
  {\bibfnamefont {A.}~\bibnamefont {{Kosowsky}}}, \bibinfo {author}
  {\bibfnamefont {T.}~\bibnamefont {{Louis}}}, \bibinfo {author} {\bibfnamefont
  {M.}~\bibnamefont {{Lungu}}}, \bibinfo {author} {\bibfnamefont
  {M.}~\bibnamefont {{Madhavacheril}}}, \bibinfo {author} {\bibfnamefont
  {L.}~\bibnamefont {{Maurin}}}, \bibinfo {author} {\bibfnamefont
  {J.}~\bibnamefont {{McMahon}}}, \bibinfo {author} {\bibfnamefont
  {K.}~\bibnamefont {{Moodley}}}, \bibinfo {author} {\bibfnamefont
  {C.}~\bibnamefont {{Munson}}}, \bibinfo {author} {\bibfnamefont
  {S.}~\bibnamefont {{Naess}}}, \bibinfo {author} {\bibfnamefont
  {F.}~\bibnamefont {{Nati}}}, \bibinfo {author} {\bibfnamefont
  {L.}~\bibnamefont {{Newburgh}}}, \bibinfo {author} {\bibfnamefont {M.~D.}\
  \bibnamefont {{Niemack}}}, \bibinfo {author} {\bibfnamefont {M.~R.}\
  \bibnamefont {{Nolta}}}, \bibinfo {author} {\bibfnamefont {L.~A.}\
  \bibnamefont {{Page}}}, \bibinfo {author} {\bibfnamefont {C.}~\bibnamefont
  {{Pappas}}}, \bibinfo {author} {\bibfnamefont {B.}~\bibnamefont
  {{Partridge}}}, \bibinfo {author} {\bibfnamefont {B.~L.}\ \bibnamefont
  {{Schmitt}}}, \bibinfo {author} {\bibfnamefont {J.~L.}\ \bibnamefont
  {{Sievers}}}, \bibinfo {author} {\bibfnamefont {S.}~\bibnamefont {{Simon}}},
  \bibinfo {author} {\bibfnamefont {D.~N.}\ \bibnamefont {{Spergel}}}, \bibinfo
  {author} {\bibfnamefont {S.~T.}\ \bibnamefont {{Staggs}}}, \bibinfo {author}
  {\bibfnamefont {E.~R.}\ \bibnamefont {{Switzer}}}, \bibinfo {author}
  {\bibfnamefont {J.~T.}\ \bibnamefont {{Ward}}}, \ and\ \bibinfo {author}
  {\bibfnamefont {E.~J.}\ \bibnamefont {{Wollack}}},\ }\href {\doibase
  10.1088/0004-637X/808/1/7} {\bibfield  {journal} {\bibinfo  {journal} {\apj}\
  }\textbf {\bibinfo {volume} {808}},\ \bibinfo {eid} {7} (\bibinfo {year}
  {2015})},\ \Eprint {http://arxiv.org/abs/1412.0626} {arXiv:1412.0626
  [astro-ph.CO]} \BibitemShut {NoStop}%
\bibitem [{\citenamefont {{BICEP2 Collaboration}}\ \emph
  {et~al.}(2016)\citenamefont {{BICEP2 Collaboration}}, \citenamefont {{Keck
  Array Collaboration}}, \citenamefont {{Ade}}, \citenamefont {{Ahmed}},
  \citenamefont {{Aikin}}, \citenamefont {{Alexander}}, \citenamefont
  {{Barkats}}, \citenamefont {{Benton}}, \citenamefont {{Bischoff}},
  \citenamefont {{Bock}}, \citenamefont {{Bowens-Rubin}}, \citenamefont
  {{Brevik}}, \citenamefont {{Buder}}, \citenamefont {{Bullock}}, \citenamefont
  {{Buza}}, \citenamefont {{Connors}}, \citenamefont {{Crill}}, \citenamefont
  {{Duband}}, \citenamefont {{Dvorkin}}, \citenamefont {{Filippini}},
  \citenamefont {{Fliescher}}, \citenamefont {{Grayson}}, \citenamefont
  {{Halpern}}, \citenamefont {{Harrison}}, \citenamefont {{Hildebrandt}},
  \citenamefont {{Hilton}}, \citenamefont {{Hui}}, \citenamefont {{Irwin}},
  \citenamefont {{Kang}}, \citenamefont {{Karkare}}, \citenamefont {{Karpel}},
  \citenamefont {{Kaufman}}, \citenamefont {{Keating}}, \citenamefont
  {{Kefeli}}, \citenamefont {{Kernasovskiy}}, \citenamefont {{Kovac}},
  \citenamefont {{Kuo}}, \citenamefont {{Leitch}}, \citenamefont {{Lueker}},
  \citenamefont {{Megerian}}, \citenamefont {{Namikawa}}, \citenamefont
  {{Netterfield}}, \citenamefont {{Nguyen}}, \citenamefont {{O'Brient}},
  \citenamefont {{Ogburn}}, \citenamefont {{Orlando}}, \citenamefont {{Pryke}},
  \citenamefont {{Richter}}, \citenamefont {{Schwarz}}, \citenamefont
  {{Sheehy}}, \citenamefont {{Staniszewski}}, \citenamefont {{Steinbach}},
  \citenamefont {{Sudiwala}}, \citenamefont {{Teply}}, \citenamefont
  {{Thompson}}, \citenamefont {{Tolan}}, \citenamefont {{Tucker}},
  \citenamefont {{Turner}}, \citenamefont {{Vieregg}}, \citenamefont {{Weber}},
  \citenamefont {{Wiebe}}, \citenamefont {{Willmert}}, \citenamefont {{Wong}},
  \citenamefont {{Wu}},\ and\ \citenamefont
  {{Yoon}}}]{BICEP2Collaboration:2016}%
  \BibitemOpen
  \bibfield  {author} {\bibinfo {author} {\bibnamefont {{BICEP2
  Collaboration}}}, \bibinfo {author} {\bibnamefont {{Keck Array
  Collaboration}}}, \bibinfo {author} {\bibfnamefont {P.~A.~R.}\ \bibnamefont
  {{Ade}}}, \bibinfo {author} {\bibfnamefont {Z.}~\bibnamefont {{Ahmed}}},
  \bibinfo {author} {\bibfnamefont {R.~W.}\ \bibnamefont {{Aikin}}}, \bibinfo
  {author} {\bibfnamefont {K.~D.}\ \bibnamefont {{Alexander}}}, \bibinfo
  {author} {\bibfnamefont {D.}~\bibnamefont {{Barkats}}}, \bibinfo {author}
  {\bibfnamefont {S.~J.}\ \bibnamefont {{Benton}}}, \bibinfo {author}
  {\bibfnamefont {C.~A.}\ \bibnamefont {{Bischoff}}}, \bibinfo {author}
  {\bibfnamefont {J.~J.}\ \bibnamefont {{Bock}}}, \bibinfo {author}
  {\bibfnamefont {R.}~\bibnamefont {{Bowens-Rubin}}}, \bibinfo {author}
  {\bibfnamefont {J.~A.}\ \bibnamefont {{Brevik}}}, \bibinfo {author}
  {\bibfnamefont {I.}~\bibnamefont {{Buder}}}, \bibinfo {author} {\bibfnamefont
  {E.}~\bibnamefont {{Bullock}}}, \bibinfo {author} {\bibfnamefont
  {V.}~\bibnamefont {{Buza}}}, \bibinfo {author} {\bibfnamefont
  {J.}~\bibnamefont {{Connors}}}, \bibinfo {author} {\bibfnamefont {B.~P.}\
  \bibnamefont {{Crill}}}, \bibinfo {author} {\bibfnamefont {L.}~\bibnamefont
  {{Duband}}}, \bibinfo {author} {\bibfnamefont {C.}~\bibnamefont {{Dvorkin}}},
  \bibinfo {author} {\bibfnamefont {J.~P.}\ \bibnamefont {{Filippini}}},
  \bibinfo {author} {\bibfnamefont {S.}~\bibnamefont {{Fliescher}}}, \bibinfo
  {author} {\bibfnamefont {J.}~\bibnamefont {{Grayson}}}, \bibinfo {author}
  {\bibfnamefont {M.}~\bibnamefont {{Halpern}}}, \bibinfo {author}
  {\bibfnamefont {S.}~\bibnamefont {{Harrison}}}, \bibinfo {author}
  {\bibfnamefont {S.~R.}\ \bibnamefont {{Hildebrandt}}}, \bibinfo {author}
  {\bibfnamefont {G.~C.}\ \bibnamefont {{Hilton}}}, \bibinfo {author}
  {\bibfnamefont {H.}~\bibnamefont {{Hui}}}, \bibinfo {author} {\bibfnamefont
  {K.~D.}\ \bibnamefont {{Irwin}}}, \bibinfo {author} {\bibfnamefont
  {J.}~\bibnamefont {{Kang}}}, \bibinfo {author} {\bibfnamefont {K.~S.}\
  \bibnamefont {{Karkare}}}, \bibinfo {author} {\bibfnamefont {E.}~\bibnamefont
  {{Karpel}}}, \bibinfo {author} {\bibfnamefont {J.~P.}\ \bibnamefont
  {{Kaufman}}}, \bibinfo {author} {\bibfnamefont {B.~G.}\ \bibnamefont
  {{Keating}}}, \bibinfo {author} {\bibfnamefont {S.}~\bibnamefont {{Kefeli}}},
  \bibinfo {author} {\bibfnamefont {S.~A.}\ \bibnamefont {{Kernasovskiy}}},
  \bibinfo {author} {\bibfnamefont {J.~M.}\ \bibnamefont {{Kovac}}}, \bibinfo
  {author} {\bibfnamefont {C.~L.}\ \bibnamefont {{Kuo}}}, \bibinfo {author}
  {\bibfnamefont {E.~M.}\ \bibnamefont {{Leitch}}}, \bibinfo {author}
  {\bibfnamefont {M.}~\bibnamefont {{Lueker}}}, \bibinfo {author}
  {\bibfnamefont {K.~G.}\ \bibnamefont {{Megerian}}}, \bibinfo {author}
  {\bibfnamefont {T.}~\bibnamefont {{Namikawa}}}, \bibinfo {author}
  {\bibfnamefont {C.~B.}\ \bibnamefont {{Netterfield}}}, \bibinfo {author}
  {\bibfnamefont {H.~T.}\ \bibnamefont {{Nguyen}}}, \bibinfo {author}
  {\bibfnamefont {R.}~\bibnamefont {{O'Brient}}}, \bibinfo {author}
  {\bibfnamefont {I.}~\bibnamefont {{Ogburn}}, \bibfnamefont {R.~W.}}, \bibinfo
  {author} {\bibfnamefont {A.}~\bibnamefont {{Orlando}}}, \bibinfo {author}
  {\bibfnamefont {C.}~\bibnamefont {{Pryke}}}, \bibinfo {author} {\bibfnamefont
  {S.}~\bibnamefont {{Richter}}}, \bibinfo {author} {\bibfnamefont
  {R.}~\bibnamefont {{Schwarz}}}, \bibinfo {author} {\bibfnamefont {C.~D.}\
  \bibnamefont {{Sheehy}}}, \bibinfo {author} {\bibfnamefont {Z.~K.}\
  \bibnamefont {{Staniszewski}}}, \bibinfo {author} {\bibfnamefont
  {B.}~\bibnamefont {{Steinbach}}}, \bibinfo {author} {\bibfnamefont {R.~V.}\
  \bibnamefont {{Sudiwala}}}, \bibinfo {author} {\bibfnamefont {G.~P.}\
  \bibnamefont {{Teply}}}, \bibinfo {author} {\bibfnamefont {K.~L.}\
  \bibnamefont {{Thompson}}}, \bibinfo {author} {\bibfnamefont {J.~E.}\
  \bibnamefont {{Tolan}}}, \bibinfo {author} {\bibfnamefont {C.}~\bibnamefont
  {{Tucker}}}, \bibinfo {author} {\bibfnamefont {A.~D.}\ \bibnamefont
  {{Turner}}}, \bibinfo {author} {\bibfnamefont {A.~G.}\ \bibnamefont
  {{Vieregg}}}, \bibinfo {author} {\bibfnamefont {A.~C.}\ \bibnamefont
  {{Weber}}}, \bibinfo {author} {\bibfnamefont {D.~V.}\ \bibnamefont
  {{Wiebe}}}, \bibinfo {author} {\bibfnamefont {J.}~\bibnamefont {{Willmert}}},
  \bibinfo {author} {\bibfnamefont {C.~L.}\ \bibnamefont {{Wong}}}, \bibinfo
  {author} {\bibfnamefont {W.~L.~K.}\ \bibnamefont {{Wu}}}, \ and\ \bibinfo
  {author} {\bibfnamefont {K.~W.}\ \bibnamefont {{Yoon}}},\ }\href {\doibase
  10.3847/1538-4357/833/2/228} {\bibfield  {journal} {\bibinfo  {journal}
  {\apj}\ }\textbf {\bibinfo {volume} {833}},\ \bibinfo {eid} {228} (\bibinfo
  {year} {2016})},\ \Eprint {http://arxiv.org/abs/1606.01968} {arXiv:1606.01968
  [astro-ph.CO]} \BibitemShut {NoStop}%
\bibitem [{\citenamefont {{Polarbear Collaboration}}\ \emph
  {et~al.}(2014)\citenamefont {{Polarbear Collaboration}}, \citenamefont
  {{Ade}}, \citenamefont {{Akiba}}, \citenamefont {{Anthony}}, \citenamefont
  {{Arnold}}, \citenamefont {{Atlas}}, \citenamefont {{Barron}}, \citenamefont
  {{Boettger}}, \citenamefont {{Borrill}}, \citenamefont {{Chapman}},
  \citenamefont {{Chinone}}, \citenamefont {{Dobbs}}, \citenamefont
  {{Elleflot}}, \citenamefont {{Errard}}, \citenamefont {{Fabbian}},
  \citenamefont {{Feng}}, \citenamefont {{Flanigan}}, \citenamefont
  {{Gilbert}}, \citenamefont {{Grainger}}, \citenamefont {{Halverson}},
  \citenamefont {{Hasegawa}}, \citenamefont {{Hattori}}, \citenamefont
  {{Hazumi}}, \citenamefont {{Holzapfel}}, \citenamefont {{Hori}},
  \citenamefont {{Howard}}, \citenamefont {{Hyland}}, \citenamefont {{Inoue}},
  \citenamefont {{Jaehnig}}, \citenamefont {{Jaffe}}, \citenamefont
  {{Keating}}, \citenamefont {{Kermish}}, \citenamefont {{Keskitalo}},
  \citenamefont {{Kisner}}, \citenamefont {{Le Jeune}}, \citenamefont {{Lee}},
  \citenamefont {{Leitch}}, \citenamefont {{Linder}}, \citenamefont {{Lungu}},
  \citenamefont {{Matsuda}}, \citenamefont {{Matsumura}}, \citenamefont
  {{Meng}}, \citenamefont {{Miller}}, \citenamefont {{Morii}}, \citenamefont
  {{Moyerman}}, \citenamefont {{Myers}}, \citenamefont {{Navaroli}},
  \citenamefont {{Nishino}}, \citenamefont {{Orlando}}, \citenamefont {{Paar}},
  \citenamefont {{Peloton}}, \citenamefont {{Poletti}}, \citenamefont
  {{Quealy}}, \citenamefont {{Rebeiz}}, \citenamefont {{Reichardt}},
  \citenamefont {{Richards}}, \citenamefont {{Ross}}, \citenamefont
  {{Schanning}}, \citenamefont {{Schenck}}, \citenamefont {{Sherwin}},
  \citenamefont {{Shimizu}}, \citenamefont {{Shimmin}}, \citenamefont
  {{Shimon}}, \citenamefont {{Siritanasak}}, \citenamefont {{Smecher}},
  \citenamefont {{Spieler}}, \citenamefont {{Stebor}}, \citenamefont
  {{Steinbach}}, \citenamefont {{Stompor}}, \citenamefont {{Suzuki}},
  \citenamefont {{Takakura}}, \citenamefont {{Tomaru}}, \citenamefont
  {{Wilson}}, \citenamefont {{Yadav}},\ and\ \citenamefont
  {{Zahn}}}]{PolarbearCollaboration:2014}%
  \BibitemOpen
  \bibfield  {author} {\bibinfo {author} {\bibnamefont {{Polarbear
  Collaboration}}}, \bibinfo {author} {\bibfnamefont {P.~A.~R.}\ \bibnamefont
  {{Ade}}}, \bibinfo {author} {\bibfnamefont {Y.}~\bibnamefont {{Akiba}}},
  \bibinfo {author} {\bibfnamefont {A.~E.}\ \bibnamefont {{Anthony}}}, \bibinfo
  {author} {\bibfnamefont {K.}~\bibnamefont {{Arnold}}}, \bibinfo {author}
  {\bibfnamefont {M.}~\bibnamefont {{Atlas}}}, \bibinfo {author} {\bibfnamefont
  {D.}~\bibnamefont {{Barron}}}, \bibinfo {author} {\bibfnamefont
  {D.}~\bibnamefont {{Boettger}}}, \bibinfo {author} {\bibfnamefont
  {J.}~\bibnamefont {{Borrill}}}, \bibinfo {author} {\bibfnamefont
  {S.}~\bibnamefont {{Chapman}}}, \bibinfo {author} {\bibfnamefont
  {Y.}~\bibnamefont {{Chinone}}}, \bibinfo {author} {\bibfnamefont
  {M.}~\bibnamefont {{Dobbs}}}, \bibinfo {author} {\bibfnamefont
  {T.}~\bibnamefont {{Elleflot}}}, \bibinfo {author} {\bibfnamefont
  {J.}~\bibnamefont {{Errard}}}, \bibinfo {author} {\bibfnamefont
  {G.}~\bibnamefont {{Fabbian}}}, \bibinfo {author} {\bibfnamefont
  {C.}~\bibnamefont {{Feng}}}, \bibinfo {author} {\bibfnamefont
  {D.}~\bibnamefont {{Flanigan}}}, \bibinfo {author} {\bibfnamefont
  {A.}~\bibnamefont {{Gilbert}}}, \bibinfo {author} {\bibfnamefont
  {W.}~\bibnamefont {{Grainger}}}, \bibinfo {author} {\bibfnamefont {N.~W.}\
  \bibnamefont {{Halverson}}}, \bibinfo {author} {\bibfnamefont
  {M.}~\bibnamefont {{Hasegawa}}}, \bibinfo {author} {\bibfnamefont
  {K.}~\bibnamefont {{Hattori}}}, \bibinfo {author} {\bibfnamefont
  {M.}~\bibnamefont {{Hazumi}}}, \bibinfo {author} {\bibfnamefont {W.~L.}\
  \bibnamefont {{Holzapfel}}}, \bibinfo {author} {\bibfnamefont
  {Y.}~\bibnamefont {{Hori}}}, \bibinfo {author} {\bibfnamefont
  {J.}~\bibnamefont {{Howard}}}, \bibinfo {author} {\bibfnamefont
  {P.}~\bibnamefont {{Hyland}}}, \bibinfo {author} {\bibfnamefont
  {Y.}~\bibnamefont {{Inoue}}}, \bibinfo {author} {\bibfnamefont {G.~C.}\
  \bibnamefont {{Jaehnig}}}, \bibinfo {author} {\bibfnamefont {A.~H.}\
  \bibnamefont {{Jaffe}}}, \bibinfo {author} {\bibfnamefont {B.}~\bibnamefont
  {{Keating}}}, \bibinfo {author} {\bibfnamefont {Z.}~\bibnamefont
  {{Kermish}}}, \bibinfo {author} {\bibfnamefont {R.}~\bibnamefont
  {{Keskitalo}}}, \bibinfo {author} {\bibfnamefont {T.}~\bibnamefont
  {{Kisner}}}, \bibinfo {author} {\bibfnamefont {M.}~\bibnamefont {{Le
  Jeune}}}, \bibinfo {author} {\bibfnamefont {A.~T.}\ \bibnamefont {{Lee}}},
  \bibinfo {author} {\bibfnamefont {E.~M.}\ \bibnamefont {{Leitch}}}, \bibinfo
  {author} {\bibfnamefont {E.}~\bibnamefont {{Linder}}}, \bibinfo {author}
  {\bibfnamefont {M.}~\bibnamefont {{Lungu}}}, \bibinfo {author} {\bibfnamefont
  {F.}~\bibnamefont {{Matsuda}}}, \bibinfo {author} {\bibfnamefont
  {T.}~\bibnamefont {{Matsumura}}}, \bibinfo {author} {\bibfnamefont
  {X.}~\bibnamefont {{Meng}}}, \bibinfo {author} {\bibfnamefont {N.~J.}\
  \bibnamefont {{Miller}}}, \bibinfo {author} {\bibfnamefont {H.}~\bibnamefont
  {{Morii}}}, \bibinfo {author} {\bibfnamefont {S.}~\bibnamefont {{Moyerman}}},
  \bibinfo {author} {\bibfnamefont {M.~J.}\ \bibnamefont {{Myers}}}, \bibinfo
  {author} {\bibfnamefont {M.}~\bibnamefont {{Navaroli}}}, \bibinfo {author}
  {\bibfnamefont {H.}~\bibnamefont {{Nishino}}}, \bibinfo {author}
  {\bibfnamefont {A.}~\bibnamefont {{Orlando}}}, \bibinfo {author}
  {\bibfnamefont {H.}~\bibnamefont {{Paar}}}, \bibinfo {author} {\bibfnamefont
  {J.}~\bibnamefont {{Peloton}}}, \bibinfo {author} {\bibfnamefont
  {D.}~\bibnamefont {{Poletti}}}, \bibinfo {author} {\bibfnamefont
  {E.}~\bibnamefont {{Quealy}}}, \bibinfo {author} {\bibfnamefont
  {G.}~\bibnamefont {{Rebeiz}}}, \bibinfo {author} {\bibfnamefont {C.~L.}\
  \bibnamefont {{Reichardt}}}, \bibinfo {author} {\bibfnamefont {P.~L.}\
  \bibnamefont {{Richards}}}, \bibinfo {author} {\bibfnamefont
  {C.}~\bibnamefont {{Ross}}}, \bibinfo {author} {\bibfnamefont
  {I.}~\bibnamefont {{Schanning}}}, \bibinfo {author} {\bibfnamefont {D.~E.}\
  \bibnamefont {{Schenck}}}, \bibinfo {author} {\bibfnamefont {B.~D.}\
  \bibnamefont {{Sherwin}}}, \bibinfo {author} {\bibfnamefont {A.}~\bibnamefont
  {{Shimizu}}}, \bibinfo {author} {\bibfnamefont {C.}~\bibnamefont
  {{Shimmin}}}, \bibinfo {author} {\bibfnamefont {M.}~\bibnamefont {{Shimon}}},
  \bibinfo {author} {\bibfnamefont {P.}~\bibnamefont {{Siritanasak}}}, \bibinfo
  {author} {\bibfnamefont {G.}~\bibnamefont {{Smecher}}}, \bibinfo {author}
  {\bibfnamefont {H.}~\bibnamefont {{Spieler}}}, \bibinfo {author}
  {\bibfnamefont {N.}~\bibnamefont {{Stebor}}}, \bibinfo {author}
  {\bibfnamefont {B.}~\bibnamefont {{Steinbach}}}, \bibinfo {author}
  {\bibfnamefont {R.}~\bibnamefont {{Stompor}}}, \bibinfo {author}
  {\bibfnamefont {A.}~\bibnamefont {{Suzuki}}}, \bibinfo {author}
  {\bibfnamefont {S.}~\bibnamefont {{Takakura}}}, \bibinfo {author}
  {\bibfnamefont {T.}~\bibnamefont {{Tomaru}}}, \bibinfo {author}
  {\bibfnamefont {B.}~\bibnamefont {{Wilson}}}, \bibinfo {author}
  {\bibfnamefont {A.}~\bibnamefont {{Yadav}}}, \ and\ \bibinfo {author}
  {\bibfnamefont {O.}~\bibnamefont {{Zahn}}},\ }\href {\doibase
  10.1088/0004-637X/794/2/171} {\bibfield  {journal} {\bibinfo  {journal}
  {\apj}\ }\textbf {\bibinfo {volume} {794}},\ \bibinfo {eid} {171} (\bibinfo
  {year} {2014})},\ \Eprint {http://arxiv.org/abs/1403.2369} {arXiv:1403.2369
  [astro-ph.CO]} \BibitemShut {NoStop}%
\bibitem [{\citenamefont {{POLARBEAR Collaboration}}\ \emph
  {et~al.}(2017)\citenamefont {{POLARBEAR Collaboration}}, \citenamefont
  {{Ade}}, \citenamefont {{Aguilar}}, \citenamefont {{Akiba}}, \citenamefont
  {{Arnold}}, \citenamefont {{Baccigalupi}}, \citenamefont {{Barron}},
  \citenamefont {{Beck}}, \citenamefont {{Bianchini}}, \citenamefont
  {{Boettger}}, \citenamefont {{Borrill}}, \citenamefont {{Chapman}},
  \citenamefont {{Chinone}}, \citenamefont {{Crowley}}, \citenamefont
  {{Cukierman}}, \citenamefont {{D{\"u}nner}}, \citenamefont {{Dobbs}},
  \citenamefont {{Ducout}}, \citenamefont {{Elleflot}}, \citenamefont
  {{Errard}}, \citenamefont {{Fabbian}}, \citenamefont {{Feeney}},
  \citenamefont {{Feng}}, \citenamefont {{Fujino}}, \citenamefont {{Galitzki}},
  \citenamefont {{Gilbert}}, \citenamefont {{Goeckner-Wald}}, \citenamefont
  {{Groh}}, \citenamefont {{Hall}}, \citenamefont {{Halverson}}, \citenamefont
  {{Hamada}}, \citenamefont {{Hasegawa}}, \citenamefont {{Hazumi}},
  \citenamefont {{Hill}}, \citenamefont {{Howe}}, \citenamefont {{Inoue}},
  \citenamefont {{Jaehnig}}, \citenamefont {{Jaffe}}, \citenamefont {{Jeong}},
  \citenamefont {{Kaneko}}, \citenamefont {{Katayama}}, \citenamefont
  {{Keating}}, \citenamefont {{Keskitalo}}, \citenamefont {{Kisner}},
  \citenamefont {{Krachmalnicoff}}, \citenamefont {{Kusaka}}, \citenamefont
  {{Le Jeune}}, \citenamefont {{Lee}}, \citenamefont {{Leitch}}, \citenamefont
  {{Leon}}, \citenamefont {{Linder}}, \citenamefont {{Lowry}}, \citenamefont
  {{Matsuda}}, \citenamefont {{Matsumura}}, \citenamefont {{Minami}},
  \citenamefont {{Montgomery}}, \citenamefont {{Navaroli}}, \citenamefont
  {{Nishino}}, \citenamefont {{Paar}}, \citenamefont {{Peloton}}, \citenamefont
  {{Pham}}, \citenamefont {{Poletti}}, \citenamefont {{Puglisi}}, \citenamefont
  {{Reichardt}}, \citenamefont {{Richards}}, \citenamefont {{Ross}},
  \citenamefont {{Segawa}}, \citenamefont {{Sherwin}}, \citenamefont
  {{Silva-Feaver}}, \citenamefont {{Siritanasak}}, \citenamefont {{Stebor}},
  \citenamefont {{Stompor}}, \citenamefont {{Suzuki}}, \citenamefont
  {{Tajima}}, \citenamefont {{Takakura}}, \citenamefont {{Takatori}},
  \citenamefont {{Tanabe}}, \citenamefont {{Teply}}, \citenamefont {{Tomaru}},
  \citenamefont {{Tucker}}, \citenamefont {{Whitehorn}},\ and\ \citenamefont
  {{Zahn}}}]{POLARBEARCollaboration:2017}%
  \BibitemOpen
  \bibfield  {author} {\bibinfo {author} {\bibnamefont {{POLARBEAR
  Collaboration}}}, \bibinfo {author} {\bibfnamefont {P.~A.~R.}\ \bibnamefont
  {{Ade}}}, \bibinfo {author} {\bibfnamefont {M.}~\bibnamefont {{Aguilar}}},
  \bibinfo {author} {\bibfnamefont {Y.}~\bibnamefont {{Akiba}}}, \bibinfo
  {author} {\bibfnamefont {K.}~\bibnamefont {{Arnold}}}, \bibinfo {author}
  {\bibfnamefont {C.}~\bibnamefont {{Baccigalupi}}}, \bibinfo {author}
  {\bibfnamefont {D.}~\bibnamefont {{Barron}}}, \bibinfo {author}
  {\bibfnamefont {D.}~\bibnamefont {{Beck}}}, \bibinfo {author} {\bibfnamefont
  {F.}~\bibnamefont {{Bianchini}}}, \bibinfo {author} {\bibfnamefont
  {D.}~\bibnamefont {{Boettger}}}, \bibinfo {author} {\bibfnamefont
  {J.}~\bibnamefont {{Borrill}}}, \bibinfo {author} {\bibfnamefont
  {S.}~\bibnamefont {{Chapman}}}, \bibinfo {author} {\bibfnamefont
  {Y.}~\bibnamefont {{Chinone}}}, \bibinfo {author} {\bibfnamefont
  {K.}~\bibnamefont {{Crowley}}}, \bibinfo {author} {\bibfnamefont
  {A.}~\bibnamefont {{Cukierman}}}, \bibinfo {author} {\bibfnamefont
  {R.}~\bibnamefont {{D{\"u}nner}}}, \bibinfo {author} {\bibfnamefont
  {M.}~\bibnamefont {{Dobbs}}}, \bibinfo {author} {\bibfnamefont
  {A.}~\bibnamefont {{Ducout}}}, \bibinfo {author} {\bibfnamefont
  {T.}~\bibnamefont {{Elleflot}}}, \bibinfo {author} {\bibfnamefont
  {J.}~\bibnamefont {{Errard}}}, \bibinfo {author} {\bibfnamefont
  {G.}~\bibnamefont {{Fabbian}}}, \bibinfo {author} {\bibfnamefont {S.~M.}\
  \bibnamefont {{Feeney}}}, \bibinfo {author} {\bibfnamefont {C.}~\bibnamefont
  {{Feng}}}, \bibinfo {author} {\bibfnamefont {T.}~\bibnamefont {{Fujino}}},
  \bibinfo {author} {\bibfnamefont {N.}~\bibnamefont {{Galitzki}}}, \bibinfo
  {author} {\bibfnamefont {A.}~\bibnamefont {{Gilbert}}}, \bibinfo {author}
  {\bibfnamefont {N.}~\bibnamefont {{Goeckner-Wald}}}, \bibinfo {author}
  {\bibfnamefont {J.~C.}\ \bibnamefont {{Groh}}}, \bibinfo {author}
  {\bibfnamefont {G.}~\bibnamefont {{Hall}}}, \bibinfo {author} {\bibfnamefont
  {N.}~\bibnamefont {{Halverson}}}, \bibinfo {author} {\bibfnamefont
  {T.}~\bibnamefont {{Hamada}}}, \bibinfo {author} {\bibfnamefont
  {M.}~\bibnamefont {{Hasegawa}}}, \bibinfo {author} {\bibfnamefont
  {M.}~\bibnamefont {{Hazumi}}}, \bibinfo {author} {\bibfnamefont {C.~A.}\
  \bibnamefont {{Hill}}}, \bibinfo {author} {\bibfnamefont {L.}~\bibnamefont
  {{Howe}}}, \bibinfo {author} {\bibfnamefont {Y.}~\bibnamefont {{Inoue}}},
  \bibinfo {author} {\bibfnamefont {G.}~\bibnamefont {{Jaehnig}}}, \bibinfo
  {author} {\bibfnamefont {A.~H.}\ \bibnamefont {{Jaffe}}}, \bibinfo {author}
  {\bibfnamefont {O.}~\bibnamefont {{Jeong}}}, \bibinfo {author} {\bibfnamefont
  {D.}~\bibnamefont {{Kaneko}}}, \bibinfo {author} {\bibfnamefont
  {N.}~\bibnamefont {{Katayama}}}, \bibinfo {author} {\bibfnamefont
  {B.}~\bibnamefont {{Keating}}}, \bibinfo {author} {\bibfnamefont
  {R.}~\bibnamefont {{Keskitalo}}}, \bibinfo {author} {\bibfnamefont
  {T.}~\bibnamefont {{Kisner}}}, \bibinfo {author} {\bibfnamefont
  {N.}~\bibnamefont {{Krachmalnicoff}}}, \bibinfo {author} {\bibfnamefont
  {A.}~\bibnamefont {{Kusaka}}}, \bibinfo {author} {\bibfnamefont
  {M.}~\bibnamefont {{Le Jeune}}}, \bibinfo {author} {\bibfnamefont {A.~T.}\
  \bibnamefont {{Lee}}}, \bibinfo {author} {\bibfnamefont {E.~M.}\ \bibnamefont
  {{Leitch}}}, \bibinfo {author} {\bibfnamefont {D.}~\bibnamefont {{Leon}}},
  \bibinfo {author} {\bibfnamefont {E.}~\bibnamefont {{Linder}}}, \bibinfo
  {author} {\bibfnamefont {L.}~\bibnamefont {{Lowry}}}, \bibinfo {author}
  {\bibfnamefont {F.}~\bibnamefont {{Matsuda}}}, \bibinfo {author}
  {\bibfnamefont {T.}~\bibnamefont {{Matsumura}}}, \bibinfo {author}
  {\bibfnamefont {Y.}~\bibnamefont {{Minami}}}, \bibinfo {author}
  {\bibfnamefont {J.}~\bibnamefont {{Montgomery}}}, \bibinfo {author}
  {\bibfnamefont {M.}~\bibnamefont {{Navaroli}}}, \bibinfo {author}
  {\bibfnamefont {H.}~\bibnamefont {{Nishino}}}, \bibinfo {author}
  {\bibfnamefont {H.}~\bibnamefont {{Paar}}}, \bibinfo {author} {\bibfnamefont
  {J.}~\bibnamefont {{Peloton}}}, \bibinfo {author} {\bibfnamefont {A.~T.~P.}\
  \bibnamefont {{Pham}}}, \bibinfo {author} {\bibfnamefont {D.}~\bibnamefont
  {{Poletti}}}, \bibinfo {author} {\bibfnamefont {G.}~\bibnamefont
  {{Puglisi}}}, \bibinfo {author} {\bibfnamefont {C.~L.}\ \bibnamefont
  {{Reichardt}}}, \bibinfo {author} {\bibfnamefont {P.~L.}\ \bibnamefont
  {{Richards}}}, \bibinfo {author} {\bibfnamefont {C.}~\bibnamefont {{Ross}}},
  \bibinfo {author} {\bibfnamefont {Y.}~\bibnamefont {{Segawa}}}, \bibinfo
  {author} {\bibfnamefont {B.~D.}\ \bibnamefont {{Sherwin}}}, \bibinfo {author}
  {\bibfnamefont {M.}~\bibnamefont {{Silva-Feaver}}}, \bibinfo {author}
  {\bibfnamefont {P.}~\bibnamefont {{Siritanasak}}}, \bibinfo {author}
  {\bibfnamefont {N.}~\bibnamefont {{Stebor}}}, \bibinfo {author}
  {\bibfnamefont {R.}~\bibnamefont {{Stompor}}}, \bibinfo {author}
  {\bibfnamefont {A.}~\bibnamefont {{Suzuki}}}, \bibinfo {author}
  {\bibfnamefont {O.}~\bibnamefont {{Tajima}}}, \bibinfo {author}
  {\bibfnamefont {S.}~\bibnamefont {{Takakura}}}, \bibinfo {author}
  {\bibfnamefont {S.}~\bibnamefont {{Takatori}}}, \bibinfo {author}
  {\bibfnamefont {D.}~\bibnamefont {{Tanabe}}}, \bibinfo {author}
  {\bibfnamefont {G.~P.}\ \bibnamefont {{Teply}}}, \bibinfo {author}
  {\bibfnamefont {T.}~\bibnamefont {{Tomaru}}}, \bibinfo {author}
  {\bibfnamefont {C.}~\bibnamefont {{Tucker}}}, \bibinfo {author}
  {\bibfnamefont {N.}~\bibnamefont {{Whitehorn}}}, \ and\ \bibinfo {author}
  {\bibfnamefont {A.}~\bibnamefont {{Zahn}}},\ }\href {\doibase
  10.3847/1538-4357/aa8e9f} {\bibfield  {journal} {\bibinfo  {journal} {\apj}\
  }\textbf {\bibinfo {volume} {848}},\ \bibinfo {eid} {121} (\bibinfo {year}
  {2017})},\ \Eprint {http://arxiv.org/abs/1705.02907} {arXiv:1705.02907
  [astro-ph.CO]} \BibitemShut {NoStop}%
\bibitem [{\citenamefont {{van Engelen}}\ \emph {et~al.}(2012)\citenamefont
  {{van Engelen}}, \citenamefont {{Keisler}}, \citenamefont {{Zahn}},
  \citenamefont {{Aird}}, \citenamefont {{Benson}}, \citenamefont {{Bleem}},
  \citenamefont {{Carlstrom}}, \citenamefont {{Chang}}, \citenamefont {{Cho}},
  \citenamefont {{Crawford}}, \citenamefont {{Crites}}, \citenamefont {{de
  Haan}}, \citenamefont {{Dobbs}}, \citenamefont {{Dudley}}, \citenamefont
  {{George}}, \citenamefont {{Halverson}}, \citenamefont {{Holder}},
  \citenamefont {{Holzapfel}}, \citenamefont {{Hoover}}, \citenamefont {{Hou}},
  \citenamefont {{Hrubes}}, \citenamefont {{Joy}}, \citenamefont {{Knox}},
  \citenamefont {{Lee}}, \citenamefont {{Leitch}}, \citenamefont {{Lueker}},
  \citenamefont {{Luong-Van}}, \citenamefont {{McMahon}}, \citenamefont
  {{Mehl}}, \citenamefont {{Meyer}}, \citenamefont {{Millea}}, \citenamefont
  {{Mohr}}, \citenamefont {{Montroy}}, \citenamefont {{Natoli}}, \citenamefont
  {{Padin}}, \citenamefont {{Plagge}}, \citenamefont {{Pryke}}, \citenamefont
  {{Reichardt}}, \citenamefont {{Ruhl}}, \citenamefont {{Sayre}}, \citenamefont
  {{Schaffer}}, \citenamefont {{Shaw}}, \citenamefont {{Shirokoff}},
  \citenamefont {{Spieler}}, \citenamefont {{Staniszewski}}, \citenamefont
  {{Stark}}, \citenamefont {{Story}}, \citenamefont {{Vanderlinde}},
  \citenamefont {{Vieira}},\ and\ \citenamefont
  {{Williamson}}}]{vanEngelen:2012}%
  \BibitemOpen
  \bibfield  {author} {\bibinfo {author} {\bibfnamefont {A.}~\bibnamefont {{van
  Engelen}}}, \bibinfo {author} {\bibfnamefont {R.}~\bibnamefont {{Keisler}}},
  \bibinfo {author} {\bibfnamefont {O.}~\bibnamefont {{Zahn}}}, \bibinfo
  {author} {\bibfnamefont {K.~A.}\ \bibnamefont {{Aird}}}, \bibinfo {author}
  {\bibfnamefont {B.~A.}\ \bibnamefont {{Benson}}}, \bibinfo {author}
  {\bibfnamefont {L.~E.}\ \bibnamefont {{Bleem}}}, \bibinfo {author}
  {\bibfnamefont {J.~E.}\ \bibnamefont {{Carlstrom}}}, \bibinfo {author}
  {\bibfnamefont {C.~L.}\ \bibnamefont {{Chang}}}, \bibinfo {author}
  {\bibfnamefont {H.~M.}\ \bibnamefont {{Cho}}}, \bibinfo {author}
  {\bibfnamefont {T.~M.}\ \bibnamefont {{Crawford}}}, \bibinfo {author}
  {\bibfnamefont {A.~T.}\ \bibnamefont {{Crites}}}, \bibinfo {author}
  {\bibfnamefont {T.}~\bibnamefont {{de Haan}}}, \bibinfo {author}
  {\bibfnamefont {M.~A.}\ \bibnamefont {{Dobbs}}}, \bibinfo {author}
  {\bibfnamefont {J.}~\bibnamefont {{Dudley}}}, \bibinfo {author}
  {\bibfnamefont {E.~M.}\ \bibnamefont {{George}}}, \bibinfo {author}
  {\bibfnamefont {N.~W.}\ \bibnamefont {{Halverson}}}, \bibinfo {author}
  {\bibfnamefont {G.~P.}\ \bibnamefont {{Holder}}}, \bibinfo {author}
  {\bibfnamefont {W.~L.}\ \bibnamefont {{Holzapfel}}}, \bibinfo {author}
  {\bibfnamefont {S.}~\bibnamefont {{Hoover}}}, \bibinfo {author}
  {\bibfnamefont {Z.}~\bibnamefont {{Hou}}}, \bibinfo {author} {\bibfnamefont
  {J.~D.}\ \bibnamefont {{Hrubes}}}, \bibinfo {author} {\bibfnamefont
  {M.}~\bibnamefont {{Joy}}}, \bibinfo {author} {\bibfnamefont
  {L.}~\bibnamefont {{Knox}}}, \bibinfo {author} {\bibfnamefont {A.~T.}\
  \bibnamefont {{Lee}}}, \bibinfo {author} {\bibfnamefont {E.~M.}\ \bibnamefont
  {{Leitch}}}, \bibinfo {author} {\bibfnamefont {M.}~\bibnamefont {{Lueker}}},
  \bibinfo {author} {\bibfnamefont {D.}~\bibnamefont {{Luong-Van}}}, \bibinfo
  {author} {\bibfnamefont {J.~J.}\ \bibnamefont {{McMahon}}}, \bibinfo {author}
  {\bibfnamefont {J.}~\bibnamefont {{Mehl}}}, \bibinfo {author} {\bibfnamefont
  {S.~S.}\ \bibnamefont {{Meyer}}}, \bibinfo {author} {\bibfnamefont
  {M.}~\bibnamefont {{Millea}}}, \bibinfo {author} {\bibfnamefont {J.~J.}\
  \bibnamefont {{Mohr}}}, \bibinfo {author} {\bibfnamefont {T.~E.}\
  \bibnamefont {{Montroy}}}, \bibinfo {author} {\bibfnamefont {T.}~\bibnamefont
  {{Natoli}}}, \bibinfo {author} {\bibfnamefont {S.}~\bibnamefont {{Padin}}},
  \bibinfo {author} {\bibfnamefont {T.}~\bibnamefont {{Plagge}}}, \bibinfo
  {author} {\bibfnamefont {C.}~\bibnamefont {{Pryke}}}, \bibinfo {author}
  {\bibfnamefont {C.~L.}\ \bibnamefont {{Reichardt}}}, \bibinfo {author}
  {\bibfnamefont {J.~E.}\ \bibnamefont {{Ruhl}}}, \bibinfo {author}
  {\bibfnamefont {J.~T.}\ \bibnamefont {{Sayre}}}, \bibinfo {author}
  {\bibfnamefont {K.~K.}\ \bibnamefont {{Schaffer}}}, \bibinfo {author}
  {\bibfnamefont {L.}~\bibnamefont {{Shaw}}}, \bibinfo {author} {\bibfnamefont
  {E.}~\bibnamefont {{Shirokoff}}}, \bibinfo {author} {\bibfnamefont {H.~G.}\
  \bibnamefont {{Spieler}}}, \bibinfo {author} {\bibfnamefont {Z.}~\bibnamefont
  {{Staniszewski}}}, \bibinfo {author} {\bibfnamefont {A.~A.}\ \bibnamefont
  {{Stark}}}, \bibinfo {author} {\bibfnamefont {K.}~\bibnamefont {{Story}}},
  \bibinfo {author} {\bibfnamefont {K.}~\bibnamefont {{Vanderlinde}}}, \bibinfo
  {author} {\bibfnamefont {J.~D.}\ \bibnamefont {{Vieira}}}, \ and\ \bibinfo
  {author} {\bibfnamefont {R.}~\bibnamefont {{Williamson}}},\ }\href {\doibase
  10.1088/0004-637X/756/2/142} {\bibfield  {journal} {\bibinfo  {journal}
  {\apj}\ }\textbf {\bibinfo {volume} {756}},\ \bibinfo {eid} {142} (\bibinfo
  {year} {2012})},\ \Eprint {http://arxiv.org/abs/1202.0546} {arXiv:1202.0546
  [astro-ph.CO]} \BibitemShut {NoStop}%
\bibitem [{\citenamefont {{Story}}\ \emph {et~al.}(2015)\citenamefont
  {{Story}}, \citenamefont {{Hanson}}, \citenamefont {{Ade}}, \citenamefont
  {{Aird}}, \citenamefont {{Austermann}}, \citenamefont {{Beall}},
  \citenamefont {{Bender}}, \citenamefont {{Benson}}, \citenamefont {{Bleem}},
  \citenamefont {{Carlstrom}}, \citenamefont {{Chang}}, \citenamefont
  {{Chiang}}, \citenamefont {{Cho}}, \citenamefont {{Citron}}, \citenamefont
  {{Crawford}}, \citenamefont {{Crites}}, \citenamefont {{de Haan}},
  \citenamefont {{Dobbs}}, \citenamefont {{Everett}}, \citenamefont
  {{Gallicchio}}, \citenamefont {{Gao}}, \citenamefont {{George}},
  \citenamefont {{Gilbert}}, \citenamefont {{Halverson}}, \citenamefont
  {{Harrington}}, \citenamefont {{Henning}}, \citenamefont {{Hilton}},
  \citenamefont {{Holder}}, \citenamefont {{Holzapfel}}, \citenamefont
  {{Hoover}}, \citenamefont {{Hou}}, \citenamefont {{Hrubes}}, \citenamefont
  {{Huang}}, \citenamefont {{Hubmayr}}, \citenamefont {{Irwin}}, \citenamefont
  {{Keisler}}, \citenamefont {{Knox}}, \citenamefont {{Lee}}, \citenamefont
  {{Leitch}}, \citenamefont {{Li}}, \citenamefont {{Liang}}, \citenamefont
  {{Luong-Van}}, \citenamefont {{McMahon}}, \citenamefont {{Mehl}},
  \citenamefont {{Meyer}}, \citenamefont {{Mocanu}}, \citenamefont {{Montroy}},
  \citenamefont {{Natoli}}, \citenamefont {{Nibarger}}, \citenamefont
  {{Novosad}}, \citenamefont {{Padin}}, \citenamefont {{Pryke}}, \citenamefont
  {{Reichardt}}, \citenamefont {{Ruhl}}, \citenamefont {{Saliwanchik}},
  \citenamefont {{Sayre}}, \citenamefont {{Schaffer}}, \citenamefont
  {{Smecher}}, \citenamefont {{Stark}}, \citenamefont {{Tucker}}, \citenamefont
  {{Vanderlinde}}, \citenamefont {{Vieira}}, \citenamefont {{Wang}},
  \citenamefont {{Whitehorn}}, \citenamefont {{Yefremenko}},\ and\
  \citenamefont {{Zahn}}}]{Story:2015}%
  \BibitemOpen
  \bibfield  {author} {\bibinfo {author} {\bibfnamefont {K.~T.}\ \bibnamefont
  {{Story}}}, \bibinfo {author} {\bibfnamefont {D.}~\bibnamefont {{Hanson}}},
  \bibinfo {author} {\bibfnamefont {P.~A.~R.}\ \bibnamefont {{Ade}}}, \bibinfo
  {author} {\bibfnamefont {K.~A.}\ \bibnamefont {{Aird}}}, \bibinfo {author}
  {\bibfnamefont {J.~E.}\ \bibnamefont {{Austermann}}}, \bibinfo {author}
  {\bibfnamefont {J.~A.}\ \bibnamefont {{Beall}}}, \bibinfo {author}
  {\bibfnamefont {A.~N.}\ \bibnamefont {{Bender}}}, \bibinfo {author}
  {\bibfnamefont {B.~A.}\ \bibnamefont {{Benson}}}, \bibinfo {author}
  {\bibfnamefont {L.~E.}\ \bibnamefont {{Bleem}}}, \bibinfo {author}
  {\bibfnamefont {J.~E.}\ \bibnamefont {{Carlstrom}}}, \bibinfo {author}
  {\bibfnamefont {C.~L.}\ \bibnamefont {{Chang}}}, \bibinfo {author}
  {\bibfnamefont {H.~C.}\ \bibnamefont {{Chiang}}}, \bibinfo {author}
  {\bibfnamefont {H.~M.}\ \bibnamefont {{Cho}}}, \bibinfo {author}
  {\bibfnamefont {R.}~\bibnamefont {{Citron}}}, \bibinfo {author}
  {\bibfnamefont {T.~M.}\ \bibnamefont {{Crawford}}}, \bibinfo {author}
  {\bibfnamefont {A.~T.}\ \bibnamefont {{Crites}}}, \bibinfo {author}
  {\bibfnamefont {T.}~\bibnamefont {{de Haan}}}, \bibinfo {author}
  {\bibfnamefont {M.~A.}\ \bibnamefont {{Dobbs}}}, \bibinfo {author}
  {\bibfnamefont {W.}~\bibnamefont {{Everett}}}, \bibinfo {author}
  {\bibfnamefont {J.}~\bibnamefont {{Gallicchio}}}, \bibinfo {author}
  {\bibfnamefont {J.}~\bibnamefont {{Gao}}}, \bibinfo {author} {\bibfnamefont
  {E.~M.}\ \bibnamefont {{George}}}, \bibinfo {author} {\bibfnamefont
  {A.}~\bibnamefont {{Gilbert}}}, \bibinfo {author} {\bibfnamefont {N.~W.}\
  \bibnamefont {{Halverson}}}, \bibinfo {author} {\bibfnamefont
  {N.}~\bibnamefont {{Harrington}}}, \bibinfo {author} {\bibfnamefont {J.~W.}\
  \bibnamefont {{Henning}}}, \bibinfo {author} {\bibfnamefont {G.~C.}\
  \bibnamefont {{Hilton}}}, \bibinfo {author} {\bibfnamefont {G.~P.}\
  \bibnamefont {{Holder}}}, \bibinfo {author} {\bibfnamefont {W.~L.}\
  \bibnamefont {{Holzapfel}}}, \bibinfo {author} {\bibfnamefont
  {S.}~\bibnamefont {{Hoover}}}, \bibinfo {author} {\bibfnamefont
  {Z.}~\bibnamefont {{Hou}}}, \bibinfo {author} {\bibfnamefont {J.~D.}\
  \bibnamefont {{Hrubes}}}, \bibinfo {author} {\bibfnamefont {N.}~\bibnamefont
  {{Huang}}}, \bibinfo {author} {\bibfnamefont {J.}~\bibnamefont {{Hubmayr}}},
  \bibinfo {author} {\bibfnamefont {K.~D.}\ \bibnamefont {{Irwin}}}, \bibinfo
  {author} {\bibfnamefont {R.}~\bibnamefont {{Keisler}}}, \bibinfo {author}
  {\bibfnamefont {L.}~\bibnamefont {{Knox}}}, \bibinfo {author} {\bibfnamefont
  {A.~T.}\ \bibnamefont {{Lee}}}, \bibinfo {author} {\bibfnamefont {E.~M.}\
  \bibnamefont {{Leitch}}}, \bibinfo {author} {\bibfnamefont {D.}~\bibnamefont
  {{Li}}}, \bibinfo {author} {\bibfnamefont {C.}~\bibnamefont {{Liang}}},
  \bibinfo {author} {\bibfnamefont {D.}~\bibnamefont {{Luong-Van}}}, \bibinfo
  {author} {\bibfnamefont {J.~J.}\ \bibnamefont {{McMahon}}}, \bibinfo {author}
  {\bibfnamefont {J.}~\bibnamefont {{Mehl}}}, \bibinfo {author} {\bibfnamefont
  {S.~S.}\ \bibnamefont {{Meyer}}}, \bibinfo {author} {\bibfnamefont
  {L.}~\bibnamefont {{Mocanu}}}, \bibinfo {author} {\bibfnamefont {T.~E.}\
  \bibnamefont {{Montroy}}}, \bibinfo {author} {\bibfnamefont {T.}~\bibnamefont
  {{Natoli}}}, \bibinfo {author} {\bibfnamefont {J.~P.}\ \bibnamefont
  {{Nibarger}}}, \bibinfo {author} {\bibfnamefont {V.}~\bibnamefont
  {{Novosad}}}, \bibinfo {author} {\bibfnamefont {S.}~\bibnamefont {{Padin}}},
  \bibinfo {author} {\bibfnamefont {C.}~\bibnamefont {{Pryke}}}, \bibinfo
  {author} {\bibfnamefont {C.~L.}\ \bibnamefont {{Reichardt}}}, \bibinfo
  {author} {\bibfnamefont {J.~E.}\ \bibnamefont {{Ruhl}}}, \bibinfo {author}
  {\bibfnamefont {B.~R.}\ \bibnamefont {{Saliwanchik}}}, \bibinfo {author}
  {\bibfnamefont {J.~T.}\ \bibnamefont {{Sayre}}}, \bibinfo {author}
  {\bibfnamefont {K.~K.}\ \bibnamefont {{Schaffer}}}, \bibinfo {author}
  {\bibfnamefont {G.}~\bibnamefont {{Smecher}}}, \bibinfo {author}
  {\bibfnamefont {A.~A.}\ \bibnamefont {{Stark}}}, \bibinfo {author}
  {\bibfnamefont {C.}~\bibnamefont {{Tucker}}}, \bibinfo {author}
  {\bibfnamefont {K.}~\bibnamefont {{Vanderlinde}}}, \bibinfo {author}
  {\bibfnamefont {J.~D.}\ \bibnamefont {{Vieira}}}, \bibinfo {author}
  {\bibfnamefont {G.}~\bibnamefont {{Wang}}}, \bibinfo {author} {\bibfnamefont
  {N.}~\bibnamefont {{Whitehorn}}}, \bibinfo {author} {\bibfnamefont
  {V.}~\bibnamefont {{Yefremenko}}}, \ and\ \bibinfo {author} {\bibfnamefont
  {O.}~\bibnamefont {{Zahn}}},\ }\href {\doibase 10.1088/0004-637X/810/1/50}
  {\bibfield  {journal} {\bibinfo  {journal} {\apj}\ }\textbf {\bibinfo
  {volume} {810}},\ \bibinfo {eid} {50} (\bibinfo {year} {2015})},\ \Eprint
  {http://arxiv.org/abs/1412.4760} {arXiv:1412.4760 [astro-ph.CO]} \BibitemShut
  {NoStop}%
\bibitem [{\citenamefont {Ade}\ \emph {et~al.}(2019)\citenamefont {Ade},
  \citenamefont {Aguirre}, \citenamefont {Ahmed}, \citenamefont {Aiola},
  \citenamefont {Ali} \emph {et~al.}}]{SO:2019:SciGoal}%
  \BibitemOpen
  \bibfield  {author} {\bibinfo {author} {\bibfnamefont {P.}~\bibnamefont
  {Ade}}, \bibinfo {author} {\bibfnamefont {J.}~\bibnamefont {Aguirre}},
  \bibinfo {author} {\bibfnamefont {Z.}~\bibnamefont {Ahmed}}, \bibinfo
  {author} {\bibfnamefont {S.}~\bibnamefont {Aiola}}, \bibinfo {author}
  {\bibfnamefont {A.}~\bibnamefont {Ali}},  \emph {et~al.},\ }\href {\doibase
  10.1088/1475-7516/2019/02/056} {\bibfield  {journal} {\bibinfo  {journal}
  {Journal of Cosmology and Astroparticle Physics}\ }\textbf {\bibinfo {volume}
  {2019}} (\bibinfo {year} {2019}),\ 10.1088/1475-7516/2019/02/056},\ \Eprint
  {http://arxiv.org/abs/1808.07445} {arXiv:1808.07445} \BibitemShut {NoStop}%
\bibitem [{\citenamefont {Abazajian}\ \emph {et~al.}(2016)\citenamefont
  {Abazajian}, \citenamefont {Adshead}, \citenamefont {Ahmed}, \citenamefont
  {Allen}, \citenamefont {Alonso} \emph {et~al.}}]{S4:2016:SciBook}%
  \BibitemOpen
  \bibfield  {author} {\bibinfo {author} {\bibfnamefont {K.~N.}\ \bibnamefont
  {Abazajian}}, \bibinfo {author} {\bibfnamefont {P.}~\bibnamefont {Adshead}},
  \bibinfo {author} {\bibfnamefont {Z.}~\bibnamefont {Ahmed}}, \bibinfo
  {author} {\bibfnamefont {S.~W.}\ \bibnamefont {Allen}}, \bibinfo {author}
  {\bibfnamefont {D.}~\bibnamefont {Alonso}},  \emph {et~al.},\ }\href
  {http://arxiv.org/abs/1610.02743} {\  (\bibinfo {year} {2016})},\ \Eprint
  {http://arxiv.org/abs/1610.02743} {arXiv:1610.02743} \BibitemShut {NoStop}%
\bibitem [{\citenamefont {{Abazajian}}\ \emph {et~al.}(2016)\citenamefont
  {{Abazajian}}, \citenamefont {{Adshead}}, \citenamefont {{Ahmed}},
  \citenamefont {{Allen}}, \citenamefont {{Alonso}}, \citenamefont {{Arnold}},
  \citenamefont {{Baccigalupi}}, \citenamefont {{Bartlett}}, \citenamefont
  {{Battaglia}}, \citenamefont {{Benson}}, \citenamefont {{Bischoff}},
  \citenamefont {{Borrill}}, \citenamefont {{Buza}}, \citenamefont
  {{Calabrese}}, \citenamefont {{Caldwell}}, \citenamefont {{Carlstrom}},
  \citenamefont {{Chang}}, \citenamefont {{Crawford}}, \citenamefont
  {{Cyr-Racine}}, \citenamefont {{De Bernardis}}, \citenamefont {{de Haan}},
  \citenamefont {{di Serego Alighieri}}, \citenamefont {{Dunkley}},
  \citenamefont {{Dvorkin}}, \citenamefont {{Errard}}, \citenamefont
  {{Fabbian}}, \citenamefont {{Feeney}}, \citenamefont {{Ferraro}},
  \citenamefont {{Filippini}}, \citenamefont {{Flauger}}, \citenamefont
  {{Fuller}}, \citenamefont {{Gluscevic}}, \citenamefont {{Green}},
  \citenamefont {{Grin}}, \citenamefont {{Grohs}}, \citenamefont {{Henning}},
  \citenamefont {{Hill}}, \citenamefont {{Hlozek}}, \citenamefont {{Holder}},
  \citenamefont {{Holzapfel}}, \citenamefont {{Hu}}, \citenamefont
  {{Huffenberger}}, \citenamefont {{Keskitalo}}, \citenamefont {{Knox}},
  \citenamefont {{Kosowsky}}, \citenamefont {{Kovac}}, \citenamefont
  {{Kovetz}}, \citenamefont {{Kuo}}, \citenamefont {{Kusaka}}, \citenamefont
  {{Le Jeune}}, \citenamefont {{Lee}}, \citenamefont {{Lilley}}, \citenamefont
  {{Loverde}}, \citenamefont {{Madhavacheril}}, \citenamefont {{Mantz}},
  \citenamefont {{Marsh}}, \citenamefont {{McMahon}}, \citenamefont
  {{Meerburg}}, \citenamefont {{Meyers}}, \citenamefont {{Miller}},
  \citenamefont {{Munoz}}, \citenamefont {{Nguyen}}, \citenamefont {{Niemack}},
  \citenamefont {{Peloso}}, \citenamefont {{Peloton}}, \citenamefont
  {{Pogosian}}, \citenamefont {{Pryke}}, \citenamefont {{Raveri}},
  \citenamefont {{Reichardt}}, \citenamefont {{Rocha}}, \citenamefont
  {{Rotti}}, \citenamefont {{Schaan}}, \citenamefont {{Schmittfull}},
  \citenamefont {{Scott}}, \citenamefont {{Sehgal}}, \citenamefont
  {{Shandera}}, \citenamefont {{Sherwin}}, \citenamefont {{Smith}},
  \citenamefont {{Sorbo}}, \citenamefont {{Starkman}}, \citenamefont {{Story}},
  \citenamefont {{van Engelen}}, \citenamefont {{Vieira}}, \citenamefont
  {{Watson}}, \citenamefont {{Whitehorn}},\ and\ \citenamefont {{Kimmy
  Wu}}}]{2016arXiv161002743A}%
  \BibitemOpen
  \bibfield  {author} {\bibinfo {author} {\bibfnamefont {K.~N.}\ \bibnamefont
  {{Abazajian}}}, \bibinfo {author} {\bibfnamefont {P.}~\bibnamefont
  {{Adshead}}}, \bibinfo {author} {\bibfnamefont {Z.}~\bibnamefont {{Ahmed}}},
  \bibinfo {author} {\bibfnamefont {S.~W.}\ \bibnamefont {{Allen}}}, \bibinfo
  {author} {\bibfnamefont {D.}~\bibnamefont {{Alonso}}}, \bibinfo {author}
  {\bibfnamefont {K.~S.}\ \bibnamefont {{Arnold}}}, \bibinfo {author}
  {\bibfnamefont {C.}~\bibnamefont {{Baccigalupi}}}, \bibinfo {author}
  {\bibfnamefont {J.~G.}\ \bibnamefont {{Bartlett}}}, \bibinfo {author}
  {\bibfnamefont {N.}~\bibnamefont {{Battaglia}}}, \bibinfo {author}
  {\bibfnamefont {B.~A.}\ \bibnamefont {{Benson}}}, \bibinfo {author}
  {\bibfnamefont {C.~A.}\ \bibnamefont {{Bischoff}}}, \bibinfo {author}
  {\bibfnamefont {J.}~\bibnamefont {{Borrill}}}, \bibinfo {author}
  {\bibfnamefont {V.}~\bibnamefont {{Buza}}}, \bibinfo {author} {\bibfnamefont
  {E.}~\bibnamefont {{Calabrese}}}, \bibinfo {author} {\bibfnamefont
  {R.}~\bibnamefont {{Caldwell}}}, \bibinfo {author} {\bibfnamefont {J.~E.}\
  \bibnamefont {{Carlstrom}}}, \bibinfo {author} {\bibfnamefont {C.~L.}\
  \bibnamefont {{Chang}}}, \bibinfo {author} {\bibfnamefont {T.~M.}\
  \bibnamefont {{Crawford}}}, \bibinfo {author} {\bibfnamefont {F.-Y.}\
  \bibnamefont {{Cyr-Racine}}}, \bibinfo {author} {\bibfnamefont
  {F.}~\bibnamefont {{De Bernardis}}}, \bibinfo {author} {\bibfnamefont
  {T.}~\bibnamefont {{de Haan}}}, \bibinfo {author} {\bibfnamefont
  {S.}~\bibnamefont {{di Serego Alighieri}}}, \bibinfo {author} {\bibfnamefont
  {J.}~\bibnamefont {{Dunkley}}}, \bibinfo {author} {\bibfnamefont
  {C.}~\bibnamefont {{Dvorkin}}}, \bibinfo {author} {\bibfnamefont
  {J.}~\bibnamefont {{Errard}}}, \bibinfo {author} {\bibfnamefont
  {G.}~\bibnamefont {{Fabbian}}}, \bibinfo {author} {\bibfnamefont
  {S.}~\bibnamefont {{Feeney}}}, \bibinfo {author} {\bibfnamefont
  {S.}~\bibnamefont {{Ferraro}}}, \bibinfo {author} {\bibfnamefont {J.~P.}\
  \bibnamefont {{Filippini}}}, \bibinfo {author} {\bibfnamefont
  {R.}~\bibnamefont {{Flauger}}}, \bibinfo {author} {\bibfnamefont {G.~M.}\
  \bibnamefont {{Fuller}}}, \bibinfo {author} {\bibfnamefont {V.}~\bibnamefont
  {{Gluscevic}}}, \bibinfo {author} {\bibfnamefont {D.}~\bibnamefont
  {{Green}}}, \bibinfo {author} {\bibfnamefont {D.}~\bibnamefont {{Grin}}},
  \bibinfo {author} {\bibfnamefont {E.}~\bibnamefont {{Grohs}}}, \bibinfo
  {author} {\bibfnamefont {J.~W.}\ \bibnamefont {{Henning}}}, \bibinfo {author}
  {\bibfnamefont {J.~C.}\ \bibnamefont {{Hill}}}, \bibinfo {author}
  {\bibfnamefont {R.}~\bibnamefont {{Hlozek}}}, \bibinfo {author}
  {\bibfnamefont {G.}~\bibnamefont {{Holder}}}, \bibinfo {author}
  {\bibfnamefont {W.}~\bibnamefont {{Holzapfel}}}, \bibinfo {author}
  {\bibfnamefont {W.}~\bibnamefont {{Hu}}}, \bibinfo {author} {\bibfnamefont
  {K.~M.}\ \bibnamefont {{Huffenberger}}}, \bibinfo {author} {\bibfnamefont
  {R.}~\bibnamefont {{Keskitalo}}}, \bibinfo {author} {\bibfnamefont
  {L.}~\bibnamefont {{Knox}}}, \bibinfo {author} {\bibfnamefont
  {A.}~\bibnamefont {{Kosowsky}}}, \bibinfo {author} {\bibfnamefont
  {J.}~\bibnamefont {{Kovac}}}, \bibinfo {author} {\bibfnamefont {E.~D.}\
  \bibnamefont {{Kovetz}}}, \bibinfo {author} {\bibfnamefont {C.-L.}\
  \bibnamefont {{Kuo}}}, \bibinfo {author} {\bibfnamefont {A.}~\bibnamefont
  {{Kusaka}}}, \bibinfo {author} {\bibfnamefont {M.}~\bibnamefont {{Le
  Jeune}}}, \bibinfo {author} {\bibfnamefont {A.~T.}\ \bibnamefont {{Lee}}},
  \bibinfo {author} {\bibfnamefont {M.}~\bibnamefont {{Lilley}}}, \bibinfo
  {author} {\bibfnamefont {M.}~\bibnamefont {{Loverde}}}, \bibinfo {author}
  {\bibfnamefont {M.~S.}\ \bibnamefont {{Madhavacheril}}}, \bibinfo {author}
  {\bibfnamefont {A.}~\bibnamefont {{Mantz}}}, \bibinfo {author} {\bibfnamefont
  {D.~J.~E.}\ \bibnamefont {{Marsh}}}, \bibinfo {author} {\bibfnamefont
  {J.}~\bibnamefont {{McMahon}}}, \bibinfo {author} {\bibfnamefont {P.~D.}\
  \bibnamefont {{Meerburg}}}, \bibinfo {author} {\bibfnamefont
  {J.}~\bibnamefont {{Meyers}}}, \bibinfo {author} {\bibfnamefont {A.~D.}\
  \bibnamefont {{Miller}}}, \bibinfo {author} {\bibfnamefont {J.~B.}\
  \bibnamefont {{Munoz}}}, \bibinfo {author} {\bibfnamefont {H.~N.}\
  \bibnamefont {{Nguyen}}}, \bibinfo {author} {\bibfnamefont {M.~D.}\
  \bibnamefont {{Niemack}}}, \bibinfo {author} {\bibfnamefont {M.}~\bibnamefont
  {{Peloso}}}, \bibinfo {author} {\bibfnamefont {J.}~\bibnamefont {{Peloton}}},
  \bibinfo {author} {\bibfnamefont {L.}~\bibnamefont {{Pogosian}}}, \bibinfo
  {author} {\bibfnamefont {C.}~\bibnamefont {{Pryke}}}, \bibinfo {author}
  {\bibfnamefont {M.}~\bibnamefont {{Raveri}}}, \bibinfo {author}
  {\bibfnamefont {C.~L.}\ \bibnamefont {{Reichardt}}}, \bibinfo {author}
  {\bibfnamefont {G.}~\bibnamefont {{Rocha}}}, \bibinfo {author} {\bibfnamefont
  {A.}~\bibnamefont {{Rotti}}}, \bibinfo {author} {\bibfnamefont
  {E.}~\bibnamefont {{Schaan}}}, \bibinfo {author} {\bibfnamefont {M.~M.}\
  \bibnamefont {{Schmittfull}}}, \bibinfo {author} {\bibfnamefont
  {D.}~\bibnamefont {{Scott}}}, \bibinfo {author} {\bibfnamefont
  {N.}~\bibnamefont {{Sehgal}}}, \bibinfo {author} {\bibfnamefont
  {S.}~\bibnamefont {{Shandera}}}, \bibinfo {author} {\bibfnamefont {B.~D.}\
  \bibnamefont {{Sherwin}}}, \bibinfo {author} {\bibfnamefont {T.~L.}\
  \bibnamefont {{Smith}}}, \bibinfo {author} {\bibfnamefont {L.}~\bibnamefont
  {{Sorbo}}}, \bibinfo {author} {\bibfnamefont {G.~D.}\ \bibnamefont
  {{Starkman}}}, \bibinfo {author} {\bibfnamefont {K.~T.}\ \bibnamefont
  {{Story}}}, \bibinfo {author} {\bibfnamefont {A.}~\bibnamefont {{van
  Engelen}}}, \bibinfo {author} {\bibfnamefont {J.~D.}\ \bibnamefont
  {{Vieira}}}, \bibinfo {author} {\bibfnamefont {S.}~\bibnamefont {{Watson}}},
  \bibinfo {author} {\bibfnamefont {N.}~\bibnamefont {{Whitehorn}}}, \ and\
  \bibinfo {author} {\bibfnamefont {W.~L.}\ \bibnamefont {{Kimmy Wu}}},\
  }\href@noop {} {\bibfield  {journal} {\bibinfo  {journal} {arXiv e-prints}\
  ,\ \bibinfo {eid} {arXiv:1610.02743}} (\bibinfo {year} {2016})},\ \Eprint
  {http://arxiv.org/abs/1610.02743} {arXiv:1610.02743 [astro-ph.CO]}
  \BibitemShut {NoStop}%
\bibitem [{\citenamefont {Ferraro}\ and\ \citenamefont
  {Hill}(2018)}]{ferraro18_bias_to_cmb_lensin_recon}%
  \BibitemOpen
  \bibfield  {author} {\bibinfo {author} {\bibfnamefont {S.}~\bibnamefont
  {Ferraro}}\ and\ \bibinfo {author} {\bibfnamefont {J.~C.}\ \bibnamefont
  {Hill}},\ }\href {\doibase 10.1103/physrevd.97.023512} {\bibfield  {journal}
  {\bibinfo  {journal} {Physical Review D}\ }\textbf {\bibinfo {volume} {97}},\
  \bibinfo {pages} {023512} (\bibinfo {year} {2018})}\BibitemShut {NoStop}%
\bibitem [{\citenamefont {Cai}\ \emph {et~al.}(2022)\citenamefont {Cai},
  \citenamefont {Madhavacheril}, \citenamefont {Hill},\ and\ \citenamefont
  {Kosowsky}}]{Cai:2021hnb}%
  \BibitemOpen
  \bibfield  {author} {\bibinfo {author} {\bibfnamefont {H.}~\bibnamefont
  {Cai}}, \bibinfo {author} {\bibfnamefont {M.~S.}\ \bibnamefont
  {Madhavacheril}}, \bibinfo {author} {\bibfnamefont {J.~C.}\ \bibnamefont
  {Hill}}, \ and\ \bibinfo {author} {\bibfnamefont {A.}~\bibnamefont
  {Kosowsky}},\ }\href {\doibase 10.1103/PhysRevD.105.043516} {\bibfield
  {journal} {\bibinfo  {journal} {Phys. Rev. D}\ }\textbf {\bibinfo {volume}
  {105}},\ \bibinfo {pages} {043516} (\bibinfo {year} {2022})},\ \Eprint
  {http://arxiv.org/abs/2111.01944} {arXiv:2111.01944 [astro-ph.CO]}
  \BibitemShut {NoStop}%
\bibitem [{\citenamefont {Mirmelstein}\ \emph {et~al.}(2021)\citenamefont
  {Mirmelstein}, \citenamefont {Fabbian}, \citenamefont {Lewis},\ and\
  \citenamefont {Peloton}}]{Mirmelstein:2020pfk}%
  \BibitemOpen
  \bibfield  {author} {\bibinfo {author} {\bibfnamefont {M.}~\bibnamefont
  {Mirmelstein}}, \bibinfo {author} {\bibfnamefont {G.}~\bibnamefont
  {Fabbian}}, \bibinfo {author} {\bibfnamefont {A.}~\bibnamefont {Lewis}}, \
  and\ \bibinfo {author} {\bibfnamefont {J.}~\bibnamefont {Peloton}},\ }\href
  {\doibase 10.1103/PhysRevD.103.123540} {\bibfield  {journal} {\bibinfo
  {journal} {Phys. Rev. D}\ }\textbf {\bibinfo {volume} {103}},\ \bibinfo
  {pages} {123540} (\bibinfo {year} {2021})},\ \Eprint
  {http://arxiv.org/abs/2011.13910} {arXiv:2011.13910 [astro-ph.CO]}
  \BibitemShut {NoStop}%
\bibitem [{\citenamefont {{Nagata}}\ and\ \citenamefont
  {{Namikawa}}(2021)}]{Nagata:2021}%
  \BibitemOpen
  \bibfield  {author} {\bibinfo {author} {\bibfnamefont {R.}~\bibnamefont
  {{Nagata}}}\ and\ \bibinfo {author} {\bibfnamefont {T.}~\bibnamefont
  {{Namikawa}}},\ }\href {\doibase 10.1093/ptep/ptab040} {\bibfield  {journal}
  {\bibinfo  {journal} {Progress of Theoretical and Experimental Physics}\
  }\textbf {\bibinfo {volume} {2021}},\ \bibinfo {eid} {053E01} (\bibinfo
  {year} {2021})},\ \Eprint {http://arxiv.org/abs/2102.00133} {arXiv:2102.00133
  [astro-ph.CO]} \BibitemShut {NoStop}%
\bibitem [{\citenamefont {Minami}\ and\ \citenamefont
  {Komatsu}(2020)}]{Minami:2020odp}%
  \BibitemOpen
  \bibfield  {author} {\bibinfo {author} {\bibfnamefont {Y.}~\bibnamefont
  {Minami}}\ and\ \bibinfo {author} {\bibfnamefont {E.}~\bibnamefont
  {Komatsu}},\ }\href {\doibase 10.1103/PhysRevLett.125.221301} {\bibfield
  {journal} {\bibinfo  {journal} {Phys. Rev. Lett.}\ }\textbf {\bibinfo
  {volume} {125}},\ \bibinfo {pages} {221301} (\bibinfo {year} {2020})},\
  \Eprint {http://arxiv.org/abs/2011.11254} {arXiv:2011.11254 [astro-ph.CO]}
  \BibitemShut {NoStop}%
\bibitem [{\citenamefont {Diego-Palazuelos}\ \emph {et~al.}(2022)\citenamefont
  {Diego-Palazuelos} \emph {et~al.}}]{Diego-Palazuelos:2022dsq}%
  \BibitemOpen
  \bibfield  {author} {\bibinfo {author} {\bibfnamefont {P.}~\bibnamefont
  {Diego-Palazuelos}} \emph {et~al.},\ }\href {\doibase
  10.1103/PhysRevLett.128.091302} {\bibfield  {journal} {\bibinfo  {journal}
  {Phys. Rev. Lett.}\ }\textbf {\bibinfo {volume} {128}},\ \bibinfo {pages}
  {091302} (\bibinfo {year} {2022})},\ \Eprint
  {http://arxiv.org/abs/2201.07682} {arXiv:2201.07682 [astro-ph.CO]}
  \BibitemShut {NoStop}%
\bibitem [{\citenamefont {Eskilt}\ and\ \citenamefont
  {Komatsu}(2022)}]{Eskilt:2022cff}%
  \BibitemOpen
  \bibfield  {author} {\bibinfo {author} {\bibfnamefont {J.~R.}\ \bibnamefont
  {Eskilt}}\ and\ \bibinfo {author} {\bibfnamefont {E.}~\bibnamefont
  {Komatsu}},\ }\href@noop {} {\  (\bibinfo {year} {2022})},\ \Eprint
  {http://arxiv.org/abs/2205.13962} {arXiv:2205.13962 [astro-ph.CO]}
  \BibitemShut {NoStop}%
\bibitem [{\citenamefont {Komatsu}(2022)}]{Komatsu:2022nvu}%
  \BibitemOpen
  \bibfield  {author} {\bibinfo {author} {\bibfnamefont {E.}~\bibnamefont
  {Komatsu}},\ }\href {\doibase 10.1038/s42254-022-00452-4} {\bibfield
  {journal} {\bibinfo  {journal} {Nature Rev. Phys.}\ }\textbf {\bibinfo
  {volume} {4}},\ \bibinfo {pages} {452} (\bibinfo {year} {2022})},\ \Eprint
  {http://arxiv.org/abs/2202.13919} {arXiv:2202.13919 [astro-ph.CO]}
  \BibitemShut {NoStop}%
\bibitem [{\citenamefont {{Sherwin}}\ and\ \citenamefont
  {{Namikawa}}(2021)}]{Sherwin:2021}%
  \BibitemOpen
  \bibfield  {author} {\bibinfo {author} {\bibfnamefont {B.~D.}\ \bibnamefont
  {{Sherwin}}}\ and\ \bibinfo {author} {\bibfnamefont {T.}~\bibnamefont
  {{Namikawa}}},\ }\href@noop {} {\bibfield  {journal} {\bibinfo  {journal}
  {arXiv e-prints}\ ,\ \bibinfo {eid} {arXiv:2108.09287}} (\bibinfo {year}
  {2021})},\ \Eprint {http://arxiv.org/abs/2108.09287} {arXiv:2108.09287
  [astro-ph.CO]} \BibitemShut {NoStop}%
\bibitem [{\citenamefont {Nakatsuka}\ \emph {et~al.}(2022)\citenamefont
  {Nakatsuka}, \citenamefont {Namikawa},\ and\ \citenamefont
  {Komatsu}}]{Nakatsuka:2022epj}%
  \BibitemOpen
  \bibfield  {author} {\bibinfo {author} {\bibfnamefont {H.}~\bibnamefont
  {Nakatsuka}}, \bibinfo {author} {\bibfnamefont {T.}~\bibnamefont {Namikawa}},
  \ and\ \bibinfo {author} {\bibfnamefont {E.}~\bibnamefont {Komatsu}},\ }\href
  {\doibase 10.1103/PhysRevD.105.123509} {\bibfield  {journal} {\bibinfo
  {journal} {Phys. Rev. D}\ }\textbf {\bibinfo {volume} {105}},\ \bibinfo
  {pages} {123509} (\bibinfo {year} {2022})},\ \Eprint
  {http://arxiv.org/abs/2203.08560} {arXiv:2203.08560 [astro-ph.CO]}
  \BibitemShut {NoStop}%
\bibitem [{\citenamefont {Lee}\ \emph {et~al.}(2022)\citenamefont {Lee},
  \citenamefont {Hotinli},\ and\ \citenamefont {Kamionkowski}}]{Lee:2022udm}%
  \BibitemOpen
  \bibfield  {author} {\bibinfo {author} {\bibfnamefont {N.}~\bibnamefont
  {Lee}}, \bibinfo {author} {\bibfnamefont {S.~C.}\ \bibnamefont {Hotinli}}, \
  and\ \bibinfo {author} {\bibfnamefont {M.}~\bibnamefont {Kamionkowski}},\
  }\href@noop {} {\  (\bibinfo {year} {2022})},\ \Eprint
  {http://arxiv.org/abs/2207.05687} {arXiv:2207.05687 [astro-ph.CO]}
  \BibitemShut {NoStop}%
\bibitem [{\citenamefont {Gasparotto}\ and\ \citenamefont
  {Obata}(2022)}]{Gasparotto:2022uqo}%
  \BibitemOpen
  \bibfield  {author} {\bibinfo {author} {\bibfnamefont {S.}~\bibnamefont
  {Gasparotto}}\ and\ \bibinfo {author} {\bibfnamefont {I.}~\bibnamefont
  {Obata}},\ }\href {\doibase 10.1088/1475-7516/2022/08/025} {\bibfield
  {journal} {\bibinfo  {journal} {JCAP}\ }\textbf {\bibinfo {volume} {08}},\
  \bibinfo {pages} {025} (\bibinfo {year} {2022})},\ \Eprint
  {http://arxiv.org/abs/2203.09409} {arXiv:2203.09409 [astro-ph.CO]}
  \BibitemShut {NoStop}%
\bibitem [{\citenamefont {{Capparelli}}\ \emph {et~al.}(2020)\citenamefont
  {{Capparelli}}, \citenamefont {{Caldwell}},\ and\ \citenamefont
  {{Melchiorri}}}]{Capparelli:2020}%
  \BibitemOpen
  \bibfield  {author} {\bibinfo {author} {\bibfnamefont {L.~M.}\ \bibnamefont
  {{Capparelli}}}, \bibinfo {author} {\bibfnamefont {R.~R.}\ \bibnamefont
  {{Caldwell}}}, \ and\ \bibinfo {author} {\bibfnamefont {A.}~\bibnamefont
  {{Melchiorri}}},\ }\href {\doibase 10.1103/PhysRevD.101.123529} {\bibfield
  {journal} {\bibinfo  {journal} {\prd}\ }\textbf {\bibinfo {volume} {101}},\
  \bibinfo {eid} {123529} (\bibinfo {year} {2020})},\ \Eprint
  {http://arxiv.org/abs/1909.04621} {arXiv:1909.04621 [astro-ph.CO]}
  \BibitemShut {NoStop}%
\bibitem [{\citenamefont {{Fujita}}\ \emph {et~al.}(2020)\citenamefont
  {{Fujita}}, \citenamefont {{Minami}}, \citenamefont {{Murai}},\ and\
  \citenamefont {{Nakatsuka}}}]{Fujita:2020}%
  \BibitemOpen
  \bibfield  {author} {\bibinfo {author} {\bibfnamefont {T.}~\bibnamefont
  {{Fujita}}}, \bibinfo {author} {\bibfnamefont {Y.}~\bibnamefont {{Minami}}},
  \bibinfo {author} {\bibfnamefont {K.}~\bibnamefont {{Murai}}}, \ and\
  \bibinfo {author} {\bibfnamefont {H.}~\bibnamefont {{Nakatsuka}}},\
  }\href@noop {} {\bibfield  {journal} {\bibinfo  {journal} {arXiv e-prints}\
  ,\ \bibinfo {eid} {arXiv:2008.02473}} (\bibinfo {year} {2020})},\ \Eprint
  {http://arxiv.org/abs/2008.02473} {arXiv:2008.02473 [astro-ph.CO]}
  \BibitemShut {NoStop}%
\bibitem [{\citenamefont {Kitajima}\ \emph {et~al.}(2022)\citenamefont
  {Kitajima}, \citenamefont {Kozai}, \citenamefont {Takahashi},\ and\
  \citenamefont {Yin}}]{Kitajima:2022jzz}%
  \BibitemOpen
  \bibfield  {author} {\bibinfo {author} {\bibfnamefont {N.}~\bibnamefont
  {Kitajima}}, \bibinfo {author} {\bibfnamefont {F.}~\bibnamefont {Kozai}},
  \bibinfo {author} {\bibfnamefont {F.}~\bibnamefont {Takahashi}}, \ and\
  \bibinfo {author} {\bibfnamefont {W.}~\bibnamefont {Yin}},\ }\href@noop {} {\
   (\bibinfo {year} {2022})},\ \Eprint {http://arxiv.org/abs/2205.05083}
  {arXiv:2205.05083 [astro-ph.CO]} \BibitemShut {NoStop}%
\bibitem [{\citenamefont {Jain}\ \emph {et~al.}(2022)\citenamefont {Jain},
  \citenamefont {Hagimoto}, \citenamefont {Long},\ and\ \citenamefont
  {Amin}}]{Jain:2022jrp}%
  \BibitemOpen
  \bibfield  {author} {\bibinfo {author} {\bibfnamefont {M.}~\bibnamefont
  {Jain}}, \bibinfo {author} {\bibfnamefont {R.}~\bibnamefont {Hagimoto}},
  \bibinfo {author} {\bibfnamefont {A.~J.}\ \bibnamefont {Long}}, \ and\
  \bibinfo {author} {\bibfnamefont {M.~A.}\ \bibnamefont {Amin}},\ }\href@noop
  {} {\  (\bibinfo {year} {2022})},\ \Eprint {http://arxiv.org/abs/2208.08391}
  {arXiv:2208.08391 [astro-ph.CO]} \BibitemShut {NoStop}%
\bibitem [{\citenamefont {{Kamionkowski}}(2009)}]{Kamionkowski:2009}%
  \BibitemOpen
  \bibfield  {author} {\bibinfo {author} {\bibfnamefont {M.}~\bibnamefont
  {{Kamionkowski}}},\ }\href {\doibase 10.1103/PhysRevLett.102.111302}
  {\bibfield  {journal} {\bibinfo  {journal} {\prl}\ }\textbf {\bibinfo
  {volume} {102}},\ \bibinfo {eid} {111302} (\bibinfo {year} {2009})},\ \Eprint
  {http://arxiv.org/abs/0810.1286} {arXiv:0810.1286 [astro-ph]} \BibitemShut
  {NoStop}%
\bibitem [{\citenamefont {Goldberg}\ \emph {et~al.}(1967)\citenamefont
  {Goldberg}, \citenamefont {Macfarlane}, \citenamefont {Newman}, \citenamefont
  {Rohrlich},\ and\ \citenamefont {Sudarshan}}]{Goldberg:1967:spin}%
  \BibitemOpen
  \bibfield  {author} {\bibinfo {author} {\bibfnamefont {J.~N.}\ \bibnamefont
  {Goldberg}}, \bibinfo {author} {\bibfnamefont {A.~J.}\ \bibnamefont
  {Macfarlane}}, \bibinfo {author} {\bibfnamefont {E.~T.}\ \bibnamefont
  {Newman}}, \bibinfo {author} {\bibfnamefont {F.}~\bibnamefont {Rohrlich}}, \
  and\ \bibinfo {author} {\bibfnamefont {E.~C.}\ \bibnamefont {Sudarshan}},\
  }\href {\doibase 10.1063/1.1705135} {\bibfield  {journal} {\bibinfo
  {journal} {Journal of Mathematical Physics}\ }\textbf {\bibinfo {volume}
  {8}},\ \bibinfo {pages} {2155} (\bibinfo {year} {1967})}\BibitemShut
  {NoStop}%
\bibitem [{\citenamefont {Kamionkowski}\ \emph {et~al.}(1997)\citenamefont
  {Kamionkowski}, \citenamefont {Kosowsky},\ and\ \citenamefont
  {Stebbins}}]{Kamionkowski:1996zd}%
  \BibitemOpen
  \bibfield  {author} {\bibinfo {author} {\bibfnamefont {M.}~\bibnamefont
  {Kamionkowski}}, \bibinfo {author} {\bibfnamefont {A.}~\bibnamefont
  {Kosowsky}}, \ and\ \bibinfo {author} {\bibfnamefont {A.}~\bibnamefont
  {Stebbins}},\ }\href {\doibase 10.1103/PhysRevLett.78.2058} {\bibfield
  {journal} {\bibinfo  {journal} {Phys. Rev. Lett.}\ }\textbf {\bibinfo
  {volume} {78}},\ \bibinfo {pages} {2058} (\bibinfo {year} {1997})},\ \Eprint
  {http://arxiv.org/abs/astro-ph/9609132} {arXiv:astro-ph/9609132} \BibitemShut
  {NoStop}%
\bibitem [{\citenamefont {{Zaldarriaga}}\ and\ \citenamefont
  {{Seljak}}(1997)}]{Zaldarriaga:1997}%
  \BibitemOpen
  \bibfield  {author} {\bibinfo {author} {\bibfnamefont {M.}~\bibnamefont
  {{Zaldarriaga}}}\ and\ \bibinfo {author} {\bibfnamefont {U.}~\bibnamefont
  {{Seljak}}},\ }\href {\doibase 10.1103/PhysRevD.55.1830} {\bibfield
  {journal} {\bibinfo  {journal} {\prd}\ }\textbf {\bibinfo {volume} {55}},\
  \bibinfo {pages} {1830} (\bibinfo {year} {1997})},\ \Eprint
  {http://arxiv.org/abs/astro-ph/9609170} {arXiv:astro-ph/9609170 [astro-ph]}
  \BibitemShut {NoStop}%
\bibitem [{\citenamefont {Lewis}\ and\ \citenamefont
  {Challinor}(2006)}]{Lewis:2006fu}%
  \BibitemOpen
  \bibfield  {author} {\bibinfo {author} {\bibfnamefont {A.}~\bibnamefont
  {Lewis}}\ and\ \bibinfo {author} {\bibfnamefont {A.}~\bibnamefont
  {Challinor}},\ }\href {\doibase 10.1016/j.physrep.2006.03.002} {\bibfield
  {journal} {\bibinfo  {journal} {Phys. Rept.}\ }\textbf {\bibinfo {volume}
  {429}},\ \bibinfo {pages} {1} (\bibinfo {year} {2006})},\ \Eprint
  {http://arxiv.org/abs/astro-ph/0601594} {arXiv:astro-ph/0601594} \BibitemShut
  {NoStop}%
\bibitem [{\citenamefont {{Sherwin}}\ \emph {et~al.}(2017)\citenamefont
  {{Sherwin}}, \citenamefont {{van Engelen}}, \citenamefont {{Sehgal}},
  \citenamefont {{Madhavacheril}}, \citenamefont {{Addison}}, \citenamefont
  {{Aiola}}, \citenamefont {{Allison}}, \citenamefont {{Battaglia}},
  \citenamefont {{Becker}}, \citenamefont {{Beall}}, \citenamefont {{Bond}},
  \citenamefont {{Calabrese}}, \citenamefont {{Datta}}, \citenamefont
  {{Devlin}}, \citenamefont {{D{\"u}nner}}, \citenamefont {{Dunkley}},
  \citenamefont {{Fox}}, \citenamefont {{Gallardo}}, \citenamefont {{Halpern}},
  \citenamefont {{Hasselfield}}, \citenamefont {{Henderson}}, \citenamefont
  {{Hill}}, \citenamefont {{Hilton}}, \citenamefont {{Hubmayr}}, \citenamefont
  {{Hughes}}, \citenamefont {{Hincks}}, \citenamefont {{Hlozek}}, \citenamefont
  {{Huffenberger}}, \citenamefont {{Koopman}}, \citenamefont {{Kosowsky}},
  \citenamefont {{Louis}}, \citenamefont {{Maurin}}, \citenamefont {{McMahon}},
  \citenamefont {{Moodley}}, \citenamefont {{Naess}}, \citenamefont {{Nati}},
  \citenamefont {{Newburgh}}, \citenamefont {{Niemack}}, \citenamefont
  {{Page}}, \citenamefont {{Sievers}}, \citenamefont {{Spergel}}, \citenamefont
  {{Staggs}}, \citenamefont {{Thornton}}, \citenamefont {{Van Lanen}},
  \citenamefont {{Vavagiakis}},\ and\ \citenamefont
  {{Wollack}}}]{Sherwin:2017}%
  \BibitemOpen
  \bibfield  {author} {\bibinfo {author} {\bibfnamefont {B.~D.}\ \bibnamefont
  {{Sherwin}}}, \bibinfo {author} {\bibfnamefont {A.}~\bibnamefont {{van
  Engelen}}}, \bibinfo {author} {\bibfnamefont {N.}~\bibnamefont {{Sehgal}}},
  \bibinfo {author} {\bibfnamefont {M.}~\bibnamefont {{Madhavacheril}}},
  \bibinfo {author} {\bibfnamefont {G.~E.}\ \bibnamefont {{Addison}}}, \bibinfo
  {author} {\bibfnamefont {S.}~\bibnamefont {{Aiola}}}, \bibinfo {author}
  {\bibfnamefont {R.}~\bibnamefont {{Allison}}}, \bibinfo {author}
  {\bibfnamefont {N.}~\bibnamefont {{Battaglia}}}, \bibinfo {author}
  {\bibfnamefont {D.~T.}\ \bibnamefont {{Becker}}}, \bibinfo {author}
  {\bibfnamefont {J.~A.}\ \bibnamefont {{Beall}}}, \bibinfo {author}
  {\bibfnamefont {J.~R.}\ \bibnamefont {{Bond}}}, \bibinfo {author}
  {\bibfnamefont {E.}~\bibnamefont {{Calabrese}}}, \bibinfo {author}
  {\bibfnamefont {R.}~\bibnamefont {{Datta}}}, \bibinfo {author} {\bibfnamefont
  {M.~J.}\ \bibnamefont {{Devlin}}}, \bibinfo {author} {\bibfnamefont
  {R.}~\bibnamefont {{D{\"u}nner}}}, \bibinfo {author} {\bibfnamefont
  {J.}~\bibnamefont {{Dunkley}}}, \bibinfo {author} {\bibfnamefont {A.~E.}\
  \bibnamefont {{Fox}}}, \bibinfo {author} {\bibfnamefont {P.}~\bibnamefont
  {{Gallardo}}}, \bibinfo {author} {\bibfnamefont {M.}~\bibnamefont
  {{Halpern}}}, \bibinfo {author} {\bibfnamefont {M.}~\bibnamefont
  {{Hasselfield}}}, \bibinfo {author} {\bibfnamefont {S.}~\bibnamefont
  {{Henderson}}}, \bibinfo {author} {\bibfnamefont {J.~C.}\ \bibnamefont
  {{Hill}}}, \bibinfo {author} {\bibfnamefont {G.~C.}\ \bibnamefont
  {{Hilton}}}, \bibinfo {author} {\bibfnamefont {J.}~\bibnamefont {{Hubmayr}}},
  \bibinfo {author} {\bibfnamefont {J.~P.}\ \bibnamefont {{Hughes}}}, \bibinfo
  {author} {\bibfnamefont {A.~D.}\ \bibnamefont {{Hincks}}}, \bibinfo {author}
  {\bibfnamefont {R.}~\bibnamefont {{Hlozek}}}, \bibinfo {author}
  {\bibfnamefont {K.~M.}\ \bibnamefont {{Huffenberger}}}, \bibinfo {author}
  {\bibfnamefont {B.}~\bibnamefont {{Koopman}}}, \bibinfo {author}
  {\bibfnamefont {A.}~\bibnamefont {{Kosowsky}}}, \bibinfo {author}
  {\bibfnamefont {T.}~\bibnamefont {{Louis}}}, \bibinfo {author} {\bibfnamefont
  {L.}~\bibnamefont {{Maurin}}}, \bibinfo {author} {\bibfnamefont
  {J.}~\bibnamefont {{McMahon}}}, \bibinfo {author} {\bibfnamefont
  {K.}~\bibnamefont {{Moodley}}}, \bibinfo {author} {\bibfnamefont
  {S.}~\bibnamefont {{Naess}}}, \bibinfo {author} {\bibfnamefont
  {F.}~\bibnamefont {{Nati}}}, \bibinfo {author} {\bibfnamefont
  {L.}~\bibnamefont {{Newburgh}}}, \bibinfo {author} {\bibfnamefont {M.~D.}\
  \bibnamefont {{Niemack}}}, \bibinfo {author} {\bibfnamefont {L.~A.}\
  \bibnamefont {{Page}}}, \bibinfo {author} {\bibfnamefont {J.}~\bibnamefont
  {{Sievers}}}, \bibinfo {author} {\bibfnamefont {D.~N.}\ \bibnamefont
  {{Spergel}}}, \bibinfo {author} {\bibfnamefont {S.~T.}\ \bibnamefont
  {{Staggs}}}, \bibinfo {author} {\bibfnamefont {R.~J.}\ \bibnamefont
  {{Thornton}}}, \bibinfo {author} {\bibfnamefont {J.}~\bibnamefont {{Van
  Lanen}}}, \bibinfo {author} {\bibfnamefont {E.}~\bibnamefont {{Vavagiakis}}},
  \ and\ \bibinfo {author} {\bibfnamefont {E.~J.}\ \bibnamefont {{Wollack}}},\
  }\href {\doibase 10.1103/PhysRevD.95.123529} {\bibfield  {journal} {\bibinfo
  {journal} {\prd}\ }\textbf {\bibinfo {volume} {95}},\ \bibinfo {eid} {123529}
  (\bibinfo {year} {2017})},\ \Eprint {http://arxiv.org/abs/1611.09753}
  {arXiv:1611.09753 [astro-ph.CO]} \BibitemShut {NoStop}%
\bibitem [{\citenamefont {{Carroll}}(1998)}]{Carroll:1998}%
  \BibitemOpen
  \bibfield  {author} {\bibinfo {author} {\bibfnamefont {S.~M.}\ \bibnamefont
  {{Carroll}}},\ }\href {\doibase 10.1103/PhysRevLett.81.3067} {\bibfield
  {journal} {\bibinfo  {journal} {\prl}\ }\textbf {\bibinfo {volume} {81}},\
  \bibinfo {pages} {3067} (\bibinfo {year} {1998})},\ \Eprint
  {http://arxiv.org/abs/astro-ph/9806099} {arXiv:astro-ph/9806099 [astro-ph]}
  \BibitemShut {NoStop}%
\bibitem [{\citenamefont {Li}\ and\ \citenamefont {Zhang}(2008)}]{Li:2008tma}%
  \BibitemOpen
  \bibfield  {author} {\bibinfo {author} {\bibfnamefont {M.}~\bibnamefont
  {Li}}\ and\ \bibinfo {author} {\bibfnamefont {X.}~\bibnamefont {Zhang}},\
  }\href {\doibase 10.1103/PhysRevD.78.103516} {\bibfield  {journal} {\bibinfo
  {journal} {Phys. Rev. D}\ }\textbf {\bibinfo {volume} {78}},\ \bibinfo
  {pages} {103516} (\bibinfo {year} {2008})},\ \Eprint
  {http://arxiv.org/abs/0810.0403} {arXiv:0810.0403 [astro-ph]} \BibitemShut
  {NoStop}%
\bibitem [{\citenamefont {Marsh}(2016)}]{Marsh:2015xka}%
  \BibitemOpen
  \bibfield  {author} {\bibinfo {author} {\bibfnamefont {D.~J.~E.}\
  \bibnamefont {Marsh}},\ }\href {\doibase 10.1016/j.physrep.2016.06.005}
  {\bibfield  {journal} {\bibinfo  {journal} {Phys. Rept.}\ }\textbf {\bibinfo
  {volume} {643}},\ \bibinfo {pages} {1} (\bibinfo {year} {2016})},\ \Eprint
  {http://arxiv.org/abs/1510.07633} {arXiv:1510.07633 [astro-ph.CO]}
  \BibitemShut {NoStop}%
\bibitem [{\citenamefont {Leon}\ \emph {et~al.}(2016)\citenamefont {Leon},
  \citenamefont {Kaufman}, \citenamefont {Keating},\ and\ \citenamefont
  {Mewes}}]{Leon:2016kvt}%
  \BibitemOpen
  \bibfield  {author} {\bibinfo {author} {\bibfnamefont {D.}~\bibnamefont
  {Leon}}, \bibinfo {author} {\bibfnamefont {J.}~\bibnamefont {Kaufman}},
  \bibinfo {author} {\bibfnamefont {B.}~\bibnamefont {Keating}}, \ and\
  \bibinfo {author} {\bibfnamefont {M.}~\bibnamefont {Mewes}},\ }\href
  {\doibase 10.1142/S0217732317300026} {\bibfield  {journal} {\bibinfo
  {journal} {Mod. Phys. Lett. A}\ }\textbf {\bibinfo {volume} {32}},\ \bibinfo
  {pages} {1730002} (\bibinfo {year} {2016})},\ \Eprint
  {http://arxiv.org/abs/1611.00418} {arXiv:1611.00418 [astro-ph.CO]}
  \BibitemShut {NoStop}%
\bibitem [{\citenamefont {Kosowsky}\ and\ \citenamefont
  {Loeb}(1996)}]{Kosowsky:1996yc}%
  \BibitemOpen
  \bibfield  {author} {\bibinfo {author} {\bibfnamefont {A.}~\bibnamefont
  {Kosowsky}}\ and\ \bibinfo {author} {\bibfnamefont {A.}~\bibnamefont
  {Loeb}},\ }\href {\doibase 10.1086/177751} {\bibfield  {journal} {\bibinfo
  {journal} {Astrophys. J.}\ }\textbf {\bibinfo {volume} {469}},\ \bibinfo
  {pages} {1} (\bibinfo {year} {1996})},\ \Eprint
  {http://arxiv.org/abs/astro-ph/9601055} {arXiv:astro-ph/9601055} \BibitemShut
  {NoStop}%
\bibitem [{\citenamefont {Harari}\ \emph {et~al.}(1997)\citenamefont {Harari},
  \citenamefont {Hayward},\ and\ \citenamefont {Zaldarriaga}}]{Harari:1996ac}%
  \BibitemOpen
  \bibfield  {author} {\bibinfo {author} {\bibfnamefont {D.~D.}\ \bibnamefont
  {Harari}}, \bibinfo {author} {\bibfnamefont {J.~D.}\ \bibnamefont {Hayward}},
  \ and\ \bibinfo {author} {\bibfnamefont {M.}~\bibnamefont {Zaldarriaga}},\
  }\href {\doibase 10.1103/PhysRevD.55.1841} {\bibfield  {journal} {\bibinfo
  {journal} {Phys. Rev. D}\ }\textbf {\bibinfo {volume} {55}},\ \bibinfo
  {pages} {1841} (\bibinfo {year} {1997})},\ \Eprint
  {http://arxiv.org/abs/astro-ph/9608098} {arXiv:astro-ph/9608098} \BibitemShut
  {NoStop}%
\bibitem [{\citenamefont {Kosowsky}\ \emph {et~al.}(2005)\citenamefont
  {Kosowsky}, \citenamefont {Kahniashvili}, \citenamefont {Lavrelashvili},\
  and\ \citenamefont {Ratra}}]{Kosowsky:2004zh}%
  \BibitemOpen
  \bibfield  {author} {\bibinfo {author} {\bibfnamefont {A.}~\bibnamefont
  {Kosowsky}}, \bibinfo {author} {\bibfnamefont {T.}~\bibnamefont
  {Kahniashvili}}, \bibinfo {author} {\bibfnamefont {G.}~\bibnamefont
  {Lavrelashvili}}, \ and\ \bibinfo {author} {\bibfnamefont {B.}~\bibnamefont
  {Ratra}},\ }\href {\doibase 10.1103/PhysRevD.71.043006} {\bibfield  {journal}
  {\bibinfo  {journal} {Phys. Rev. D}\ }\textbf {\bibinfo {volume} {71}},\
  \bibinfo {pages} {043006} (\bibinfo {year} {2005})},\ \Eprint
  {http://arxiv.org/abs/astro-ph/0409767} {arXiv:astro-ph/0409767} \BibitemShut
  {NoStop}%
\bibitem [{\citenamefont {{Yadav}}\ \emph {et~al.}(2012)\citenamefont
  {{Yadav}}, \citenamefont {{Pogosian}},\ and\ \citenamefont
  {{Vachaspati}}}]{2012PhRvD..86l3009Y}%
  \BibitemOpen
  \bibfield  {author} {\bibinfo {author} {\bibfnamefont {A.}~\bibnamefont
  {{Yadav}}}, \bibinfo {author} {\bibfnamefont {L.}~\bibnamefont {{Pogosian}}},
  \ and\ \bibinfo {author} {\bibfnamefont {T.}~\bibnamefont {{Vachaspati}}},\
  }\href {\doibase 10.1103/PhysRevD.86.123009} {\bibfield  {journal} {\bibinfo
  {journal} {\prd}\ }\textbf {\bibinfo {volume} {86}},\ \bibinfo {eid} {123009}
  (\bibinfo {year} {2012})},\ \Eprint {http://arxiv.org/abs/1207.3356}
  {arXiv:1207.3356 [astro-ph.CO]} \BibitemShut {NoStop}%
\bibitem [{\citenamefont {De}\ \emph {et~al.}(2013)\citenamefont {De},
  \citenamefont {Pogosian},\ and\ \citenamefont {Vachaspati}}]{De:2013dra}%
  \BibitemOpen
  \bibfield  {author} {\bibinfo {author} {\bibfnamefont {S.}~\bibnamefont
  {De}}, \bibinfo {author} {\bibfnamefont {L.}~\bibnamefont {Pogosian}}, \ and\
  \bibinfo {author} {\bibfnamefont {T.}~\bibnamefont {Vachaspati}},\ }\href
  {\doibase 10.1103/PhysRevD.88.063527} {\bibfield  {journal} {\bibinfo
  {journal} {Phys. Rev. D}\ }\textbf {\bibinfo {volume} {88}},\ \bibinfo
  {pages} {063527} (\bibinfo {year} {2013})},\ \Eprint
  {http://arxiv.org/abs/1305.7225} {arXiv:1305.7225 [astro-ph.CO]} \BibitemShut
  {NoStop}%
\bibitem [{\citenamefont {Pogosian}(2014)}]{Pogosian:2013dya}%
  \BibitemOpen
  \bibfield  {author} {\bibinfo {author} {\bibfnamefont {L.}~\bibnamefont
  {Pogosian}},\ }\href {\doibase 10.1093/mnras/stt2378} {\bibfield  {journal}
  {\bibinfo  {journal} {Mon. Not. Roy. Astron. Soc.}\ }\textbf {\bibinfo
  {volume} {438}},\ \bibinfo {pages} {2508} (\bibinfo {year} {2014})},\ \Eprint
  {http://arxiv.org/abs/1311.2926} {arXiv:1311.2926 [astro-ph.CO]} \BibitemShut
  {NoStop}%
\bibitem [{\citenamefont {{Colladay}}\ and\ \citenamefont
  {{Kosteleck{\'y}}}(1997)}]{1997PhRvD..55.6760C}%
  \BibitemOpen
  \bibfield  {author} {\bibinfo {author} {\bibfnamefont {D.}~\bibnamefont
  {{Colladay}}}\ and\ \bibinfo {author} {\bibfnamefont {V.~A.}\ \bibnamefont
  {{Kosteleck{\'y}}}},\ }\href {\doibase 10.1103/PhysRevD.55.6760} {\bibfield
  {journal} {\bibinfo  {journal} {\prd}\ }\textbf {\bibinfo {volume} {55}},\
  \bibinfo {pages} {6760} (\bibinfo {year} {1997})},\ \Eprint
  {http://arxiv.org/abs/hep-ph/9703464} {arXiv:hep-ph/9703464 [hep-ph]}
  \BibitemShut {NoStop}%
\bibitem [{\citenamefont {{Colladay}}\ and\ \citenamefont
  {{Kosteleck{\'y}}}(1998)}]{1998PhRvD..58k6002C}%
  \BibitemOpen
  \bibfield  {author} {\bibinfo {author} {\bibfnamefont {D.}~\bibnamefont
  {{Colladay}}}\ and\ \bibinfo {author} {\bibfnamefont {V.~A.}\ \bibnamefont
  {{Kosteleck{\'y}}}},\ }\href {\doibase 10.1103/PhysRevD.58.116002} {\bibfield
   {journal} {\bibinfo  {journal} {\prd}\ }\textbf {\bibinfo {volume} {58}},\
  \bibinfo {eid} {116002} (\bibinfo {year} {1998})},\ \Eprint
  {http://arxiv.org/abs/hep-ph/9809521} {arXiv:hep-ph/9809521 [astro-ph]}
  \BibitemShut {NoStop}%
\bibitem [{\citenamefont {{Leon}}\ \emph {et~al.}(2017)\citenamefont {{Leon}},
  \citenamefont {{Kaufman}}, \citenamefont {{Keating}},\ and\ \citenamefont
  {{Mewes}}}]{Leon:2017}%
  \BibitemOpen
  \bibfield  {author} {\bibinfo {author} {\bibfnamefont {D.}~\bibnamefont
  {{Leon}}}, \bibinfo {author} {\bibfnamefont {J.}~\bibnamefont {{Kaufman}}},
  \bibinfo {author} {\bibfnamefont {B.}~\bibnamefont {{Keating}}}, \ and\
  \bibinfo {author} {\bibfnamefont {M.}~\bibnamefont {{Mewes}}},\ }\href
  {\doibase 10.1142/S0217732317300026} {\bibfield  {journal} {\bibinfo
  {journal} {Modern Physics Letters A}\ }\textbf {\bibinfo {volume} {32}},\
  \bibinfo {eid} {1730002} (\bibinfo {year} {2017})},\ \Eprint
  {http://arxiv.org/abs/1611.00418} {arXiv:1611.00418 [astro-ph.CO]}
  \BibitemShut {NoStop}%
\bibitem [{\citenamefont {{Caldwell}}\ \emph {et~al.}(2011)\citenamefont
  {{Caldwell}}, \citenamefont {{Gluscevic}},\ and\ \citenamefont
  {{Kamionkowski}}}]{Caldwell:2011}%
  \BibitemOpen
  \bibfield  {author} {\bibinfo {author} {\bibfnamefont {R.~R.}\ \bibnamefont
  {{Caldwell}}}, \bibinfo {author} {\bibfnamefont {V.}~\bibnamefont
  {{Gluscevic}}}, \ and\ \bibinfo {author} {\bibfnamefont {M.}~\bibnamefont
  {{Kamionkowski}}},\ }\href {\doibase 10.1103/PhysRevD.84.043504} {\bibfield
  {journal} {\bibinfo  {journal} {\prd}\ }\textbf {\bibinfo {volume} {84}},\
  \bibinfo {eid} {043504} (\bibinfo {year} {2011})},\ \Eprint
  {http://arxiv.org/abs/1104.1634} {arXiv:1104.1634 [astro-ph.CO]} \BibitemShut
  {NoStop}%
\bibitem [{\citenamefont {{Gluscevic}}\ \emph {et~al.}(2009)\citenamefont
  {{Gluscevic}}, \citenamefont {{Kamionkowski}},\ and\ \citenamefont
  {{Cooray}}}]{Gluscevic:2009}%
  \BibitemOpen
  \bibfield  {author} {\bibinfo {author} {\bibfnamefont {V.}~\bibnamefont
  {{Gluscevic}}}, \bibinfo {author} {\bibfnamefont {M.}~\bibnamefont
  {{Kamionkowski}}}, \ and\ \bibinfo {author} {\bibfnamefont {A.}~\bibnamefont
  {{Cooray}}},\ }\href {\doibase 10.1103/PhysRevD.80.023510} {\bibfield
  {journal} {\bibinfo  {journal} {\prd}\ }\textbf {\bibinfo {volume} {80}},\
  \bibinfo {eid} {023510} (\bibinfo {year} {2009})},\ \Eprint
  {http://arxiv.org/abs/0905.1687} {arXiv:0905.1687 [astro-ph.CO]} \BibitemShut
  {NoStop}%
\bibitem [{\citenamefont {{Namikawa}}\ \emph {et~al.}(2020)\citenamefont
  {{Namikawa}}, \citenamefont {{Guan}}, \citenamefont {{Darwish}},
  \citenamefont {{Sherwin}}, \citenamefont {{Aiola}}, \citenamefont
  {{Battaglia}}, \citenamefont {{Beall}}, \citenamefont {{Becker}},
  \citenamefont {{Bond}}, \citenamefont {{Calabrese}}, \citenamefont
  {{Chesmore}}, \citenamefont {{Choi}}, \citenamefont {{Devlin}}, \citenamefont
  {{Dunkley}}, \citenamefont {{D{\"u}nner}}, \citenamefont {{Fox}},
  \citenamefont {{Gallardo}}, \citenamefont {{Gluscevic}}, \citenamefont
  {{Han}}, \citenamefont {{Hasselfield}}, \citenamefont {{Hilton}},
  \citenamefont {{Hincks}}, \citenamefont {{Hlo{\v{z}}ek}}, \citenamefont
  {{Hubmayr}}, \citenamefont {{Huffenberger}}, \citenamefont {{Hughes}},
  \citenamefont {{Koopman}}, \citenamefont {{Kosowsky}}, \citenamefont
  {{Louis}}, \citenamefont {{Lungu}}, \citenamefont {{MacInnis}}, \citenamefont
  {{Madhavacheril}}, \citenamefont {{Mallaby-Kay}}, \citenamefont {{Maurin}},
  \citenamefont {{McMahon}}, \citenamefont {{Moodley}}, \citenamefont
  {{Naess}}, \citenamefont {{Nati}}, \citenamefont {{Newburgh}}, \citenamefont
  {{Nibarger}}, \citenamefont {{Niemack}}, \citenamefont {{Page}},
  \citenamefont {{Qu}}, \citenamefont {{Robertson}}, \citenamefont
  {{Schillaci}}, \citenamefont {{Sehgal}}, \citenamefont {{Sif{\'o}n}},
  \citenamefont {{Simon}}, \citenamefont {{Spergel}}, \citenamefont {{Staggs}},
  \citenamefont {{Storer}}, \citenamefont {{van Engelen}}, \citenamefont {{van
  Lanen}},\ and\ \citenamefont {{Wollack}}}]{Namikawa:2020}%
  \BibitemOpen
  \bibfield  {author} {\bibinfo {author} {\bibfnamefont {T.}~\bibnamefont
  {{Namikawa}}}, \bibinfo {author} {\bibfnamefont {Y.}~\bibnamefont {{Guan}}},
  \bibinfo {author} {\bibfnamefont {O.}~\bibnamefont {{Darwish}}}, \bibinfo
  {author} {\bibfnamefont {B.~D.}\ \bibnamefont {{Sherwin}}}, \bibinfo {author}
  {\bibfnamefont {S.}~\bibnamefont {{Aiola}}}, \bibinfo {author} {\bibfnamefont
  {N.}~\bibnamefont {{Battaglia}}}, \bibinfo {author} {\bibfnamefont {J.~A.}\
  \bibnamefont {{Beall}}}, \bibinfo {author} {\bibfnamefont {D.~T.}\
  \bibnamefont {{Becker}}}, \bibinfo {author} {\bibfnamefont {J.~R.}\
  \bibnamefont {{Bond}}}, \bibinfo {author} {\bibfnamefont {E.}~\bibnamefont
  {{Calabrese}}}, \bibinfo {author} {\bibfnamefont {G.~E.}\ \bibnamefont
  {{Chesmore}}}, \bibinfo {author} {\bibfnamefont {S.~K.}\ \bibnamefont
  {{Choi}}}, \bibinfo {author} {\bibfnamefont {M.~J.}\ \bibnamefont
  {{Devlin}}}, \bibinfo {author} {\bibfnamefont {J.}~\bibnamefont {{Dunkley}}},
  \bibinfo {author} {\bibfnamefont {R.}~\bibnamefont {{D{\"u}nner}}}, \bibinfo
  {author} {\bibfnamefont {A.~E.}\ \bibnamefont {{Fox}}}, \bibinfo {author}
  {\bibfnamefont {P.~A.}\ \bibnamefont {{Gallardo}}}, \bibinfo {author}
  {\bibfnamefont {V.}~\bibnamefont {{Gluscevic}}}, \bibinfo {author}
  {\bibfnamefont {D.}~\bibnamefont {{Han}}}, \bibinfo {author} {\bibfnamefont
  {M.}~\bibnamefont {{Hasselfield}}}, \bibinfo {author} {\bibfnamefont {G.~C.}\
  \bibnamefont {{Hilton}}}, \bibinfo {author} {\bibfnamefont {A.~D.}\
  \bibnamefont {{Hincks}}}, \bibinfo {author} {\bibfnamefont {R.}~\bibnamefont
  {{Hlo{\v{z}}ek}}}, \bibinfo {author} {\bibfnamefont {J.}~\bibnamefont
  {{Hubmayr}}}, \bibinfo {author} {\bibfnamefont {K.}~\bibnamefont
  {{Huffenberger}}}, \bibinfo {author} {\bibfnamefont {J.~P.}\ \bibnamefont
  {{Hughes}}}, \bibinfo {author} {\bibfnamefont {B.~J.}\ \bibnamefont
  {{Koopman}}}, \bibinfo {author} {\bibfnamefont {A.}~\bibnamefont
  {{Kosowsky}}}, \bibinfo {author} {\bibfnamefont {T.}~\bibnamefont {{Louis}}},
  \bibinfo {author} {\bibfnamefont {M.}~\bibnamefont {{Lungu}}}, \bibinfo
  {author} {\bibfnamefont {A.}~\bibnamefont {{MacInnis}}}, \bibinfo {author}
  {\bibfnamefont {M.~S.}\ \bibnamefont {{Madhavacheril}}}, \bibinfo {author}
  {\bibfnamefont {M.}~\bibnamefont {{Mallaby-Kay}}}, \bibinfo {author}
  {\bibfnamefont {L.}~\bibnamefont {{Maurin}}}, \bibinfo {author}
  {\bibfnamefont {J.}~\bibnamefont {{McMahon}}}, \bibinfo {author}
  {\bibfnamefont {K.}~\bibnamefont {{Moodley}}}, \bibinfo {author}
  {\bibfnamefont {S.}~\bibnamefont {{Naess}}}, \bibinfo {author} {\bibfnamefont
  {F.}~\bibnamefont {{Nati}}}, \bibinfo {author} {\bibfnamefont {L.~B.}\
  \bibnamefont {{Newburgh}}}, \bibinfo {author} {\bibfnamefont {J.~P.}\
  \bibnamefont {{Nibarger}}}, \bibinfo {author} {\bibfnamefont {M.~D.}\
  \bibnamefont {{Niemack}}}, \bibinfo {author} {\bibfnamefont {L.~A.}\
  \bibnamefont {{Page}}}, \bibinfo {author} {\bibfnamefont {F.~J.}\
  \bibnamefont {{Qu}}}, \bibinfo {author} {\bibfnamefont {N.}~\bibnamefont
  {{Robertson}}}, \bibinfo {author} {\bibfnamefont {A.}~\bibnamefont
  {{Schillaci}}}, \bibinfo {author} {\bibfnamefont {N.}~\bibnamefont
  {{Sehgal}}}, \bibinfo {author} {\bibfnamefont {C.}~\bibnamefont
  {{Sif{\'o}n}}}, \bibinfo {author} {\bibfnamefont {S.~M.}\ \bibnamefont
  {{Simon}}}, \bibinfo {author} {\bibfnamefont {D.~N.}\ \bibnamefont
  {{Spergel}}}, \bibinfo {author} {\bibfnamefont {S.~T.}\ \bibnamefont
  {{Staggs}}}, \bibinfo {author} {\bibfnamefont {E.~R.}\ \bibnamefont
  {{Storer}}}, \bibinfo {author} {\bibfnamefont {A.}~\bibnamefont {{van
  Engelen}}}, \bibinfo {author} {\bibfnamefont {J.}~\bibnamefont {{van
  Lanen}}}, \ and\ \bibinfo {author} {\bibfnamefont {E.~J.}\ \bibnamefont
  {{Wollack}}},\ }\href {\doibase 10.1103/PhysRevD.101.083527} {\bibfield
  {journal} {\bibinfo  {journal} {\prd}\ }\textbf {\bibinfo {volume} {101}},\
  \bibinfo {eid} {083527} (\bibinfo {year} {2020})},\ \Eprint
  {http://arxiv.org/abs/2001.10465} {arXiv:2001.10465 [astro-ph.CO]}
  \BibitemShut {NoStop}%
\bibitem [{\citenamefont {Bianchini}\ \emph {et~al.}(2020)\citenamefont
  {Bianchini} \emph {et~al.}}]{SPT:2020cxx}%
  \BibitemOpen
  \bibfield  {author} {\bibinfo {author} {\bibfnamefont {F.}~\bibnamefont
  {Bianchini}} \emph {et~al.} (\bibinfo {collaboration} {SPT}),\ }\href
  {\doibase 10.1103/PhysRevD.102.083504} {\bibfield  {journal} {\bibinfo
  {journal} {Phys. Rev. D}\ }\textbf {\bibinfo {volume} {102}},\ \bibinfo
  {pages} {083504} (\bibinfo {year} {2020})},\ \Eprint
  {http://arxiv.org/abs/2006.08061} {arXiv:2006.08061 [astro-ph.CO]}
  \BibitemShut {NoStop}%
\bibitem [{\citenamefont {Aiola}\ \emph {et~al.}(2022)\citenamefont {Aiola}
  \emph {et~al.}}]{CMB-HD:2022bsz}%
  \BibitemOpen
  \bibfield  {author} {\bibinfo {author} {\bibfnamefont {S.}~\bibnamefont
  {Aiola}} \emph {et~al.} (\bibinfo {collaboration} {CMB-HD}),\ }\href@noop {}
  {\  (\bibinfo {year} {2022})},\ \Eprint {http://arxiv.org/abs/2203.05728}
  {arXiv:2203.05728 [astro-ph.CO]} \BibitemShut {NoStop}%
\bibitem [{\citenamefont {Pogosian}\ \emph {et~al.}(2019)\citenamefont
  {Pogosian}, \citenamefont {Shimon}, \citenamefont {Mewes},\ and\
  \citenamefont {Keating}}]{Pogosian:2019jbt}%
  \BibitemOpen
  \bibfield  {author} {\bibinfo {author} {\bibfnamefont {L.}~\bibnamefont
  {Pogosian}}, \bibinfo {author} {\bibfnamefont {M.}~\bibnamefont {Shimon}},
  \bibinfo {author} {\bibfnamefont {M.}~\bibnamefont {Mewes}}, \ and\ \bibinfo
  {author} {\bibfnamefont {B.}~\bibnamefont {Keating}},\ }\href {\doibase
  10.1103/PhysRevD.100.023507} {\bibfield  {journal} {\bibinfo  {journal}
  {Phys. Rev. D}\ }\textbf {\bibinfo {volume} {100}},\ \bibinfo {pages}
  {023507} (\bibinfo {year} {2019})},\ \Eprint
  {http://arxiv.org/abs/1904.07855} {arXiv:1904.07855 [astro-ph.CO]}
  \BibitemShut {NoStop}%
\bibitem [{\citenamefont {Mandal}\ \emph {et~al.}(2022)\citenamefont {Mandal},
  \citenamefont {Sehgal},\ and\ \citenamefont {Namikawa}}]{Mandal:2022tqu}%
  \BibitemOpen
  \bibfield  {author} {\bibinfo {author} {\bibfnamefont {S.}~\bibnamefont
  {Mandal}}, \bibinfo {author} {\bibfnamefont {N.}~\bibnamefont {Sehgal}}, \
  and\ \bibinfo {author} {\bibfnamefont {T.}~\bibnamefont {Namikawa}},\ }\href
  {\doibase 10.1103/PhysRevD.105.063537} {\bibfield  {journal} {\bibinfo
  {journal} {Phys. Rev. D}\ }\textbf {\bibinfo {volume} {105}},\ \bibinfo
  {pages} {063537} (\bibinfo {year} {2022})},\ \Eprint
  {http://arxiv.org/abs/2201.02204} {arXiv:2201.02204 [astro-ph.CO]}
  \BibitemShut {NoStop}%
\bibitem [{\citenamefont {Gluscevic}\ \emph {et~al.}(2009)\citenamefont
  {Gluscevic}, \citenamefont {Kamionkowski},\ and\ \citenamefont
  {Cooray}}]{Gluscevic_2009}%
  \BibitemOpen
  \bibfield  {author} {\bibinfo {author} {\bibfnamefont {V.}~\bibnamefont
  {Gluscevic}}, \bibinfo {author} {\bibfnamefont {M.}~\bibnamefont
  {Kamionkowski}}, \ and\ \bibinfo {author} {\bibfnamefont {A.}~\bibnamefont
  {Cooray}},\ }\href {\doibase 10.1103/physrevd.80.023510} {\bibfield
  {journal} {\bibinfo  {journal} {Physical Review D}\ }\textbf {\bibinfo
  {volume} {80}} (\bibinfo {year} {2009}),\
  10.1103/physrevd.80.023510}\BibitemShut {NoStop}%
\bibitem [{\citenamefont {Li}\ and\ \citenamefont {Yu}(2013)}]{Li:2013vga}%
  \BibitemOpen
  \bibfield  {author} {\bibinfo {author} {\bibfnamefont {M.}~\bibnamefont
  {Li}}\ and\ \bibinfo {author} {\bibfnamefont {B.}~\bibnamefont {Yu}},\ }\href
  {\doibase 10.1088/1475-7516/2013/06/016} {\bibfield  {journal} {\bibinfo
  {journal} {JCAP}\ }\textbf {\bibinfo {volume} {06}},\ \bibinfo {pages} {016}
  (\bibinfo {year} {2013})},\ \Eprint {http://arxiv.org/abs/1303.1881}
  {arXiv:1303.1881 [astro-ph.CO]} \BibitemShut {NoStop}%
\bibitem [{\citenamefont {Cai}\ and\ \citenamefont {Guan}(2022)}]{Cai:2021zbb}%
  \BibitemOpen
  \bibfield  {author} {\bibinfo {author} {\bibfnamefont {H.}~\bibnamefont
  {Cai}}\ and\ \bibinfo {author} {\bibfnamefont {Y.}~\bibnamefont {Guan}},\
  }\href {\doibase 10.1103/PhysRevD.105.063536} {\bibfield  {journal} {\bibinfo
   {journal} {Phys. Rev. D}\ }\textbf {\bibinfo {volume} {105}},\ \bibinfo
  {pages} {063536} (\bibinfo {year} {2022})},\ \Eprint
  {http://arxiv.org/abs/2111.14199} {arXiv:2111.14199 [astro-ph.CO]}
  \BibitemShut {NoStop}%
\bibitem [{\citenamefont {Namikawa}\ \emph {et~al.}(2013)\citenamefont
  {Namikawa}, \citenamefont {Hanson},\ and\ \citenamefont
  {Takahashi}}]{Namikawa_2013}%
  \BibitemOpen
  \bibfield  {author} {\bibinfo {author} {\bibfnamefont {T.}~\bibnamefont
  {Namikawa}}, \bibinfo {author} {\bibfnamefont {D.}~\bibnamefont {Hanson}}, \
  and\ \bibinfo {author} {\bibfnamefont {R.}~\bibnamefont {Takahashi}},\ }\href
  {\doibase 10.1093/mnras/stt195} {\bibfield  {journal} {\bibinfo  {journal}
  {Monthly Notices of the Royal Astronomical Society}\ }\textbf {\bibinfo
  {volume} {431}},\ \bibinfo {pages} {609–620} (\bibinfo {year}
  {2013})}\BibitemShut {NoStop}%
\bibitem [{\citenamefont {Madhavacheril}\ \emph {et~al.}(2020)\citenamefont
  {Madhavacheril}, \citenamefont {Smith}, \citenamefont {Sherwin},\ and\
  \citenamefont {Naess}}]{Madhavacheril:2020ido}%
  \BibitemOpen
  \bibfield  {author} {\bibinfo {author} {\bibfnamefont {M.~S.}\ \bibnamefont
  {Madhavacheril}}, \bibinfo {author} {\bibfnamefont {K.~M.}\ \bibnamefont
  {Smith}}, \bibinfo {author} {\bibfnamefont {B.~D.}\ \bibnamefont {Sherwin}},
  \ and\ \bibinfo {author} {\bibfnamefont {S.}~\bibnamefont {Naess}},\
  }\href@noop {} {\  (\bibinfo {year} {2020})},\ \Eprint
  {http://arxiv.org/abs/2011.02475} {arXiv:2011.02475 [astro-ph.CO]}
  \BibitemShut {NoStop}%
\bibitem [{\citenamefont {{Naess}}\ \emph {et~al.}(2021)\citenamefont
  {{Naess}}, \citenamefont {{Madhavacheril}},\ and\ \citenamefont
  {{Hasselfield}}}]{2021ascl.soft02003N}%
  \BibitemOpen
  \bibfield  {author} {\bibinfo {author} {\bibfnamefont {S.}~\bibnamefont
  {{Naess}}}, \bibinfo {author} {\bibfnamefont {M.}~\bibnamefont
  {{Madhavacheril}}}, \ and\ \bibinfo {author} {\bibfnamefont {M.}~\bibnamefont
  {{Hasselfield}}},\ }\href@noop {} {\enquote {\bibinfo {title} {{Pixell:
  Rectangular pixel map manipulation and harmonic analysis library}},}\ }
  (\bibinfo {year} {2021}),\ \Eprint {http://arxiv.org/abs/2102.003}
  {ascl:2102.003} \BibitemShut {NoStop}%
\bibitem [{\citenamefont {{Planck Collaboration}}\ \emph
  {et~al.}(2018)\citenamefont {{Planck Collaboration}}, \citenamefont
  {Aghanim}, \citenamefont {Akrami}, \citenamefont {Ashdown}, \citenamefont
  {Aumont} \emph {et~al.}}]{Planck2018:VI:CP}%
  \BibitemOpen
  \bibfield  {author} {\bibinfo {author} {\bibnamefont {{Planck
  Collaboration}}}, \bibinfo {author} {\bibfnamefont {N.}~\bibnamefont
  {Aghanim}}, \bibinfo {author} {\bibfnamefont {Y.}~\bibnamefont {Akrami}},
  \bibinfo {author} {\bibfnamefont {M.}~\bibnamefont {Ashdown}}, \bibinfo
  {author} {\bibfnamefont {J.}~\bibnamefont {Aumont}},  \emph {et~al.},\ }\href
  {http://arxiv.org/abs/1807.06209} {\  (\bibinfo {year} {2018})},\ \Eprint
  {http://arxiv.org/abs/1807.06209} {arXiv:1807.06209} \BibitemShut {NoStop}%
\bibitem [{\citenamefont {{Pogosian}}\ \emph {et~al.}(2019)\citenamefont
  {{Pogosian}}, \citenamefont {{Shimon}}, \citenamefont {{Mewes}},\ and\
  \citenamefont {{Keating}}}]{2019PhRvD.100b3507P}%
  \BibitemOpen
  \bibfield  {author} {\bibinfo {author} {\bibfnamefont {L.}~\bibnamefont
  {{Pogosian}}}, \bibinfo {author} {\bibfnamefont {M.}~\bibnamefont
  {{Shimon}}}, \bibinfo {author} {\bibfnamefont {M.}~\bibnamefont {{Mewes}}}, \
  and\ \bibinfo {author} {\bibfnamefont {B.}~\bibnamefont {{Keating}}},\ }\href
  {\doibase 10.1103/PhysRevD.100.023507} {\bibfield  {journal} {\bibinfo
  {journal} {\prd}\ }\textbf {\bibinfo {volume} {100}},\ \bibinfo {eid}
  {023507} (\bibinfo {year} {2019})},\ \Eprint
  {http://arxiv.org/abs/1904.07855} {arXiv:1904.07855 [astro-ph.CO]}
  \BibitemShut {NoStop}%
\bibitem [{\citenamefont {{Namikawa}}(2021)}]{2021ascl.soft04021N}%
  \BibitemOpen
  \bibfield  {author} {\bibinfo {author} {\bibfnamefont {T.}~\bibnamefont
  {{Namikawa}}},\ }\href@noop {} {\enquote {\bibinfo {title} {{cmblensplus:
  Cosmic microwave background tools}},}\ }\bibinfo {howpublished} {Astrophysics
  Source Code Library, record ascl:2104.021} (\bibinfo {year} {2021}),\ \Eprint
  {http://arxiv.org/abs/2104.021} {ascl:2104.021} \BibitemShut {NoStop}%
\bibitem [{\citenamefont {Knox}(1995)}]{PhysRevD.52.4307}%
  \BibitemOpen
  \bibfield  {author} {\bibinfo {author} {\bibfnamefont {L.}~\bibnamefont
  {Knox}},\ }\href {\doibase 10.1103/PhysRevD.52.4307} {\bibfield  {journal}
  {\bibinfo  {journal} {Phys. Rev. D}\ }\textbf {\bibinfo {volume} {52}},\
  \bibinfo {pages} {4307} (\bibinfo {year} {1995})}\BibitemShut {NoStop}%
\bibitem [{\citenamefont {{Kesden}}\ \emph {et~al.}(2003)\citenamefont
  {{Kesden}}, \citenamefont {{Cooray}},\ and\ \citenamefont
  {{Kamionkowski}}}]{Kesden:2003}%
  \BibitemOpen
  \bibfield  {author} {\bibinfo {author} {\bibfnamefont {M.}~\bibnamefont
  {{Kesden}}}, \bibinfo {author} {\bibfnamefont {A.}~\bibnamefont {{Cooray}}},
  \ and\ \bibinfo {author} {\bibfnamefont {M.}~\bibnamefont {{Kamionkowski}}},\
  }\href {\doibase 10.1103/PhysRevD.67.123507} {\bibfield  {journal} {\bibinfo
  {journal} {\prd}\ }\textbf {\bibinfo {volume} {67}},\ \bibinfo {eid} {123507}
  (\bibinfo {year} {2003})},\ \Eprint {http://arxiv.org/abs/astro-ph/0302536}
  {arXiv:astro-ph/0302536 [astro-ph]} \BibitemShut {NoStop}%
\bibitem [{\citenamefont {Hanson}\ \emph {et~al.}(2011)\citenamefont {Hanson},
  \citenamefont {Challinor}, \citenamefont {Efstathiou},\ and\ \citenamefont
  {Bielewicz}}]{PhysRevD.83.043005}%
  \BibitemOpen
  \bibfield  {author} {\bibinfo {author} {\bibfnamefont {D.}~\bibnamefont
  {Hanson}}, \bibinfo {author} {\bibfnamefont {A.}~\bibnamefont {Challinor}},
  \bibinfo {author} {\bibfnamefont {G.}~\bibnamefont {Efstathiou}}, \ and\
  \bibinfo {author} {\bibfnamefont {P.}~\bibnamefont {Bielewicz}},\ }\href
  {\doibase 10.1103/PhysRevD.83.043005} {\bibfield  {journal} {\bibinfo
  {journal} {Phys. Rev. D}\ }\textbf {\bibinfo {volume} {83}},\ \bibinfo
  {pages} {043005} (\bibinfo {year} {2011})}\BibitemShut {NoStop}%
\bibitem [{\citenamefont {Namikawa}\ and\ \citenamefont
  {Takahashi}(2014{\natexlab{a}})}]{Namikawa:2013:BHE-pol}%
  \BibitemOpen
  \bibfield  {author} {\bibinfo {author} {\bibfnamefont {T.}~\bibnamefont
  {Namikawa}}\ and\ \bibinfo {author} {\bibfnamefont {R.}~\bibnamefont
  {Takahashi}},\ }\href {\doibase 10.1093/mnras/stt2290} {\bibfield  {journal}
  {\bibinfo  {journal} {Mon. Not. Roy. Astron. Soc.}\ }\textbf {\bibinfo
  {volume} {438}},\ \bibinfo {pages} {1507} (\bibinfo {year}
  {2014}{\natexlab{a}})},\ \Eprint {http://arxiv.org/abs/1310.2372}
  {arXiv:1310.2372 [astro-ph.CO]} \BibitemShut {NoStop}%
\bibitem [{\citenamefont {Osborne}\ \emph {et~al.}(2014)\citenamefont
  {Osborne}, \citenamefont {Hanson},\ and\ \citenamefont
  {Dor\'e}}]{Osborne:2013nna}%
  \BibitemOpen
  \bibfield  {author} {\bibinfo {author} {\bibfnamefont {S.~J.}\ \bibnamefont
  {Osborne}}, \bibinfo {author} {\bibfnamefont {D.}~\bibnamefont {Hanson}}, \
  and\ \bibinfo {author} {\bibfnamefont {O.}~\bibnamefont {Dor\'e}},\ }\href
  {\doibase 10.1088/1475-7516/2014/03/024} {\bibfield  {journal} {\bibinfo
  {journal} {JCAP}\ }\textbf {\bibinfo {volume} {03}},\ \bibinfo {pages} {024}
  (\bibinfo {year} {2014})},\ \Eprint {http://arxiv.org/abs/1310.7547}
  {arXiv:1310.7547 [astro-ph.CO]} \BibitemShut {NoStop}%
\bibitem [{\citenamefont {Sailer}\ \emph {et~al.}(2020)\citenamefont {Sailer},
  \citenamefont {Schaan},\ and\ \citenamefont {Ferraro}}]{Sailer:2020lal}%
  \BibitemOpen
  \bibfield  {author} {\bibinfo {author} {\bibfnamefont {N.}~\bibnamefont
  {Sailer}}, \bibinfo {author} {\bibfnamefont {E.}~\bibnamefont {Schaan}}, \
  and\ \bibinfo {author} {\bibfnamefont {S.}~\bibnamefont {Ferraro}},\ }\href
  {\doibase 10.1103/PhysRevD.102.063517} {\bibfield  {journal} {\bibinfo
  {journal} {Phys. Rev. D}\ }\textbf {\bibinfo {volume} {102}},\ \bibinfo
  {pages} {063517} (\bibinfo {year} {2020})},\ \Eprint
  {http://arxiv.org/abs/2007.04325} {arXiv:2007.04325 [astro-ph.CO]}
  \BibitemShut {NoStop}%
\bibitem [{\citenamefont {{Yadav}}\ \emph {et~al.}(2010)\citenamefont
  {{Yadav}}, \citenamefont {{Su}},\ and\ \citenamefont
  {{Zaldarriaga}}}]{Yadav:2010}%
  \BibitemOpen
  \bibfield  {author} {\bibinfo {author} {\bibfnamefont {A.~P.~S.}\
  \bibnamefont {{Yadav}}}, \bibinfo {author} {\bibfnamefont {M.}~\bibnamefont
  {{Su}}}, \ and\ \bibinfo {author} {\bibfnamefont {M.}~\bibnamefont
  {{Zaldarriaga}}},\ }\href {\doibase 10.1103/PhysRevD.81.063512} {\bibfield
  {journal} {\bibinfo  {journal} {\prd}\ }\textbf {\bibinfo {volume} {81}},\
  \bibinfo {eid} {063512} (\bibinfo {year} {2010})},\ \Eprint
  {http://arxiv.org/abs/0912.3532} {arXiv:0912.3532 [astro-ph.CO]} \BibitemShut
  {NoStop}%
\bibitem [{\citenamefont {De~Bernardis}\ \emph {et~al.}(2009)\citenamefont
  {De~Bernardis}, \citenamefont {Kitching}, \citenamefont {Heavens},\ and\
  \citenamefont {Melchiorri}}]{DeBernardis:2009di}%
  \BibitemOpen
  \bibfield  {author} {\bibinfo {author} {\bibfnamefont {F.}~\bibnamefont
  {De~Bernardis}}, \bibinfo {author} {\bibfnamefont {T.~D.}\ \bibnamefont
  {Kitching}}, \bibinfo {author} {\bibfnamefont {A.}~\bibnamefont {Heavens}}, \
  and\ \bibinfo {author} {\bibfnamefont {A.}~\bibnamefont {Melchiorri}},\
  }\href {\doibase 10.1103/PhysRevD.80.123509} {\bibfield  {journal} {\bibinfo
  {journal} {Phys. Rev. D}\ }\textbf {\bibinfo {volume} {80}},\ \bibinfo
  {pages} {123509} (\bibinfo {year} {2009})},\ \Eprint
  {http://arxiv.org/abs/0907.1917} {arXiv:0907.1917 [astro-ph.CO]} \BibitemShut
  {NoStop}%
\bibitem [{\citenamefont {Abazajian}\ \emph {et~al.}(2015)\citenamefont
  {Abazajian} \emph
  {et~al.}}]{TopicalConvenersKNAbazajianJECarlstromATLee:2013bxd}%
  \BibitemOpen
  \bibfield  {author} {\bibinfo {author} {\bibfnamefont {K.~N.}\ \bibnamefont
  {Abazajian}} \emph {et~al.} (\bibinfo {collaboration} {Topical Conveners:
  K.N. Abazajian, J.E. Carlstrom, A.T. Lee}),\ }\href {\doibase
  10.1016/j.astropartphys.2014.05.014} {\bibfield  {journal} {\bibinfo
  {journal} {Astropart. Phys.}\ }\textbf {\bibinfo {volume} {63}},\ \bibinfo
  {pages} {66} (\bibinfo {year} {2015})},\ \Eprint
  {http://arxiv.org/abs/1309.5383} {arXiv:1309.5383 [astro-ph.CO]} \BibitemShut
  {NoStop}%
\bibitem [{\citenamefont {Ade}\ \emph {et~al.}(2016)\citenamefont {Ade} \emph
  {et~al.}}]{Planck:2015mym}%
  \BibitemOpen
  \bibfield  {author} {\bibinfo {author} {\bibfnamefont {P.~A.~R.}\
  \bibnamefont {Ade}} \emph {et~al.} (\bibinfo {collaboration} {Planck}),\
  }\href {\doibase 10.1051/0004-6361/201525941} {\bibfield  {journal} {\bibinfo
   {journal} {Astron. Astrophys.}\ }\textbf {\bibinfo {volume} {594}},\
  \bibinfo {pages} {A15} (\bibinfo {year} {2016})},\ \Eprint
  {http://arxiv.org/abs/1502.01591} {arXiv:1502.01591 [astro-ph.CO]}
  \BibitemShut {NoStop}%
\bibitem [{\citenamefont {{Zonca}}\ \emph {et~al.}(2019)\citenamefont
  {{Zonca}}, \citenamefont {{Singer}}, \citenamefont {{Lenz}}, \citenamefont
  {{Reinecke}}, \citenamefont {{Rosset}}, \citenamefont {{Hivon}},\ and\
  \citenamefont {{Gorski}}}]{2019JOSS....4.1298Z}%
  \BibitemOpen
  \bibfield  {author} {\bibinfo {author} {\bibfnamefont {A.}~\bibnamefont
  {{Zonca}}}, \bibinfo {author} {\bibfnamefont {L.}~\bibnamefont {{Singer}}},
  \bibinfo {author} {\bibfnamefont {D.}~\bibnamefont {{Lenz}}}, \bibinfo
  {author} {\bibfnamefont {M.}~\bibnamefont {{Reinecke}}}, \bibinfo {author}
  {\bibfnamefont {C.}~\bibnamefont {{Rosset}}}, \bibinfo {author}
  {\bibfnamefont {E.}~\bibnamefont {{Hivon}}}, \ and\ \bibinfo {author}
  {\bibfnamefont {K.}~\bibnamefont {{Gorski}}},\ }\href {\doibase
  10.21105/joss.01298} {\bibfield  {journal} {\bibinfo  {journal} {The Journal
  of Open Source Software}\ }\textbf {\bibinfo {volume} {4}},\ \bibinfo {eid}
  {1298} (\bibinfo {year} {2019})}\BibitemShut {NoStop}%
\bibitem [{\citenamefont {Namikawa}\ and\ \citenamefont
  {Takahashi}(2014{\natexlab{b}})}]{Namikawa:2013xka}%
  \BibitemOpen
  \bibfield  {author} {\bibinfo {author} {\bibfnamefont {T.}~\bibnamefont
  {Namikawa}}\ and\ \bibinfo {author} {\bibfnamefont {R.}~\bibnamefont
  {Takahashi}},\ }\href {\doibase 10.1093/mnras/stt2290} {\bibfield  {journal}
  {\bibinfo  {journal} {Mon. Not. Roy. Astron. Soc.}\ }\textbf {\bibinfo
  {volume} {438}},\ \bibinfo {pages} {1507} (\bibinfo {year}
  {2014}{\natexlab{b}})},\ \Eprint {http://arxiv.org/abs/1310.2372}
  {arXiv:1310.2372 [astro-ph.CO]} \BibitemShut {NoStop}%
\bibitem [{\citenamefont {Amendola}(1996)}]{10.1093/mnras/283.3.983}%
  \BibitemOpen
  \bibfield  {author} {\bibinfo {author} {\bibfnamefont {L.}~\bibnamefont
  {Amendola}},\ }\href {\doibase 10.1093/mnras/283.3.983} {\bibfield  {journal}
  {\bibinfo  {journal} {Monthly Notices of the Royal Astronomical Society}\
  }\textbf {\bibinfo {volume} {283}},\ \bibinfo {pages} {983} (\bibinfo {year}
  {1996})},\ \Eprint
  {http://arxiv.org/abs/https://academic.oup.com/mnras/article-pdf/283/3/983/3305407/283-3-983.pdf}
  {https://academic.oup.com/mnras/article-pdf/283/3/983/3305407/283-3-983.pdf}
  \BibitemShut {NoStop}%
\bibitem [{\citenamefont {Regan}\ \emph {et~al.}(2010)\citenamefont {Regan},
  \citenamefont {Shellard},\ and\ \citenamefont
  {Fergusson}}]{PhysRevD.82.023520}%
  \BibitemOpen
  \bibfield  {author} {\bibinfo {author} {\bibfnamefont {D.~M.}\ \bibnamefont
  {Regan}}, \bibinfo {author} {\bibfnamefont {E.~P.~S.}\ \bibnamefont
  {Shellard}}, \ and\ \bibinfo {author} {\bibfnamefont {J.~R.}\ \bibnamefont
  {Fergusson}},\ }\href {\doibase 10.1103/PhysRevD.82.023520} {\bibfield
  {journal} {\bibinfo  {journal} {Phys. Rev. D}\ }\textbf {\bibinfo {volume}
  {82}},\ \bibinfo {pages} {023520} (\bibinfo {year} {2010})}\BibitemShut
  {NoStop}%
\end{thebibliography}%
\end{document}